\documentclass[preprint]{aastex}

\newcommand{\logg} {\log g}
\newcommand{\halpha} {H$\alpha$}
\newcommand{\hbeta} {H$\beta$}
\newcommand{\Te} {T_{\rm eff}}
\newcommand{\hm} {$\rm{H^-}$}
\newcommand{\mv} {$M_V$}
\newcommand{\htwo} {H$_2$}
\newcommand{\msun} {$M_\odot$}

\newcommand{\lsun} {$L_\odot$}
\newcommand\gta{\lower 0.5ex\hbox{$\buildrel > \over \sim\ $}} 
\newcommand\lta{\lower 0.5ex\hbox{$\buildrel < \over \sim\ $}} 
\newcommand{\nhe} {N({\rm He})/N({\rm H})}
\newcommand{\nh} {N({\rm H})/N({\rm He})}
\shortauthors{Bergeron et al.}
\shorttitle{Analysis of Cool White Dwarfs}
\begin{document}


\title{Photometric and Spectroscopic Analysis of Cool White Dwarfs\\
with Trigonometric Parallax Measurements\altaffilmark{1}}

\author{P. Bergeron\altaffilmark{2,3}}
\affil{D\'epartement de Physique, Universit\'e de Montr\'eal, C.P.~6128, 
Succ.~Centre-Ville, 
Montr\'eal, Qu\'ebec, Canada, H3C 3J7.}
\email{bergeron@astro.umontreal.ca}
\author{S.~K. Leggett\altaffilmark{2,3}}
\affil{UKIRT, Joint Astronomy Centre, 660 North A'ohoku Place, Hilo, HI 96720.}
\email{s.leggett@jach.hawaii.edu}
\and
\author{Mar\'\i a Teresa Ruiz\altaffilmark{2,3,4}}
\affil{Departamento de Astronom\'\i a, Universidad de Chile, Casilla 36-D, 
Santiago, Chile.}
\email{mtruiz@das.uchile.cl}

\altaffiltext {1} {Based partially on data obtained with the
UK Infrared Telescope operated by the Joint Astronomy Centre, Hilo, HI,
on behalf of the UK Particle Physics and Astronomy Research Council,
and with the NASA Infrared Telescope Facility operated by the Institute
for Astronomy at the University of Hawaii under contract to NASA.}

\altaffiltext {2} {Visiting Astronomer, Cerro-Tololo
Inter-American Observatory, National Optical Astronomical Observatories,
which is operated by AURA, Inc., under contract with the National
Science Foundation.}

\altaffiltext {3} {Guest observer, Kitt Peak National Observatory, National
Optical Astronomy Observatories, operated by the Association of
Universities for Research in Astronomy, Inc., under cooperative
agreement with the National Science Foundation.}

\altaffiltext {4} {Visiting Astronomer at La Silla (ESO).}

\begin{abstract}

A detailed photometric and spectroscopic analysis of cool ($\Te\ \lta
12,000$~K) white dwarf stars is presented. The sample has been drawn
from the Yale Parallax Catalog and from a proper-motion survey in the
southern hemisphere. Optical $BVRI$ and infrared $JHK$ photometry, as
well as spectroscopy at \halpha, have been secured for a sample of 152
white dwarfs. The discovery of seven new DA white dwarfs, two new DQ
white dwarfs, one new magnetic white dwarf, and three weak magnetic
white dwarf candidates, is reported. Our sample also identifies 19
known or suspected double degenerates.

The photometric energy distributions, the \halpha\ line profiles, and
the trigonometric parallax measurements are combined and compared
against the predictions of model atmosphere calculations to determine
the effective temperature and the radius of each object in the sample,
and also to constrain the atmospheric composition. New evolutionary
sequences with carbon/oxygen cores with thin and thick hydrogen layers
are used to derive stellar masses and ages. The results are used to
improve our understanding of the chemical evolution of cool white
dwarfs.

We confirm the existence of a range in effective temperature between
$\sim$5000 and 6000~K where almost all white dwarfs have hydrogen-rich
atmospheres. Our sample shows little evidence for mixed H/He white
dwarfs, with the exception of two helium-rich DA stars, and four
(possibly five) C$_2$H white dwarfs which have been interpreted as
having mixed H/He/C atmospheres. The observed sequence of DQ stars is
found to terminate abruptly near 6500~K, below which they are believed
to turn into C$_2$H stars. True DC stars slightly above this
temperature are found to exhibit hydrogen-like energy distributions
despite the lack of \halpha\ absorption features. The mean mass of our
complete sample is 0.65 \msun\ with a dispersion of $\sigma\sim0.20$
\msun.

Attempts to interpret the chemical evolution of cool white dwarfs show
the problem to be complex. Convective mixing is called upon to account
for the increase of the non-DA to DA ratio below 12,000~K, as well as
the reappearance of helium-rich stars below $\sim 5000$~K. The
possible presence of helium in cool DA stars, the existence of the
non-DA gap, and the nature of the peculiar DC stars are also explained
in terms of convective mixing, although our understanding of how this
mechanism works needs to be revised in order to account for these
observations. Given this chemical evolution uncertainty, it is not
clear whether thick or thin hydrogen layer models should be used to
determine cooling ages. The oldest object in our sample is
$\sim$7.9~Gyr or $\sim$9.7~Gyr old depending on whether thin or thick
hydrogen layer models are used, respectively.

\end{abstract}


\section{Introduction}

Cool white dwarfs have recently gained attention due to the discovery
of faint ($V \sim 26-28.5$) objects in the Hubble Deep Field Survey
\citep[see][and references therein]{hansen98} which have been interpreted by
\citet{hansen98, hansen99} as being cool ($\Te\ \lta 3000$~K) and old 
($t\ \gta 11$ Gyr) hydrogen atmosphere white dwarfs. Hansen has shown
that the location of these objects {\it blueward} of the white dwarf
cooling sequence in the \mv\ vs $(V-I)$ color-magnitude diagram is the
result of the extremely strong \htwo\ molecular absorption features
which extend well into the optical regions of their energy
distributions. Hence, even though hydrogen-rich white dwarfs become
redder with decreasing effective temperature down to $\Te\sim 4000$~K
\citep[see, e.g., Fig.~6 of][]{bwb95}, even cooler stars actually become bluer.
Note that the most recent photometric and spectroscopic analysis of
cool white dwarfs in the local Galactic disk \citep[hereafter
BRL]{brl} has shown no evidence of white dwarfs cooler than 4000~K.

In addition, \citet{hansen99} has also discussed the importance of
using detailed stellar atmospheres as boundary conditions for
calculating cooling white dwarf models. For instance, the cooling age
of a 0.6 \msun, $\Te\sim 5000$~K white dwarf can differ by $\sim 1.5$
Gyr depending on whether the object has a hydrogen- or a
helium-dominated atmosphere. Given this sensitivity, it is important
to determine precisely the atmospheric composition of cool white
dwarfs before any attempt to determine their age is made.

To complicate things even more, white dwarfs are not expected to
preserve their hydrogen or helium atmospheric signature throughout
their evolution. Indeed, there is both theoretical and observational
evidence which indicates that the chemical composition of white dwarfs
is likely to change many times during their lifetime due to competing
mechanisms (e.g., gravitational settling, convective mixing,
convective dredge-up, accretion from the interstellar medium,
radiative acceleration) which may affect the composition of the outer
layers of these stars as they cool  \citep[see][for a review on
the subject]{fw87}. This is even more important for cool white dwarfs where
convective mixing of a thin ($M_{\rm H}/M_\star\ \lta10^{-6}$)
hydrogen atmosphere with the more massive convective helium envelope
can turn a hydrogen-rich atmosphere into a helium-dominated atmosphere
anywhere between 12,000~K and 5000~K. The exact temperature at which
convective mixing occurs --- or whether it will occur at all --- depends
on the thickness of the hydrogen layer (the thicker the hydrogen
layer, the cooler the mixing temperature). Furthermore, BRL have
clearly demonstrated that the chemical evolution of cool white dwarfs
is even more complex than previously anticipated, and that several
mechanisms --- some of them still unknown --- are at play to alter the
chemical constituent of their atmospheres as they cool.
Consequently, the age of any cool white dwarf is not only
strongly dependent on its atmospheric composition, as pointed out by
\citet{hansen99}, {\it but also on its chemical evolution
history}. For instance, a recently mixed helium-rich atmosphere white
dwarf would have an integrated cooling age of its immediate
predecessor --- i.e.~that of a hydrogen-rich atmosphere white
dwarf. Stellar ages and chemical evolution can thus not be easily
disentangled. This unfortunate consequence stresses the importance of
better understanding the chemical evolution of cool white dwarf stars.

For several years now, we have been pursuing our comprehensive
investigation of cool white dwarf evolution. We presented in BRL
photometric and spectroscopic observations of 110 cool white dwarfs,
and analyzed them with model atmosphere calculations appropriate for
cool white dwarfs with pure hydrogen and pure helium compositions, as
well as with mixed H/He compositions. Our fitting technique consists
of comparing the observed energy distributions in the optical $BVRI$
and infrared $JHK$ with those predicted from model atmospheres in
order to determine the effective temperature of the star. When available,
trigonometric parallax measurements are used to derive the radius of
the star, and hence the mass through the mass-radius relation for
white dwarfs. Finally, high signal-to-noise spectroscopy at \halpha\
is used to determine, or at least constrain, the chemical composition
of their atmospheres. The main results of the BRL analysis are
summarized here together with our current understanding of the
physical models which have been proposed to account for the
observations.

Most cool white dwarfs have energy distributions which are consistent
with either pure hydrogen or pure helium atmospheric compositions,
with little evidence for objects with mixed H/He compositions. In
general, the energy distributions of DA (hydrogen-line) stars are
consistent with pure hydrogen atmospheres, while those of non-DA stars
are consistent with pure helium atmospheres. Exceptions include a few
extremely cool non-DA stars with hydrogen-dominated atmospheres which
are simply too cool to exhibit any hydrogen absorption features, the
so-called C$_2$H stars whose peculiar energy distributions and strong
molecular absorption bands have been interpreted as the result of
mixed H/He/C compositions with $\nh \sim 10^{-2}$ \citep{schmidt95}, and
a few other objects which are most likely unresolved binaries (see BRL
and references therein). Note that hydrogen lines are observed in DA
stars as cool as $\Te\sim 4500$~K, that is, almost at the end of the
observable white dwarf cooling sequence of the local Galactic disk.

BRL have also shown that the temperature distribution of DA and non-DA
stars is not homogeneous. In particular, even though non-DA stars are
abundant above $\Te\sim 6000$~K, most white dwarfs below this
temperature, down to $\Te\sim 5000$~K, are of the DA type (the
so-called ``non-DA gap''). Moreover, BRL have identified a group of
objects which lie close to, and above, $\Te\sim 6000$~K whose energy
distributions are better reproduced with pure hydrogen models, even
though their optical spectra are completely featureless; hydrogen
atmosphere white dwarfs in this particular range of temperature should
exhibit strong hydrogen lines. BRL proposed an exotic model to account
for both the strange nature of these objects and the existence of the
non-DA gap.  A warmer, pure helium atmosphere white
dwarf slowly accretes hydrogen from the interstellar medium as
it cools. Most atomic levels of the hydrogen accreted are
quenched as a result of the high atmospheric pressures which
characterize cool, helium-rich atmospheres. In particular, the bound
level of \hm\ is destroyed, and the corresponding bound-free
\hm\ opacity --- the most important opacity source in hydrogen-rich
atmospheres --- is significantly reduced. Consequently, the accretion
of hydrogen onto a helium-dominated atmosphere would have little effect on
the opacity and thus on the atmospheric pressure. BRL proposed that
large amounts of hydrogen could be {\it hidden} through this simple
accretion/quenching mechanism. They also suggested that at some point,
the free-free \hm\ opacity would become large enough for the
atmospheric pressure to drop below the threshold value at which the
\hm\ ion starts to recombine. The star then rapidly turns into a
normal hydrogen-rich white dwarf. BRL suggested that the group of
peculiar white dwarfs discussed above could represent such objects
with a mixture of helium and quenched hydrogen, in the process of turning
into normal DA stars, and that the non-DA gap could be accounted for
by the sudden transformation of apparently helium-rich atmosphere
white dwarfs into hydrogen-dominated atmospheres.

More recently, \citet{malo99} have explored more quantitatively the
model proposed by BRL. The results of \citeauthor{malo99} indicate
that the accretion of even minute amounts of hydrogen onto a
helium-dominated atmosphere, $\log \nh\ \gta 10^{-6}$, provide enough
free electrons --- through the thermal ionization of hydrogen --- to
increase the He$^-$ opacity significantly, which in turn, reduces the
atmospheric pressure below which quenching effects on hydrogen become
negligible. The presence of hydrogen therefore has an {\it indirect}
effect on the total opacity through the free electrons it provides to
the plasma. On the basis of their analysis, \citeauthor{malo99}
rejected the model proposed by BRL. Given these results, the nature of
the peculiar objects identified by BRL and the existence of the non-DA
gap remain unexplained.

\citet{hansen99} provided an alternative explanation for the non-DA gap.
The gap is evident in his Figure 13 even though he restricted his
sample of objects to that of \citet{ldm88} rather than using the more
complete sample studied by BRL, where the gap is even more obvious.
Hansen has shown that the concentration of non-DA stars just below
$\Te\sim 5000$~K have nearly the same age as the concentration of DA
stars around 6000~K. On the basis of this conclusion alone, Hansen
suggested that the non-DA gap alluded to by BRL is only a result of
the difference of cooling timescales between hydrogen-dominated and
helium-dominated atmospheres. However, this interpretation cannot be
correct since it neglects the fact that non-DA stars are also observed
in large numbers {\it above} the gap at $\Te\ge 6000$~K.  In other
words, non-DA stars are seen both above and below the gap, and the
high-temperature end of the gap does not reflect a simple termination
of the cooling sequence of degenerates with helium-dominated
atmospheres.  Unless helium-rich white dwarfs evolve more rapidly {\it
through} the gap than above or below the gap, which is not the case,
an explanation based solely on differences in cooling rates does not
hold. Obviously, a chemical evolution from one spectral type to
another must be invoked.

Other results from the BRL analysis support the idea that some
physical mechanism, yet unidentified, is at play near the blue edge of
the non-DA gap. For instance, the most widely accepted explanation for
the presence of carbon in the photospheric layers of DQ stars is that
carbon is convectively dredged-up from the core, as demonstrated
quantitatively by \citet{pelletier86}. These calculations show that
the extent of the carbon diffusion tail is reduced at low effective
temperature due to the partial recombination of carbon, and smaller
carbon abundances are predicted in DQ stars with lower effective
temperature, in apparent agreement with the observations
\citep[see Fig.~4 of][]{pelletier86}. As pointed out by BRL, however, the
coolest DQ stars have the strongest observed C$_2$ molecular bands,
and cooler objects with even smaller abundances are well within the
observational limits and should be easily detectable. The BRL analysis
has revealed instead the existence of an abrupt cutoff in effective
temperature at $\Te\sim 6500$~K below which DQ stars are no longer
found. The presence of carbon in stars below this temperature
threshold is in the form of what has been interpreted as C$_2$H
molecular bands by \citet{schmidt95}. The most obvious explanation
proposed by BRL is that DQ stars turn into C$_2$H stars below $\Te\sim
6500$~K. The mechanism proposed by BRL by which the hydrogen required
to form the C$_2$H molecule suddenly appears in these objects relies
on their composition transition model, which was later discarded by
\citet{malo99}. Hence, even though the DQ to C$_2$H transition remains
the most likely explanation for the existence of the abrupt cutoff of
the DQ star phenomenon, the physical mechanism by which this
transition occurs needs further investigation.

BRL have also shown that below the non-DA gap near $\Te\sim 5000$~K,
non-DA stars reappear in large number. The most likely explanation for
this phenomenon is the convective mixing of the relatively thin
surface hydrogen layer with the deeper and more massive helium
envelope. Detailed envelope calculations indicate that the bottom of
the hydrogen convection layer reaches its deepest point at a certain
temperature \citep[see, e.g.,][]{tassoul90}; white dwarfs with
hydrogen layers thicker than this maximum depth point can never mix
and will remain hydrogen-rich forever. The canonical value of the
lowest possible mixing temperature is $T_{\rm mix}\sim6000$~K, which
corresponds to a hydrogen layer mass of $M_{\rm
H}/M_\star\sim10^{-6}$. This is $\sim 1000$~K hotter than the
temperature required to account for the BRL observations. New envelope
models discussed by BRL which include realistic model atmospheres as
boundary conditions reduce the mixing temperature limit well down to
5000~K, however. Hence, convective mixing may well represent the
correct interpretation for the reappearance of non-DA stars below the
non-DA gap.  The continued presence of DA stars requires that at least
some stars have hydrogen layers thicker than $M_{\rm
H}/M_\star\sim10^{-6}$.

Following the BRL analysis, \citet{lrb98} have gathered new
photometric and spectroscopic data for the 43 cool, low-luminosity,
high--proper-motion white dwarfs in the \citet{ldm88} sample.  This
sample was analyzed with the techniques developed by BRL in order to
improve the determination of the cool white dwarf luminosity function
and the estimate of the age of the local Galactic disk. BRL discuss at
length the shortcomings of the previous estimates by
\citet{ldm88}, and in particular the lack of precise atmospheric 
parameter determinations as well as accurate bolometric corrections
for the stars in the sample. The improved luminosity function of
\citet{lrb98} yields an age of the local region of the Galactic disk
of $8\pm1.5$ Gyr, with the uncertainty being mainly due to the core
composition, the effect of separation on crystallization, and the
unknown mass of the outer helium envelope.  The \citet{ldm88} sample
is being extended to a lower proper-motion limit and a new luminosity
function is being determined which supports the earlier work
\citep{ldhl99}.

As mentioned in BRL, these earlier efforts represent only the first
step towards a better understanding of the chemical evolution of cool
white dwarfs. Even though the BRL analysis has provided us with a
wealth of new observational and theoretical results, many questions
have remained unanswered, and the problems to be solved have
multiplied. The success of our analysis has prompted us to extend our
study to a larger and more complete sample of cool white dwarfs. We
thus present in this paper a detailed analysis of all cool ($\Te\ \lta
12,000$~K) white dwarfs with trigonometric parallax measurements
available to us. Such measurements are of particular importance to
derive stellar radii, and hence stellar masses. This new sample is
comprised of 152 white dwarfs for which masses can be derived. In
contrast, only 67 out of the 110 white dwarfs analyzed by BRL had
trigonometric parallax measurements. Furthermore, BRL concentrated
their effort towards cool ($\Te\ \lta 7000$~K), non-DA white dwarfs,
while our new sample is distributed over a wider range of effective
temperature and spectral types.

In \S\ 2, we discuss the improvements in the model atmospheres and
cooling models since our most recent analysis. In \S\ 3, we present
our new photometric and spectroscopic observations, which, combined
with our earlier observations, are analyzed in \S\ 4. The global
properties of our sample are examined in \S\ 5. Our conclusions follow
in \S\ 6.

\section{Theoretical Framework}\label{sbsc:thfr}

\subsection{Model Atmospheres}

The model atmospheres used in this analysis are described at length in
\citet{bsw95}, and improvements to these models are discussed in BRL
and below. These models are appropriate for cool white dwarfs with
pure hydrogen or pure helium atmospheres, as well as mixed H/He
compositions.

Our current model grid now includes self-consistently the
$\rm{He_2^+}$ ion in the ionization equilibrium calculations. As
discussed in detail by \citet{malo99}, the $\rm{He_2^+}$ ion was
treated as a trace element in our earlier calculations. Although this
assumption is valid in terms of the number of particles, $N({\rm
He_2^+})<< N({\rm He})$, the presence of the $\rm{He_2^+}$ ion affects
the ionization equilibrium significantly for temperatures below
$\Te\sim8500$~K. In cool, pure helium (or helium-rich) models,
the electron density is governed by He$_2^+$, rather than He$^+$ (or
\hm; see \citeauthor{malo99}). Hence, despite the fact that the continuous
absorption opacity of He$_2^+$ is negligible at low temperatures, the
inclusion of this ion in model atmospheres has repercussion on the
He$^-$ free-free opacity --- the dominant opacity source in
helium-rich models --- through its effect on the electron density.

The occupation probability of \citet{hm88} for the calculation of the
hydrogen populations has also been included more accurately in our
current model grid following the work of \citet{malo99}. This
formalism allows us to calculate cool white dwarf models with
arbitrarily low hydrogen abundances. Indeed, the photospheric
pressures which characterize low hydrogen abundance atmospheres are so
high that the hydrogen atomic levels (including that of the \hm\ ion)
are strongly perturbed. In the most extreme cases, pressure ionization
of hydrogen may even occur (we note that even though the pressure
ionization model used in the Hummer-Mihalas formalism is very crude,
it is sufficient for our current needs). Since the Hummer-Mihalas
formalism had not been included in our earlier investigations, BRL
were unable to compute cool models with $\nh\ \lta10^{-3}$ (see \S~5.3
of BRL). We note that the nonideal effects included in the
equation-of-state of our pure helium model grid \citep[see][and
references therein]{bsw95} are neglected in the mixed H/He models.

Our latest hydrogen (and mixed hydrogen/helium) models include the new
\htwo\ collision-induced opacity calculations of \citet{borysow97}, as 
well as the additional term discussed by \citet{saumon99} which takes
into account collisions involving three \htwo\ molecules. Our hydrogen
model grid has also been extended to $\Te=1500$~K, the coolest models
calculated by \citet{saumon99}. The non-ideal effects in the
equation-of-state introduced by Saumon and Jacobson are neglected
here, however. A more detailed discussion of these cooler models will be
presented elsewhere. 

The hydrogen and mixed hydrogen/helium models have been calculated
with the so-called ML2/$\alpha=0.6$ parameterization of the
mixing-length theory (as opposed to the ML1 calibration used in BRL),
following the suggestion of \citet{bergeronzz95} from their analysis
of ZZ Ceti stars. Although this will not affect the atmospheric
parameter determinations of the cooler objects in our sample ($\Te\
\lta 8000$~K), the emergent energy distributions of warmer models are
quite sensitive to the convective efficiency used in the calculations,
as shown by \citet{bergeronzz95}. We note finally that this new model
grid has not changed significantly the quality of the fits shown in
BRL, or the determination of the atmospheric parameters from our
previous analyses.

\subsection{On the Importance of Metals in Cool, Helium-Rich Atmospheres}

Recently, \citet{jorgensen} have studied the importance of including
metals in the atmospheres of cool, helium-rich white dwarfs. Their
results indicate that the model atmospheres and predicted energy
distributions are quite sensitive to the presence of additional
elements. They also claim that their improved models reproduce the
scatter observed in two-color diagrams better than the models used in
BRL. Here, we present our own assessment of this problem. A more
complete appraisal of the J\o rgensen et al. results will be presented
elsewhere \citep{bergeron2001}.

The main effect of including heavy elements in the model atmospheres
of cool, helium-rich white dwarfs is to increase the number of free
electrons in the plasma, which in turn affect the He$^-$ free-free opacity
considerably. As mentioned by
\citet{jorgensen}, more than 99\% of the electrons may come from
metals which have a relatively low ionization potential. As an
experiment, we calculated a pure helium model at $\Te=6000$~K,
$\logg=8$, but with an electron density artificially increased by a
factor of 1000 (only 0.1\% of the electrons come from helium directly,
i.e. the value quoted by J\o rgensen et al.). The temperature and
pressure structures of this model are compared in Figure
\ref{fg:comp_tp} with those of a standard pure helium model. As can be seen,
the temperature structure of the upper layers has changed
significantly, but that of the deeper layers where the {\it continuum
is formed} ($\tau_{\rm R}\sim1$) has remained relatively
unaffected. The gas pressure, on the other hand, has been reduced
quite dramatically throughout the atmosphere as a result of the
increase in the opacity ($P_g \propto \kappa^{-1}$), which in this
case is dominated by the He$^-$ free-free opacity. Perhaps of greater
interest is the comparison of the predicted fluxes from these two
models, displayed in Figure
\ref{fg:compflux}. Not surprisingly, the results indicate that the
observed differences are small. The reason for this is that in the
pure helium model, the He$^-$ free-free opacity already dominates at
all wavelengths where the continuum is formed. An increase in the
electron density thus increases the opacity by the same factor {\it at
all wavelengths}. Consequently, the emergent flux cannot be
redistributed any differently. The small differences observed in
Figure \ref{fg:compflux} can be measured more quantitatively by
fitting our test model with our pure helium grid. We find that the
predicted energy distribution of our test model can be fit to better
than 1\% at all wavelengths with our pure helium grid at
$\Te=6037$~K. We thus conclude that the presence of metals in
helium-rich models does not affect significantly the atmospheric
parameters derived in the BRL analysis and in this analysis. Note that
the presence of metallic lines or molecular absorption features may
affect the predicted broadband fluxes, but we do not believe that they
can affect the model structure significantly, with the exception of
the most extreme cases such as those discussed in \S~\ref{sbsc:optir}.

There are several reasons why the conclusions of \citet{jorgensen} are
different from those reached here. First they state that white dwarfs
are ``carbon rich''. This is simply not true as only a small fraction
of cool non-DA stars exhibit carbon features. It is also a well known
observational fact that the presence of carbon and metallic features
are mutually exclusive (for reasons still unknown). The authors have
also shown in their Figure 3 that the inclusion of molecules in their
model produces a substantial heating at all layers. However, this
statement is somewhat misleading since their plot shows the
temperature profile as a function of gas pressure, and as these two
parameters are not independent in a model atmosphere, it is not
possible from such a plot to determine which of the parameters
(temperature or gas pressure) has in fact changed. A more appropriate
independent parameter is the Rosseland mean optical depth, as used
here in Figure \ref{fg:comp_tp}. J\o rgensen et al.~also claim that
the new \htwo-He collision-induced opacity they have calculated ---
and not used in our own model calculations --- has a significant
influence on the predicted emergent fluxes. However, their Figure 6
shows instead that the differences are actually quite small at the
$\Te$ values considered here, and not pertinent to the present
discussion. More importantly, they have assumed a non-zero hydrogen
abundance of $\nh=10^{-5}$ on the basis of either old studies, or
analyses of objects that do show direct or indirect traces of hydrogen
(e.g., Ross 640, LHS 1126). Traces of hydrogen in these cool models
produce a strong infrared flux deficiency (see their Fig.~6) which is
simply not observed in cool non-DA stars studied by us (BRL and this
analysis), with the glaring exception of LHS 1126
\citep{bergeron94} near $\Te\sim5500$~K whose infrared flux deficiency
has been attributed to the \htwo-He molecular opacity. Surprisingly
enough, \citet{jorgensen} conclude that at this particular
temperature, the \htwo-He opacity has little effect. This is
because they have considered hydrogen abundances which are way too
small in this case --- LHS 1126 has $\nh\sim10^{-2}$. The claim by
J\o rgensen et al.~that the predicted energy distribution is not sensitive
to the helium-to-hydrogen abundance ratio is thus also inaccurate.

Their Figures 8 and 9 show their helium-rich models, as well as some
of their hydrogen-rich models, in a ($B$--$V$, $V$--$K$) two-color
diagram, together with the BRL observations. Also shown for comparison
are the hydrogen-rich and helium-rich models used in BRL. The authors
conclude that both their hydrogen-rich and helium-rich models
reproduce the observed photometric data better than the models used in
BRL.  However, the authors failed to distinguish the DA and non-DA
stars in their Figures, as was carefully done in Figure 9 of BRL. A
more careful examination of both results indicates that their
helium-rich models actually fail to match the non-DA stars close to
$B$--$V\sim0.5$, $V$--$K\sim 0.9$ altogether, while our pure helium
models go straight through the observed data. Their Figure 8 shows
instead that their helium-rich models reproduce nicely the sequence of
{\it DA stars} which have hydrogen-rich atmospheres. The authors also
claim that at the warm end of the observed sequence, their models
``follow better the tendency of a linear observed sequence that the
observations indicate, than do the BRL models which diverge toward
higher values of $B$--$V$''. Again, the authors failed to realize that
the warm end of the sequence in BRL is composed exclusively of non-DA
stars. The fact is that DA stars {\it do diverge} toward higher values
of $B$--$V$, as do the models (see our Figure \ref{fg:bmv:vmk}
below). If anything, their hydrogen-rich models fail to reproduce the
warm end of the observed DA sequence. 

Finally and more importantly,
\citet{jorgensen} claim that their helium-rich models reproduce the
scatter in the observational data seen at the cool end of diagram
better than do the BRL models which follow an almost straight
line. First of all, most of the observed scatter is due to the fact
that the coolest objects include {\it both hydrogen-rich and
helium-rich white dwarfs}. Indeed, when one considers only the
helium-rich stars (the filled circles in the right panel of Fig.~9 of
BRL), they do form an almost straight line, as opposed to the 
J\o rgensen et al.~models. Second, these authors show that their models
can reproduce the entire cool end of the diagram by changing the C/O
ratio. This is again somewhat misleading. Indeed, the observed
``hook'' in their predicted colors is entirely due to the
\htwo-He collision-induced opacity. Similar hooks can be observed in
the BRL models as well, whether they are due to \htwo-\htwo\ or
\htwo-He molecular opacities (see Fig.~9 of BRL). Since these 
collision-induced opacities are strongly dependent on the atmospheric
pressure, the predicted colors will be sensitive to surface gravity
(see the hydrogen sequences with various values of $\logg$ in Fig.~9
of BRL, middle panel), or to the hydrogen-to-helium abundance (see the
helium-rich sequences with various values of $\nhe$ in Fig.~9 of BRL,
right panel). The results shown in Figure 8 of \citet{jorgensen}
simply reflect a change in the opacity, and thus atmospheric pressure,
due to a change in the electron density for various values of
C/O. More problematic is the effective temperature near 3000~K
inferred from Figures 8 and 9 of \citet{jorgensen} for the bulk of the
coolest white dwarfs. At this temperature, the predicted energy
distribution shows a strong infrared flux deficiency which is simply
not observed in these objects, as discussed above. This is the danger
of interpreting cool white dwarfs in two-color diagrams only, and in
particular the ($B$--$V$, $V$--$K$) diagram (see \S~5.2.2 of BRL),
rather than fitting the entire energy distribution, which is the
technique favored by BRL and in this analysis.

\subsection{Evolutionary Cooling Sequences}

In order to derive stellar masses from the measured radii, and to
determine cooling ages, we must rely on detailed evolutionary models.
In BRL, we made use of the models of \citet{wood90} for the pure
helium and mixed hydrogen/helium compositions, and those of
\citet{wood95} for the pure hydrogen models. The former models
have carbon-core compositions, helium layers of $q({\rm He})\equiv
M_{\rm He}/M_{\star}=10^{-4}$, and no hydrogen layers, while the
latter have carbon-core compositions as well, but helium layers of
$q({\rm He})=10^{-2}$, and thick hydrogen layers of $q({\rm
H})=10^{-4}$. 

In this paper, we use instead two families of new evolutionary models
kindly computed for us by G. Fontaine who used the new evolutionary
code developed by P. Brassard at Universit\'e de Montr\'eal on the
basis of finite-element techniques. While details of that code will be
described elsewhere \citep{bf00}, a few remarks about it are worthy of
interest. The code is especially designed to follow in a robust and
efficient way moving discontinuities in a cooling white dwarf model,
namely, the top and base of superficial convection zones, and the
advancing crystallization front in the core and in the envelope. The
full structure of a model, from the center to the top of the
atmosphere (typically located at $\tau_R$ = $10^{-8}$), is included in
the evolutionary calculations. Upgraded input physics, including the
new model atmospheres described above, has been incorporated in the
code in order to be able to compute realistic white dwarf models
reaching the regime of very low effective temperatures that is of
interest for discussing white dwarf populations in the halo and in old
globular clusters.

The first family of evolutionary structures that we use are standard
``thick envelope'' stellar models with a helium mantle of $q({\rm
He})=10^{-2}$, and an outermost hydrogen layer of $q({\rm
H})=10^{-4}$. The core of these models consists of a uniform mixture
of carbon and oxygen in equal proportions ($X_{\rm C} = X_{\rm O} = 0.5$). No
sedimentation between carbon and oxygen upon crystallization has been
taken into account in these models. Sedimentation would add a small
delay to the cooling age, but would not change the mass-radius
relation in any significant way. On the other hand, the effects of
convective feedback on the position of the hydrogen convection zone
have been taken into account in the calculations. Likewise, neutrino
cooling has been included in a routine way, but this only affects the
evolution in the high luminosity phases only, as is well known. No
residual thermonuclear burning has been included in order to keep the
same value of the fractional mass of the outer hydrogen layer in all
models. Finally, 16 different masses in the range 0.2--1.3 \msun\ have
been considered.

The second family of models (with again 16 different sequences) is
similar to the first one, except for the assumed stratification of the
envelope. This second batch of structures are now ``thin envelope''
stellar models with a helium mantle of $q({\rm He})=10^{-2}$ as
before, but with an outermost hydrogen layer of only $q({\rm
H})=10^{-10}$. No convective mixing between the hydrogen and helium
layers is allowed in these models. We picked this particular
stratification as representative of ``thin'' hydrogen envelope models.
The small layer of hydrogen atop these models does not change in any
significant way the mass-radius relation for the cooler models that
are of interest here. From that point of view, that same mass-radius
relation can thus be used for non-DA stars. Evolutionary calculations
of helium-rich atmospheres can also be carried out, but the
cooling times show an extreme sensitivity to the presence of even very
small traces of heavy elements in the atmosphere/envelope, leading to
a continuum of ages, hence we adopted the current strategy.

\section{Observations}

\subsection{Selection of the Sample}

In the BRL study, 110 genuine cool white dwarfs were selected on the
basis of their spectral type classification from the catalog of
\citet{mccook87}, from a proper-motion survey in the southern
hemisphere \citep{ruiz93, ruiz95, ruiz96}, and from the study of
\citet{monet92}. This first sample was strongly biased against DA 
stars in order to determine whether true DC stars existed, and if they
did, what was their atmospheric composition. Our high signal-to-noise
spectroscopy actually revealed the presence of \halpha\ in 20 of the
previously classified non-DA stars, a result which stresses the
importance of large aperture telescopes for studying cool white
dwarfs, as first emphasized by \citet{greenstein86}. Several
additional DA, DQ, and DZ stars were included in the sample as well to
test the validity of our model atmosphere calculations, and to improve
our understanding of the chemical evolution of cool white dwarfs.

BRL found that even though the model energy distributions are
somewhat sensitive to the surface gravity, it is practically
impossible to determine $\logg$ from the observed photometry
alone. Only for stars with available trigonometric parallax
measurements is it possible to determine the stellar radius, and thus
the mass through the mass-radius relation. In the BRL sample, 67 white
dwarfs had trigonometric parallax measurements taken from the Yale
parallax catalog \citep[][hereafter YPC]{ypc}, and from
\citet{brl92}, \citet{ruiz89}, \citet{ruizetal95}, and
\citet{anguita96}.  Stellar masses were determined for a subset of 60
white dwarfs (6 objects had unreliable parallaxes, while one DQ star
had carbon bands too strong to be analyzed properly). BRL assumed a
surface gravity of $\logg=8.0$ for the remaining objects in their
sample. The natural step following the comprehensive study of BRL was
to gather spectroscopic and photometric data for all cool white dwarfs
with measured trigonometric parallaxes available in the literature.

A total of 161 DA and non-DA white dwarfs with a temperature index
($\theta = 50,400/\Te$) of 4 and cooler were thus selected from the
YPC, and from \citet{ruiz96}. This temperature cutoff corresponds
roughly to that below which convective mixing of DA stars is believed
to occur. Only stars with parallax uncertainties smaller than 30\%
were considered; 7 objects were rejected on this basis (0145$-$174,
1042$+$593, 1639$+$153, 1708$-$147, 2139$+$132, 2240$-$017, and
2254$+$076). Two other objects were not considered since one is a
known DC + M dwarf system (1133$+$358), while the other one has a
known spectroscopic temperature in excess of 20,000~K
(1919$+$145). Hence, 152 white dwarf candidates were retained for
photometric and spectroscopic observations. Among this sample, 67
objects are in common with BRL, and new photometric and spectroscopic
data are presented in this paper for 85 white dwarfs.

Since this new sample is based on trigonometric parallax measurements,
it should contain no bias against any specific spectral type, in
contrast with the BRL analysis which favored non-DA white dwarfs. The
number of DA and non-DA stars as a function of $\Te$ for our
trigonometric parallax sample is displayed in Figure
\ref{fg:DAvsnonDA}. Also shown is the subset of objects in the BRL
sample whose stellar masses could be determined from trigonometric
parallax measurements --- $M(\pi)$. The temperature estimates are
obtained from the detailed fits to the photometric energy
distributions (\S\ 5 of BRL and \S\ \ref{sbsc:adopt} below; 60 stars
in the BRL sample and 150 stars in this analysis), while the spectral
types are determined from our high signal-to-noise spectroscopic
observations. The comparison indicates that our new sample has
significantly increased the number of DA stars above $\Te=5000$~K and
the number of non-DA stars at all temperatures with respect to the BRL
sample. Also apparent from this figure is the paucity of non-DA stars
in the 5000-6000~K temperature range. Note that the coolest bin of the
non-DA panel may contain a significant number of hydrogen-rich white
dwarfs that are simply too cool to exhibit an \halpha\ feature.

\subsection{Optical Spectroscopy}

High signal-to-noise spectroscopy has been obtained for all 152
objects in our sample, 15 of which have been secured by
\citet{greenstein86}. Some of the spectroscopic observations have
already been presented in BRL and in \citet{lrb98}. The new spectra
reported here have been obtained during several observing runs using
the KPNO and CTIO 4 m telescopes equipped with the RC spectrograph
(1996, 1998; where the years indicate the
period the observations were obtained), and the ESO 3.6 m telescope
equipped with the EFOSC (1996, 1998). The spectral resolution varies
between 5 and 15 \AA\ (18
\AA\ for the Greenstein spectra). Details of our observing procedure
and data reduction are provided in \citet{brl92}. The spectral
coverage varies according to the detector used although all spectra
cover at least the \halpha\ region.

The new spectra --- not already shown in BRL --- which show \halpha\ are
displayed in Figure \ref{fg:plotspecDA} in order of decreasing
equivalent widths. The names of the 7 newly identified DA stars are
italicized in Figure
\ref{fg:plotspecDA}. We note that the coolest DA stars in our sample
are all new DA identifications, a result which stresses once again the
importance of large aperture telescopes for the spectral
classification of cool white dwarfs \citep[BRL,][]{greenstein86}.

Objects of particular interest in Figure \ref{fg:plotspecDA} include
two magnetic white dwarfs with discernible Zeeman splitting: GD 175
(1503$-$070) discovered in this survey, and G128$-$72 (2329+267) first
identified by \citet{moran98}. There are also two objects with
particularly broad and shallow profiles: the DZA stars Ross 640
(1626$+$368) and the DZQA star L745$-$46A (0738$-$172). Both of these
objects have helium-dominated atmospheres \citep[][this
analysis]{liebert77, koester96, koester00}, with their \halpha\ line profile
broadened by van der Waals interactions with neutral helium; we note
that such objects are rare in our sample. The
\halpha\ line profile of the G107$-$70A/B system (0727$+$482A/B) is also
broad and shallow with respect to other cool objects. In this case,
however, since the A and B components of the system are not resolved,
it is more likely that we are dealing with a normal DA star whose line
profile is diluted by a featureless white dwarf companion (see
\S\ \ref{sbsc:addobj}, however).  The central component of the Zeeman
triplet in emission is also clearly visible in the spectrum of GD 356
\citep[1639$+$537; see also][]{greenstein85}, while the $\pi$
components are outside the wavelength range displayed here. Finally,
the coolest DA star shown in Figure \ref{fg:plotspecDA} is G227$-$28
(1820$+$609) with an effective temperature of 4780~K. This is still
500~K hotter than the coolest DA star in our complete sample, LHS 239
(0747+073B), with $\Te=4210$~K (see BRL and \S\ \ref{sbsc:adopt} below).

Our non-DA spectra --- not already shown in BRL --- are displayed in
Figure \ref{fg:plotspecnonDA}. As noted in BRL, a featureless white
dwarf spectrum does not necessarily imply a helium-rich atmosphere
since a hydrogen-rich white dwarf may be too cool to show an \halpha\
absorption feature, and a full analysis of the energy distribution is
required to establish properly the atmospheric composition of these
objects.

The subset of DQ stars with blue spectral coverage is displayed in
Figure \ref{fg:plotspecDQ} in order of decreasing effective
temperature (as determined in \S\ \ref{sbsc:adopt}). These are all
well known DQ stars with the exception of two new identifications, LHS
2392 (1115$-$029) and G184$-$12 (1831$+$197). As discussed in BRL and
above, the convective dredge-up model of \citet{pelletier86} predicts
a continuous decrease of carbon abundances with decreasing effective
temperature. As seen from Figure \ref{fg:plotspecDQ}, however, the
coolest DQ stars have the strongest carbon features, and therefore,
the sudden disappearance of the DQ phenomenon at even cooler
temperatures needs to be attributed to an additional effect.

\subsection{Optical and Infrared Photometry}\label{sbsc:optir}

Optical $BVRI$ and infrared $JHK$ photometry has been secured for 150
and 148 program objects, respectively, out of our complete sample of
152 white dwarfs. Since some optical photometry is required for
fitting the energy distribution, $BV$ photometry for the 2 missing
objects 0326$-$273 and 0839$-$327 has been taken from
\citet{mccook99}.

Most of the data for the 67 objects in common with BRL were acquired
between 1991 August and 1996 February, but we have supplemented that
data with some newer photometry. Additional $BVRI$ photometry
presented here has been obtained with the 0.9 m telescopes at KPNO
(1996, 1998) and CTIO (1995), while the new $JHK$ photometry has been
obtained with the CTIO Blanco 4 m telescope using CIRIM (1995), the
NSFCAM camera on the NASA IRTF (1996), the IRCAM on UKIRT (1996,
1998), and UKIRT service time (1997 to 1999).  Details of our
observing procedure and data reduction are provided in \citet{brl92}.

The $BVRI$ and $JHK$ photometry is presented in Table 1 together with
the number of independent observations ($N$). Also given in Table 1
are the WD number and the name of each object, the measured
trigonometric parallax and uncertainty, and the \halpha\ equivalent
width ($W$). A value of $W=0$ implies a featureless spectrum near the
\halpha\ region; for the magnetic DA stars, the equivalent width is
taken as the sum of the three Zeeman components. The photometric
uncertainties are typically 3\% at $V$, $R$, and $I$, and 5\%
elsewhere, with the exception of the data marked ``:'' which indicates
a 10\% uncertainty.  The optical and infrared photometry is on the
Cousins system
\citep{bessell87} and the CIT system \citep{elias82}, respectively.

The optical and infrared photometry of the A and B components of the
G107$-$70 (0727$+$482) system is based on our combined optical
photometry and the $\Delta V=0.30$ mag difference reported by
\citet{ldm88} and references therein, as well as on our own magnitude
difference measurement of $0.3\pm0.1$ mag at $J$, $H$, and $K$. Note
that the spectrum of G107$-$70A/B displayed in Figure
\ref{fg:plotspecDA} is the {\it combined} spectrum of both components.

The \mv\ vs ($V$--$I$) color-magnitude diagram is displayed in
Figure \ref{fg:Mv:vmi} for 149 stars; 3 objects in Table 1 have no
($V$--$I$) measurement. The left panel shows all objects together with
the error measurements associated with the trigonometric parallax
measurements. In the middle and right panels, the same objects are
distinguished in terms of their DA or non-DA spectral types. The pure
hydrogen and pure helium cooling sequences, calculated as in
\citet{bwb95} but with our latest model grid and cooling models 
discussed in \S\ \ref{sbsc:thfr}, are superimposed on the observed
data.

DA and non-DA stars form well-defined narrow sequences in this
diagram, although not as narrow as the USNO data set of
\citet{monet92}, also shown in \citet{bwb95}, which is a much more
homogeneous parallax sample than the one studied here. Interestingly,
all overluminous stars with apparent masses below 0.4 \msun\ are of
the DA spectral type, with the exception of LP 31$-$40 (0324$+$738)
which is probably too cool to show \halpha\ anyway. As discussed in
BRL, most, if not all, of these objects are unresolved binaries
\citep[e.g. L870$-$2,][]{saffer88} and their luminosity is the
contribution of two white dwarfs with probably normal masses. The
color-magnitude diagram presented in Figure
\ref{fg:Mv:vmi} indicates that such systems are mostly composed of DA
white dwarfs, a result which suggests a particular evolutionary
path. We will come back to this point below (see \S\ \ref{sbsc:mdist}).

A particularly interesting object in Figure \ref{fg:Mv:vmi} is the
non-DA star ESO 439$-$26 (1136$-$286) analyzed in detail by
\citet{ruizetal95}. This object with its absolute visual magnitude of
$M_V=17.4$ was first interpreted as an extremely old --- and thus cool
--- white dwarf. However, the \citeauthor{ruizetal95} analysis convincingly
demonstrated that the low luminosity of this object could be
attributed to its small radius (or high mass). The most recent
parameters derived for this object come from the BRL analysis,
$\Te=4490$~K and $M=1.20$ \msun. However, as first demonstrated by
\citet{hansen98}, white dwarfs with $\Te\ \lta4000$~K become
increasingly {\it bluer}. An inspection of Figure \ref{fg:Mv:vmi} now
suggests an alternate interpretation for ESO 439$-$26, namely that
this star could be a $\Te\sim3200$~K, $\sim0.8$
\msun\ white dwarf; such a low temperature would imply an age of roughly 
13 Gyr, i.e.~twice as old as the age determined by BRL for this
object. The detailed analysis of the entire energy distribution
presented below will show that the infrared $J$ magnitude
is totally inconsistent with this last interpretation, and that the
earlier conclusions reached by \citet{ruizetal95} are still valid.

If the overluminous DA stars and ESO 439$-$26 are excluded, all objects in 
Figure \ref{fg:Mv:vmi} have masses in the range 0.4 \lta $M/$\msun\ \lta 1.0. 
If we assume that most DA stars have hydrogen-rich atmospheres, as
demonstrated by BRL, their mean mass inferred from Figure
\ref{fg:Mv:vmi} is somewhere around 0.7 \msun. Similarly, if we assume
that most non-DA stars have helium-rich atmospheres (with the
exception of the hydrogen-rich white dwarfs which are too cool to show
\halpha), their mean mass inferred from Figure \ref{fg:Mv:vmi} is also
somewhere near 0.7 \msun. Both of these average masses are about 0.1
\msun\ higher than those inferred from the spectroscopic analyses of
warmer DA and DB stars \citep{bsl92, bergeronzz95, beauchamp96}.

Under similar assumptions, the coolest DA and non-DA stars in Figure
\ref{fg:Mv:vmi} have temperatures near $\Te\sim4500$~K based on their location 
in this diagram with respect to the pure hydrogen and pure helium
cooling sequences, respectively. As mentioned above, however, the
coolest non-DA stars may well have a hydrogen-rich atmospheric
composition, and only a complete analysis of the combined energy
distribution will yield reliable temperature estimates. As can be seen
from Figure \ref{fg:Mv:vmi}, such an uncertainty in the atmospheric
composition could easily result in a $\sim 500$~K temperature
uncertainty for the coolest non-DA white dwarfs in our sample.

The ($B$--$V$, $V$--$K$) two-color diagram is displayed in Figure
\ref{fg:bmv:vmk} for 145 objects; 2 objects in Table 1 have no
($B$--$V$) measurement while 5 have no $K$ measurement. DA and non-DA
stars form two distinct sequences which overlap near ($V$--$K$)$\sim
0.4$. This behavior is observed in the model sequences as well. We
note in particular that both the observed and predicted DA sequences
tend to diverge toward higher values of $B$--$V$ for $V$--$K<0.4$, in
contrast with the models of \citet{jorgensen} which follow instead the
non-DA sequence in this range of $V$--$K$. Even though the observed DA
sequence is well reproduced by the 0.6 \msun\ pure hydrogen models at
the hot end of the sample, an increasing departure is observed above
($V$--$K$)$\sim 1.3$. In these regions the cool observed DA sequence
is better reproduced by the pure helium models! BRL proposed an
explanation for this departure in terms of a missing opacity source
near the $B$ filter in the pure hydrogen models, most likely due to a
pseudocontinuum opacity originating from the Lyman edge (see \S\ 5.2.2
of BRL for a complete description). Figure 13 of BRL shows indeed how
the agreement between the observations and the model predictions could
be improved by including, even approximately, such an opacity in the
model calculations. However, the lack of a more precise theoretical
framework precludes us from including this opacity in our current
model grid, and until this can be accomplished, we must refrain from
interpreting such ($B$--$V$, $V$--$K$) diagrams for DA stars any
further.

On the other hand, the non-DA observed sequence in Figure
\ref{fg:bmv:vmk} and the pure helium models are in much better
agreement. Non-DA stars, most of which have helium-rich atmospheres,
are characterized by more transparent atmospheres than hydrogen-rich
atmospheres, and as a result, their broadband colors are particularly
sensitive to the presence of strong metallic lines or molecular
bands. Consequently, several non-DA stars in Figure \ref{fg:bmv:vmk}
lie well outside the ``mean'' observed sequence. This is the case for
LP 701$-$29 (2251$-$070) whose spectrum exhibits a strong drop in the
flux shortward of $\sim4400$
\AA\ attributed to Ca I and II lines \citep{kapranidis86}, LHS 1126
(0038$-$226) which shows a strong infrared flux deficiency attributed
to \htwo-He collision-induced absorptions \citep{bergeron94}, G165$-$7
(1328$+$307) a heavily blanketed white dwarf with strong metallic
lines \citep{wehrse80}, G240$-$72 (1748$+$708) whose spectrum is
characterized by a ``yellow sag'' $\sim 15$\% deep and $\sim 2000$
\AA\ wide \citep[see, e.g., Fig.~19 of][]{wesemael93}, and the strong
DQ star BPM 27606 (2154$-$512) whose spectrum is displayed in Figure
\ref{fg:plotspecDQ}. 

Because of the theoretical problems outlined above with the pure
hydrogen models, BRL preferred to rely on the ($V$--$I$, $V$--$K$)
two-color diagram in which a much better agreement between theory and
observation could be achieved. Such a diagram is displayed in Figure
\ref{fg:vmi:vmk} for 144 objects; 3 objects in Table 1
have no ($B$--$V$) measurement while 5 have no $K$ measurement. The
results show that both DA and non-DA stars form narrow cooling
sequences, but that they are offset by $\sim 0.1$ mag. The outliers
are all of the non-DA type, some of which have been identified in the
right panel and already discussed above. G148-B4B (1215$+$323) has
most likely its $I$ magnitude contaminated by the close M dwarf
companion, while G195$-$19 (0912$+$536) is a $\sim 100$ MG magnetic
white dwarf with some undetermined spectroscopic features
\citep{schmidt94}. The non-DA stars LP 131$-$66 (1247$+$550) and ER 8
(1310$-$472) actually have hydrogen-rich atmospheres (see BRL and the
detailed analysis below).

Also shown in the DA and non-DA panels are the model sequences for
pure hydrogen, pure helium, and mixed He/H atmospheres. The departure
of the pure hydrogen and mixed He/H models from the pure helium models
is due to the collision-induced absorption by molecular hydrogen due
to collisions with \htwo\ and neutral helium, respectively. Because of
these strong infrared absorption features, the ($V$--$I$, $V$--$K$)
two-color diagram is particularly useful for detecting extremely cool
hydrogen-rich white dwarfs, as well as cool white dwarfs with mixed
hydrogen and helium atmospheres. As can be seen from Figure \ref{fg:vmi:vmk},
such objects are extremely rare, with the glaring exception of LHS
1126 which has already been analyzed in BRL ($\Te=5390$~K, $\nhe\sim
70$; see their Fig.~31).

Finally, the existence of a non-DA gap in the 5000-6000~K temperature
range is already obvious in this diagram, especially when the non-DA
sequence is contrasted with the DA sequence. The only two objects
inside the gap are LHS 2710 (1313$-$198) and ESO 292$-$43 (2345$-$447)
which have already been analyzed by BRL and identified as being
peculiar (see \S\ 6.3.1 of BRL for more details). It is again unlikely
that the explanation proposed by \citet{hansen99}, namely that
different evolutionary timescales of hydrogen-rich and helium-rich
atmosphere white dwarfs, can account for the {\it observational} fact
shown in the right panel of Figure \ref{fg:vmi:vmk}, unless non-DA
stars evolve more rapidly within the gap than outside the gap.

\section{Detailed Analysis}

\subsection{Fitting Technique}\label{sbsc:fittec}

Our fitting technique is described at length in BRL. Briefly,
magnitudes are converted into observed fluxes, and the resulting
energy distributions are fitted with those predicted from our model
atmosphere calculations using a nonlinear least-squares method. Only
$\Te$ and the solid angle $(R/D)^2$ are considered free
parameters. The distance $D$ is obtained from the trigonometric
parallax measurement, and the stellar radius $R$ is converted into
mass using the new cooling sequences described in \S\ \ref{sbsc:thfr}
with thin hydrogen layers for the pure helium and mixed
hydrogen/helium compositions, and those with thick hydrogen layers for
the pure hydrogen models; masses derived from thin hydrogen layer are
about 0.03 \msun\ smaller than those obtained from thick hydrogen
models \citep{bergeronzz95}.

As discussed in \S\ 3.3, there is a significant UV flux deficiency in
hydrogen-rich stars, most likely due to a pseudocontinuum opacity
originating from the Lyman edge. As such, the $B$ magnitude has been
omitted in the fits of hydrogen-rich stars cooler than
$\Te\sim5500$~K. For identical reasons, it has also been omitted in
the fits of strong DZ stars such as vMa 2 (0046$+$051), G165$-$7
(1328$+$307), LP 701$-$29 (2251$-$070), etc.

Some objects in our sample possess strong molecular absorption
features which are not taken into account in our model flux
calculations. Such objects include the DQ stars, the C$_2$H stars --- LHS
1126 (0038$-$226), LHS 290 (1043$-$188), and G225$-$68 (1633$+$572),
as well as G240$-$72 (1748$+$708) whose spectrum is characterized by
the yellow sag discussed above. BRL have shown that the fits to the
energy distribution of DQ stars are not affected significantly by the
presence of the C$_2$ Swan bands, with the glaring exception of BPM
27606 (2154$-$512), the strongest DQ star known. For some of the
strongest C$_2$H stars and for G240$-$72, the molecular absorption
features are so important that the quality of the fits is poor. In
these cases, we rely on a technique used in BRL where the optical
spectrum of the star is normalized to a continuum set to unity, and
folded with the grid of model spectra. The presence of the strong
molecular bands is thus taken into account, at least in first order,
in the calculations of the emergent flux distributions. This
approximate procedure improves the quality of our fits markedly.

\subsection{Sample Fits}

Several fits to cool white dwarfs with various effective temperatures,
surface gravities, and chemical compositions have already been
displayed in BRL. In particular, their \S\S~5.3 and 5.4 show how the
combined photometric energy distributions together with the high
signal-to-noise spectra at \halpha\ could be used to determine, or at
least constrain, the chemical composition of cool degenerates. The
most important discriminant between the hydrogen-rich and helium-rich
energy distributions lies near the infrared $H$ bandpass where
hydrogen-dominated atmospheres exhibit a local maximum in their
emergent flux due to the bound-free H$^-$ opacity threshold at 1.6
\micron. This bump in the emergent flux distribution is of course not observed 
in pure helium atmospheres. The chemical composition effects on our
fits will not be repeated here, and we simply refer the reader to
Figures 15 and 16 of BRL where several illustrative examples are
displayed. Here we present instead typical fits to hydrogen-rich stars,
helium-rich stars, stars with mixed compositions, confirmed or suspected
double degenerates, as well as other objects of astrophysical
interest. We attempt as much as possible to present fits for objects
which have not already been analyzed in BRL. In the following plots,
the observed fluxes and corresponding uncertainties are represented by
error bars while the model fluxes are shown as filled circles. The
atmospheric parameters of each fit are indicated in each panel.

\subsubsection{Hydrogen-rich Atmosphere White Dwarfs}\label{sbsc:hrich}

Sample fits for DA stars covering the effective temperature range of
interest are displayed in Figure \ref{fg:sampleDA}; sample fits for
hydrogen-rich non-DA white dwarfs too cool to exhibit \halpha\ have
already been shown in Figure 26 of BRL. The values of $\Te$ and
$\logg$ given in each panel are obtained from fits with pure hydrogen
atmospheric compositions. The effects of using models with small
traces of helium are discussed in detail in \S\ 5.3 of BRL.  The
spectroscopic observations at \halpha\ are not used directly in the
fitting procedure, but they serve as an internal check of our
photometric solutions. The theoretical line profiles are simply
interpolated at the values of $\Te$ and $\logg$ obtained from the
energy distribution fits and compared with the observed line profiles.
As can be seen from Figure \ref{fg:sampleDA}, the observed energy
distributions are well reproduced by our theoretical models
throughout the temperature range of interest, and the predicted line
profiles are in excellent agreement with the solution derived
from the photometric observations.

LTT 4816 (1236$-$495; top object in Fig.~\ref{fg:sampleDA}) is a known
ZZ Ceti star for which \citet{bergeronzz95} obtained a spectroscopic
solution of $\Te=11,730\pm350$~K and $\logg=8.81\pm0.05$, in good
agreement with the parameters derived here $\Te=11,550\pm470$~K and
$\logg=8.63\pm0.18$. It is worth emphasizing that the atmospheric
parameter determination of both analyses is based on completely
different techniques. It the former case, \citet{bergeronzz95} used
fits to the Balmer line profiles (H$\beta$ to H8), while here we rely
on $\Te$ and $\logg$ estimates based on photometric observations and
trigonometric parallax measurements. Note that both analyses use the
same calibration of the mixing-length theory in the model
calculations, namely the so-called ML2/$\alpha=0.6$
parameterization. Interestingly enough, our photometric fit to LTT 4816 based on
ML2/$\alpha=1.0$ models yields $\Te=11,410$~K and $\logg=8.64$ (with
the same uncertainties as above), while the spectroscopic fit to the
Balmer lines based on the same models yields $\Te=12,670$~K and
$\logg=8.66$ \citep[see Table 2 of][]{bergeronzz95}. Hence, even
though the photometric solution is not very sensitive to the value of
$\alpha$ (i.e.~the ratio of the mixing length to the pressure scale
height), the spectroscopic solution differs by 940~K in $\Te$ and 0.15
dex in $\logg$. Moreover, the internal consistency between the
spectroscopic and photometric solutions is much better with
ML2/$\alpha=0.6$ than with ML2/$\alpha=1.0$ models. A similar
conclusion was reached by \citet{bergeronzz95} based on a comparison
of the atmospheric parameters of 22 ZZ Ceti stars obtained from fits
to the ultraviolet energy distributions and fits to the Balmer line
profiles.  Our result thus brings additional support to the conclusion
that the ML2/$\alpha=0.6$ calibration of the mixing-length theory
provides an excellent internal consistency between atmospheric
parameters determined with various techniques over a wide range of
wavelengths, which now extends from the ultraviolet to the infrared.

A similar example is G74$-$7 (0208$+$396; middle panel of
Fig.~\ref{fg:sampleDA}), a well-studied DAZ white dwarf which has been
re-analyzed recently by \citet{billeres97} who presented an analysis
of optical and ultraviolet spectroscopic observations. Their
spectroscopic fit to the Balmer lines (see their Fig.~4) yields
$\Te=7260\pm40$~K and $\logg=8.03\pm0.07$, in excellent agreement with
our photometric solution $\Te=7310\pm180$~K and
$\logg=8.01\pm0.09$. More comparisons between photometric and
spectroscopic solutions are provided in \S\ \ref{sbsc:compspec}.

The coolest object in Figure \ref{fg:sampleDA} is LHS 1801 (0551$+$468)
with $\Te=5380$~K. The fit to its energy distribution suffers from the
pathological problem discussed above for the coolest hydrogen-rich
stars in our sample in which some unknown pseudocontinuum opacity
unaccounted for in our model calculations affect the emergent flux
near the $B$ bandpass. Note that this problem does not occur for LHS
2522 (1208$+$576; also shown in Fig.~\ref{fg:sampleDA}), a DA star
which is only 500~K hotter than LHS 1801. Hence, the phenomenon begins
to manifest itself in the intermediate temperature regime near
$\Te\sim5500$~K. As discussed in \S~\ref{sbsc:fittec}, the $B$ filter
has been dropped in all fits to hydrogen-rich stars below this
temperature threshold. As shown in Figure \ref{fg:sampleDA}, the
internal consistency achieved in this manner of the fit to the
energy distribution and the \halpha\ line profile is more
than satisfactory.

\subsubsection{Helium-rich Atmosphere White Dwarfs}\label{sbsc:herich}

Sample fits for helium-rich, non-DA stars covering the effective
temperature range of interest are displayed in Figure
\ref{fg:samplenonDA}. The values of $\Te$ and $\logg$ given in each
panel are obtained from fits with pure helium atmospheric
compositions. The spectroscopic fits are of course not shown here
since all objects are featureless near the \halpha\ region (see Figure
\ref{fg:plotspecnonDA}). Pure helium models provide excellent fits to
the energy distributions of all of these objects in the $\Te$ range
considered in this analysis, with the exception of G24$-$9
(2011$+$065) which is discussed below. In particular, at low effective
temperatures (see, e.g., LHS 542), the problem observed in
hydrogen-rich stars at the blue end of the energy distribution is not
observed in helium-rich stars. This supports our conclusion that some
unaccounted UV opacity from hydrogen is at the origin of this problem.

The first object in Figure \ref{fg:samplenonDA} is LP 257$-$28
(0802$+$386), a hot DZ star with a blue spectrum showing strong H and
K lines \citep{sion90}. LHS 43 (1142$-$645) is a weak DQ star (see
Figure \ref{fg:plotspecDQ}). BRL have discussed at length the validity
of the pure helium models to analyze DZ and DQ stars despite the
presence of strong metallic lines or carbon bands. LHS 2333
(1055$-$072) is classified as a DA white dwarf in
\citet{mccook99}, although a note added in their catalog reports the
star to be featureless according to
\citet{schmidtsmith95}. Our own spectroscopic observation which covers
the range $4600-7300$ \AA\ confirms the results of Schmidt \& Smith
that LHS 2333 has a featureless spectrum --- a true DC star. BRL have
shown that DC stars in this temperature range ($8000\ \gta\Te\
\gta 6000$~K) usually belong to a group of peculiar objects whose energy 
distributions are better reproduced with pure hydrogen models rather
than with pure helium models (for reasons still unknown). A good
example of such a peculiar white dwarf is the next object shown in
Figure \ref{fg:samplenonDA}, G24$-$9 (2011$+$065),  where the fits with 
both pure helium and pure hydrogen models are displayed. LHS 2333 seems 
to be an exception to the rule, unless there is something even more peculiar
with this particular object. It is tempting to suggest that LHS
2333 may have recently changed spectral type from DA to
DC. However, despite the three DA spectral classifications of this
object in the catalog of \citet{mccook99}, only that of
\citet{eggen65} is an actual spectroscopic observation (the other two
references in the catalog do not report any new observation). Given
that this classification relies on old photographic plates, it is
difficult to claim that the spectral type of this object has actually
changed over the 30 year or so period.

Finally, the last object in Figure \ref{fg:samplenonDA}, LHS 542
(2316$-$064), is a cool DC white dwarf according to our own spectrum
which reveals no significant absorption feature in the range
$\sim$3780-6890 \AA.

\subsubsection{Mixed H/He Atmosphere White Dwarfs}

There is little evidence for a large population of cool white dwarfs
with mixed hydrogen and helium atmospheres, as first demonstrated by
BRL. Such stars would show up easily in ($B$--$V$, $V$--$K$)
or ($V$--$I$, $V$--$K$) two-color diagrams due to the \htwo--He
collision induced opacity which produces a strong infrared flux
deficiency. The C$_2$H stars also have energy
distributions characterized by such an infrared flux deficiency,
the most striking example of which is LHS 1126 labeled in the right
panel of Figure \ref{fg:vmi:vmk}.  As mentioned above, such stars have 
been interpreted as having mixed H/He/C atmospheres with $\nh \sim 10^{-2}$
\citep{schmidt95}. \citet{jorgensen}
have recently cast some doubt on the Schmidt et al.~interpretation
based of their own analysis of the chemical abundance equilibrium of a
mixture of H/He/C/O and heavier elements. However they have explored
atmospheric parameters which do not match those inferred for C2H
stars: hydrogen abundances about three orders of magnitude too small
and effective temperatures much cooler ($\Te=4300$~K) than the narrow
range of $\Te\sim5400$ -- 6200 K where such stars are found.

The presence of trace amounts of hydrogen can be deduced from the
presence of a broad and shallow \halpha\ absorption feature in an
otherwise helium-rich atmosphere due to the van der Waals broadening
of the hydrogen lines by neutral helium. This is the case for Ross 640
(1626$+$368) and L745$-$46A (0738$-$172) whose fits are displayed in
Figure \ref{fg:sampleHe}. Also shown in this Figure is the fit to LHS
290 (1043$-$188), a C$_2$H star whose optical spectrum is shown in
Figure 30 of BRL but for which we had not obtained any photometry at
the time of writing the BRL analysis (see Fig.~31 of BRL for fits to
additional C$_2$H stars). In Figure
\ref{fg:sampleHe}, the effective temperature and surface gravity of
each object is derived from the fits to the energy distribution, while
the hydrogen-to-helium abundance ratio is determined from fits to the
\halpha\ line profile; only a limit to the abundance can be obtained
in the case of LHS 290. The effective temperatures and abundances we
derive for Ross 640 and L745$-$46A are in excellent agreement with
those determined by \citet{koester00} and references therein.

\subsubsection{Unresolved Double Degenerates}

Overluminous objects in the \mv\ vs ($V$--$I$) color-magnitude
diagram displayed in Figure \ref{fg:Mv:vmi} are most likely unresolved
binaries composed of two normal white dwarfs. The masses inferred for
these objects in a color-magnitude diagram are well below 0.4
\msun; note that two unresolved 0.6 \msun\ white dwarfs would appear in 
such a diagram as a single $\sim 0.3$ \msun\ white
dwarf. Alternatively, it is also possible that the overluminous
objects in Figure \ref{fg:Mv:vmi} correspond to single low-mass ($M <
0.4$ \msun) white dwarf stars.

Sample fits for five of those overluminous objects are displayed in
Figure \ref{fg:sampledoubleDA}. As mentioned above, all of these are
of the DA spectral type, and the surface gravities determined by
fitting the solid angle are well below the average for normal DA
stars. Otherwise, the fits to the energy distributions do not reveal
any additional peculiarity, and in particular, they are all consistent
with single white dwarfs. If we were dealing with double degenerates
with much different effective temperatures and luminosities, one could
expect the combined energy distribution to be somewhat distorted. The
results of Figure \ref{fg:sampledoubleDA} suggest that either these
are single low-mass white dwarfs, or double degenerates with
comparable atmospheric parameters.

The evidence that we are actually dealing with unresolved double
degenerates stems from the fits to the \halpha\ line profiles also
shown in Figure \ref{fg:sampledoubleDA}. In the case of a single
low-mass star, the fit to the \halpha\ line profile would not show any
discrepancy. Those shown here, however, are certainly at odds with the
atmospheric parameters obtained from the energy distributions
(contrast the fits shown in Figure \ref{fg:sampledoubleDA} with those
shown in Figure \ref{fg:sampleDA}). The only marginal case for which
the theoretical line profile provides a satisfactory fit is L870$-$2
(0135$-$052), a well-known double degenerate system discovered by
\citet{saffer88}! \citet{bergeron90} have actually determined that the 
effective temperatures of both components of the L870$-$2 system
differ by only 500~K ($\Te=6920$ and 7470~K), a result which explains
why the combined \halpha\ line profile can be reproduced by a single
spectrum with an intermediate effective temperature.  For the other
stars in Figure \ref{fg:sampledoubleDA}, the line profile fit
discrepancies are much more obvious. G21$-$15 (1824$+$040) is also a
known double degenerate with a period of 6.266 days
\citep{saffer98, maxted99}. \citet{maxted99} have reported a null 
result for G1$-$45 (0101$+$048) in their radial velocity survey, however.
Nothing is known about the
two remaining objects in Figure \ref{fg:sampledoubleDA}.

\subsubsection{Magnetic White Dwarfs}

Our sample includes magnetic white dwarfs, several of which
have been discovered and analyzed in the course of our survey (see BRL
and references therein). Here we present additional fits to magnetic
white dwarfs as well as magnetic candidates.

Figure \ref{fg:figMAG} shows our best fits to the magnetic white
dwarfs G128$-$72 (2329$+$267) and GD 175 (1503$-$070), both of which
exhibit Zeeman splitting at \halpha. As discussed above, the magnetic
nature of G128$-$72 has been reported by \citet{moran98} while GD 175
represents a new discovery. Here the effective temperatures and
surface gravities have been taken from our best fits to the energy
distributions. The synthetic magnetic spectra are calculated using
offset dipole models following the prescription outlined in
\citet{brl92}. Both the value of the dipole field strength, $B_d$, and
the dipole offset, $a_z$, are considered free parameters, while the
viewing angle $i$ between the dipole axis and the line of sight is
determined from a visual inspection of the fits. For G128$-$72, our
best fit is achieved with $B_d=2.2$ MG, $a_z=0.1$, and $i=50^{\circ}$,
in good agreement with the values derived by \citet{moran98},
$B_d=2.31\pm0.59$ MG and $i=60\pm5^{\circ}$, under the assumption of a
centered dipole model. Note that the shape of the shifted Zeeman
components are particularly useful to determine the viewing angle, as
shown for instance in Figure 4 of \citet{brl92}.

GD 175 is an obvious case of an unresolved double degenerate binary
containing a magnetic DA component and a featureless DC component,
similar to the G62$-$46 system analyzed by \citet[][see also the
case of LHS 2273 shown in Fig.~33 of BRL]{brl93}. In such systems, the
\halpha\ line profile predicted under the assumption of a single 
magnetic DA star is much stronger than the observed line profile, as
shown in Figure \ref{fg:figMAG} for GD 175 (dotted line). The simplest
explanation for this discrepancy is that GD 175 is an unresolved
double degenerate binary system composed of a magnetic DA star whose
emergent flux is diluted by the presence of a featureless DC
companion. The thick solid line in the bottom panel represents our
best solution with the assumption that this DC white dwarf
contributes equally to the total continuum flux near \halpha. We do
not attempt here to derive the atmospheric parameters for the
individual components of the system, as carried out for G62$-$46 by
\citet{brl93}. Instead we note that if two identical --- but different
chemical compositions --- white dwarfs are contributing to the total
flux, the mass of the magnetic DA component is near 1 \msun, as
opposed to $M\sim 0.7$ \msun\ when a single star is assumed.

Cool DA white dwarfs with much weaker magnetic fields will not show
any Zeeman splitting, at least not at the spectroscopic resolution used
here. A good example of such an object is the $\sim35$ kG white dwarf
LHS 1038 (0009$+$501) reported by \citet{schmidt94} and analyzed
photometrically by BRL (see their Fig.~33). The spectroscopic fit at
\halpha\ using non-magnetic models is reproduced in Figure 
\ref{fg:sampleMAG}. For all objects in this figure,
the atmospheric parameters are those obtained from the energy
distributions. As mentioned in BRL, the theoretical line profile
for LHS 1038 is predicted somewhat too deep, as a result of the
magnetic splitting of the \halpha\ line core into its unresolved
Zeeman components.  A similar object has also been reported by BRL
(see their Fig.~30), LHS 5064 (0257$+$080), and its fit at \halpha\ is
reproduced in Figure \ref{fg:sampleMAG} as well. BRL made the
suggestion that LHS 5064 was also a weakly magnetic ($\lta 100$ kG) DA
star. Recently, \citet{schmidt98} observed this object using the
technique of Zeeman spectropolarimetry and indeed detected a weak
($\sim 30$ kG) magnetic field at \halpha. Three additional objects in
our survey exhibit similar features --- LHS 2800 (1344$+$106),
G138$-$47 (1635$+$137), and G156$-$64 (2253$-$081), and the fits at
\halpha\ are also shown in Figure \ref{fg:sampleMAG}. \citet{schmidt98} has
reported a null result on all three of these objects, however. It is
always possible that an unfortunate orientation of the magnetic axis
prevents the detection of circular polarization. Given the resemblance
of these fits, it would be worth observing these stars at high
spectral resolution to detect any weak Zeeman splitting. Alternative
explanations for these three peculiar fits to the \halpha\ line
profile include the presence of helium in their
atmosphere. Experiments with mixed H/He models, not shown here,
indicate that by the time the predicted depth of the \halpha\ line
{\it core} agrees with the observed profile, the line {\it wings} are
predicted much too broad. Another possibility is the presence of an
unresolved DC white dwarf which would dilute the absorption profile of
the DA star. However, the trigonometric parallax measurements of these
objects make them subluminous in an absolute magnitude-color diagram,
implying larger than average masses, and inconsistent with these stars
being unresolved degenerate binaries as they would then be
superluminous. Finally, the remaining object displayed in Figure
\ref{fg:sampleMAG} is LHS 3501 (1953$-$011) recently reported to be
magnetic by \citet[][and references therein]{maxted00}, with a field
strength of $\sim 500$ kG although the Zeeman triplet at \halpha\
indicates a field strength of only $\sim 100$ kG. The line profile is
also found to be variable over a time-scale of a day or less. Our line
fit shown here appears to be consistent with a non-magnetic DA star,
however. It is possible that our spectroscopic observation was secured
during a rotational phase when Zeeman splitting was the least
apparent.

Figure \ref{fg:figG227-35} shows our best fit to the energy
distribution of the magnetic white dwarf G227$-$35 (1829$+$547). This
star has been analyzed in detail by \citet{putney95} who convincingly
demonstrated that this object is a strongly magnetic $\sim 117$ MG
{\it hydrogen-rich} white dwarf. Their spectrum shows a prominent
feature in circular polarization with a weak corresponding absorption
feature which corresponds to the 2$p$-1--3$d$-2 transition of
\halpha. However, as shown in Figure \ref{fg:figG227-35}, the observed
energy distribution is reproduced much better with a pure helium
atmosphere rather than a pure hydrogen atmosphere! The most likely
explanation in this case is that the H$^-$ continuum opacity in pure
hydrogen magnetic models is sufficently disrupted in the presence of
such a high magnetic field that the corresponding energy distribution
resembles instead that of a pure helium model. 

\subsubsection{Additional Objects of Astrophysical Interest}\label{sbsc:addobj}

{\it ESO 439$-$26}.---As discussed in \S~\ref{sbsc:optir}, with the
recent result that extremely cool, hydrogen atmosphere white dwarfs
become bluer, there are now two possible interpretations for the
location of the non-DA star ESO 439$-$26 (1136$-$286) in the \mv\ vs
($V$--$I$) color-magnitude diagram displayed in Figure
\ref{fg:Mv:vmi}. The first one is that the object is a massive
($M\sim1.20$ \msun), pure helium atmosphere white dwarf, i.e.~the
interpretation proposed by
\citet{ruizetal95}. The second one is that the object is a much 
cooler ($\Te<4000$~K), and thus older, lower mass white dwarf. Our
best fit to the complete energy distribution using our pure helium model
grid is displayed in Figure \ref{fg:figESO439-26}. Our solution, $\Te=4490$~K
and $\logg=9.02$, which corresponds to $M=1.19$ \msun, is consistent
with the location of ESO 439$-$26 in the right panel of Figure
\ref{fg:Mv:vmi} (i.e.~the pure helium sequences). Also shown in 
Figure \ref{fg:figESO439-26} is our best fit to the $V$ and $I$
measurements using our pure hydrogen model grid. This time our solution,
$\Te=3150$~K and $\logg=8.29$, which corresponds to $M\sim0.76$ \msun,
is consistent with the location of ESO 439$-$26 in the middle panel of
Figure \ref{fg:Mv:vmi} (i.e.~the pure hydrogen sequences). However, the
predicted flux at $J$ is clearly inconsistent with the observed flux,
and the pure hydrogen solution must be rejected. This result stresses
the importance of obtaining infrared photometry for such cool white
dwarfs before drawing any conclusion about their temperature and
especially their age.

{\it G107$-$70A/B}.---G107$-$70A/B (0727$+$482A/B) is a partially
resolved pair of cool degenerate stars with an orbital period of 20.5
yr \citep{strand76, harrington81}. This system has long been thought
to be composed of two DC white dwarfs, but our new combined spectrum
shown in Figure
\ref{fg:plotspecDA} clearly shows a weak \halpha\ absorption feature,
a result which indicates that at least one of the two components of
the system has some hydrogen in its atmosphere. A visual inspection of
the spectra of the coolest DA stars in Figure \ref{fg:plotspecDA}
shows that the \halpha\ profile of G107$-$70A/B is much broader and
shallower than the other DA spectra with similar line strengths. This
suggests that the system may be composed of a normal DA star whose
\halpha\ absorption line profile is diluted by the contribution of a 
featureless white dwarf component. In the following we attempt to
determine the atmospheric parameters of the individual components of
the G107$-$70A/B system. The magnitude difference between both white
dwarfs of the G107$-$70A/B system has been measured to be 0.3 mag at
$V$, $J$, $H$, and $K$; we assume the same difference at $B$, $R$, and
$I$. We have obtained fits to the A and B components under the
assumption that one star has a pure hydrogen atmosphere, while the
other one has either a pure hydrogen or pure helium atmosphere (for a
total of three possible combinations). Then the theoretical combined
spectrum is calculated as the sum of both interpolated spectra,
properly weighted by the luminosity of each star. Our results indicate
that acceptable fits to the energy distributions can always be
achieved, no matter what atmospheric composition is assumed. On the
other hand, the \halpha\ line profile is always predicted much weaker
and sharper than the observed line profile. The dilution of the
\halpha\ line profile by the DC component does not alter significantly
enough the {\it shape} of the combined line profile. Our best solution
shown in Figure \ref{fg:figG107-70} is obtained with two pure hydrogen
atmosphere white dwarfs. The discrepancy at \halpha\ cannot be
attributed to an inadequacy with the models since we are able to fit
single DA stars in the same range of effective temperature (see, e.g.,
LHS 1801 in Figure \ref{fg:sampleDA}). We have also tried to restrict
our analysis to only $V$ and the infrared where the magnitude
difference has actually been measured, but without
improvement. Experiments with mixed H/He atmospheres have not been
successful either. Perhaps the \halpha\ absorption feature of one of
the DA component, or both, is broadened by a weak magnetic field. It
is even possible that the G107$-$70 system is composed of three
degenerate components; after all, such systems are now known to exist
\citep{maxted3d}. We note finally that the combined mass of the pair with our adopted
solution is $M\sim1.19$ \msun, in excellent agreement with the
astrometric value of $1.29$ \msun\ derived by
\citet{borgman83}. 

\section{Global Properties of the Sample}

\subsection{Adopted Atmospheric Parameters}\label{sbsc:adopt}

Among the 152 white dwarfs listed in Table 1, 150 have been analyzed
with the techniques described in this paper. The stars omitted are
L481$-$60 (1544$-$377) which has a bright star nearby that
contaminates the photometric observations, and BPM 27606 (2154$-$512)
whose C$_2$ Swan bands are so strong that the color indices are
significantly affected and cannot be modeled within the present
theoretical framework.

As opposed to the BRL analysis, all white dwarfs analyzed here have
trigonometric parallax measurements which allow us to determine their
radii, and hence their masses and surface gravities through the
mass-radius relation. The results of our analysis are reported in
Table 2. For each white dwarf we give the effective temperature, the
surface gravity, the dominant atmospheric constituent, the stellar
mass, the absolute visual magnitude, the total luminosity, and the
white dwarf cooling age (not including the main sequence
lifetime). The bolometric magnitudes can be derived from the simple
expression, $M_{\rm bol} = -2.5 \log L/L_\odot + 4.75$. The numbers in
parentheses in Table 2 correspond to the uncertainties of each
parameter. The error of $\Te$ is derived directly from the fits to the
energy distributions, while the errors of log $g$ and other quantities
have been derived by propagating the error of the trigonometric
parallax uncertainty.  We have also assumed single stars for all
objects in Table 2, even for the suspected or known unresolved
degenerate binaries, and a note mentioning the binary nature of these
systems has been added.

As discussed in \S~\ref{sbsc:fittec}, new C/O core evolutionary models
with thin and thick hydrogen layers have been used for the helium- and
hydrogen-dominated atmospheres, respectively. We have assumed pure
hydrogen and pure helium compositions for the hydrogen- and
helium-rich white dwarfs, respectively, with the exception of the
C$_2$H stars, as well as Ross 640 and L745$-$46A displayed in Figure
\ref{fg:sampleHe}, for which the helium abundances have been determined
individually. Pure helium compositions were also assumed for the
group of peculiar hot ($\Te>6000$~K) non-DA white dwarfs whose
energy distributions are consistent with pure hydrogen models (see
BRL); a note in Table 2 identifies these objects. For the magnetic
white dwarf G227$-$35 shown in Figure \ref{fg:figG227-35}, we adopt the
pure helium solution since its energy distribution is better
reproduced by the pure helium model rather than the pure hydrogen
model, despite the fact it has a hydrogen-rich composition according
to \citet{putney95}.

\subsection{Comparison with Spectroscopic Determinations for DA 
Stars}\label{sbsc:compspec}

Additional blue spectroscopic observations which cover the high Balmer
lines from \hbeta\ to H9 are also available to us for a subset of 24
cool DA white dwarfs in our sample. These observations can be used to
obtain independent estimates of $\Te$ and $\logg$ using the
spectroscopic technique \citep[see, e.g.,][]{bsl92} which consists of
fitting all the hydrogen line profiles observed in each star
simultaneously.  The value of $\logg$ can be converted into mass using an
appropriate mass-radius relation for white dwarfs. These independent
determinations can be compared against the values derived here using
optical/infrared photometry and trigonometric parallaxes.

The comparison of effective temperature and mass determinations for
the 24 white dwarfs in common with our sample is displayed in Figure
\ref{fg:TspecvsTphot}.  Seven objects discussed further below are
numbered in this figure in order of increasing right ascension.
The effective temperatures agree well within
the uncertainties of both techniques throughout the entire temperature
range considered in this analysis, with the exception of the four
stars labeled in the bottom panel of Figure
\ref{fg:TspecvsTphot}.  Of these, L587$-$77A (0326$-$273; labeled 2) 
and Case 2 (1606$+$422; labeled 3)
are both low-mass ($M < 0.4$ \msun) white dwarfs, and they are most
likely unresolved degenerate binaries with individual components
having much different effective temperatures, while G21$-$15
(1824$+$040; labeled 6) is a known double degenerate \citep[][see our fit in
Fig.~\ref{fg:sampledoubleDA}]{saffer98, maxted99}. This comparison
indicates that photometry combined with spectroscopy can be used to
identify  unresolved degenerate binary candidates. The last
object, G128$-$72 (2329$+$267; labeled 7), is a magnetic white dwarf with our
best fit at \halpha\ already shown in Figure \ref{fg:figMAG}. In this
case, the spectroscopic temperature is overestimated since the blue
spectrum of this star has been fit with non-magnetic models, and
magnetic line broadening can be compensated, to some extent, by an
increase in effective temperature (and $\logg$, see below).

The agreement between the photometric masses and the spectroscopic
masses for this sample of 24 DA stars, 
shown in the top panel of Figure \ref{fg:TspecvsTphot}, is not
as good as that between the effective temperatures, even within the
mass uncertainties. In general, the spectroscopic masses are larger
than those obtained from trigonometric parallax measurements. There
are a few exceptions, however, and these are labeled in the top panel
of Figure \ref{fg:TspecvsTphot}. First, G19$-$20 (1716$+$020; labeled 5)
has a photometric mass of $0.67\pm0.08$ \msun\ and a spectroscopic
mass of $0.49\pm0.03$ \msun. An independent mass estimate for this
white dwarf star comes from the gravitational redshift measurement
which yields a value of $0.48\pm0.05$ \msun\ \citep{redshift95}, a
result which suggests that the trigonometric parallax measurement is
probably in error for this particular object. Second, G31$-$35
(0011$+$000; labeled 1), has a fairly large mass uncertainty
($M=0.85\pm0.12$ \msun) due to a $\sim15$\% error in the trigonometric
parallax measurement, and the difference from the spectroscopic
mass of $M=0.70\pm0.04$ \msun\ is not significant. The objects
whose spectroscopic masses are larger than the photometric masses
include the three low-mass white dwarfs (objects labeled 2, 3, and 6)
already identified as binaries or binary candidates in the bottom
panel of Figure \ref{fg:TspecvsTphot}, and G128$-$72
(2329$+$267; labeled 7) the magnetic white dwarf discussed above. For
the three binaries, the photometric masses are underestimated since the total
flux received from the system is assumed to originate from a single
object, while the spectroscopic masses are intermediate values of the
masses of the individual components, as demonstrated by
\citet{liebert91}. If we assume instead that two identical white
dwarfs contribute to the total flux received at Earth, we obtain the
photometric masses indicated by the three vectors in Figure
\ref{fg:TspecvsTphot}, and these agree perfectly 
with the spectroscopic masses for two of these objects, while the
third object is probably composed of two white dwarfs with different
luminosities.

As discussed in \S\ \ref{sbsc:hrich}, there are DA stars for which the
agreement between the photometric and spectroscopic solutions is
excellent, the best example of which is G74$-$7 (0208$+$396) displayed
in Figure \ref{fg:sampleDA}. Why then are there several DA white dwarfs 
whose spectroscopic masses are significantly larger than the photometric
masses? What is the missing parameter, if any? \citet{bergeron89} has
suggested that the convective mixing of a thin hydrogen atmosphere
with the deeper and more massive helium envelope, which is believed to
occur in hydrogen-rich white dwarfs cooler that $\Te\sim12,000$~K,
could result in atmospheres which are only moderately enriched with helium
rather than in helium-dominated atmospheres. Since helium is
spectroscopically invisible at these temperatures, the mixed H/He
white dwarfs would still appear as normal DA stars. \citet[][see also
\citealt{bsl91}]{bergeron89} has also demonstrated that normal gravity,
helium-enriched atmospheres are totally equivalent, {\it from a
spectroscopic point of view}, to pure hydrogen atmospheres but with
large surface gravities (and hence masses). Hydrogen line profiles at
a given temperature are affected by atmospheric pressure, which in
turn is sensitive to changes in the surface gravity and in the helium
abundance, and both effects cannot be disentangled from spectroscopy
alone.  In contrast, the photometric technique provides a direct
measure of the star radius --- which is not very sensitive to the
presence of helium. Hence, the mass discrepancies observed in Figure
\ref{fg:TspecvsTphot} could well be the result of helium enrichment in
some of the white dwarfs in our sample. 

We illustrate the effect of adding small amounts of helium to the atmospheres
of cool DA white dwarfs  in Figure \ref{fg:exampleHe} for
LHS 3254 (1655$+$215), the most discrepant case in Figure
\ref{fg:TspecvsTphot} (labeled 4 in the top panel). The top panels show
our best photometric fit (left panel) and our best spectroscopic fit
(right panel) using our pure hydrogen model grid. Even though the
effective temperatures agree within $\sim200$~K, the surface gravity
determined from spectroscopy, $\logg=8.20$ (or $M=0.73$ \msun), is
much larger than that obtained from the photometric solution,
$\logg=7.87$ (or $M=0.53$ \msun). By allowing a small amount of
helium in the model grid, in this case $\log \nhe = -0.20$, we achieve
the solutions displayed in the bottom panels of Figure
\ref{fg:exampleHe}. First, the photometric fit has changed little,
both qualitatively and quantitatively, a result which implies that the
overall results of our analysis remain unaffected by this new
parameter. Second, the spectroscopic fit has not changed
qualitatively, but the surface gravity has decreased significantly,
and most importantly, the photometric and spectroscopic $\logg$ values
(and hence masses) are now in perfect agreement. The effective
temperatures are also in better agreement. Hence, our results
reinforce the conclusions of \citet{bergeron89} that several --- but
probably not all --- cool DA stars have helium-enriched atmospheres,
with helium abundances as high as $\nhe\sim1$.  We note that (see 
the following section) the spectroscopically determined masses
for hot DB stars can also be brought into better agreement with our
photometrically determined values if very small amounts of hydrogen are 
added to the pure helium model grid.  At least trace amounts of both
hydrogen and helium are therefore likely to exist
in many cool white dwarf atmospheres.

\subsection{Mass Distributions}\label{sbsc:mdist}

The mass of all white dwarfs in our sample is shown as a function of
effective temperature in Figure \ref{fg:spec_evol}. Each object is
represented by a different symbol indicating its atmospheric
composition and/or spectral type. Filled and open symbols represent
hydrogen-rich and helium-rich atmospheric compositions,
respectively. The more specific symbols are discussed in the next
section. A comparison of this Figure with Figure 35 of BRL indicates
that our new sample has increased in size considerably, especially
above $\Te\sim7000$~K where the BRL sample contains only a few
objects. Furthermore, it is possible here to determine stellar masses
for all white dwarfs in our sample, in contrast with BRL who could
only determine masses for about 50\%  of their sample.

While massive white dwarfs are found with both hydrogen and helium
atmospheres, low mass white dwarfs have mostly hydrogen-rich
atmospheres, with the exceptions of the DQ star G184$-$12 (1831$+$197)
at $\Te=7590$~K, and LP 31$-$40 (0324$+$738) at $\Te=4650$~K. The mass
of the former object could be as large as 0.56
\msun\ due to the large parallax uncertainty, while the latter object lacks
the crucial $JHK$ photometry which would allow us to determine its
atmospheric constituent more precisely, and this object may well have
a hydrogen-rich atmosphere. Consequently, it is probable that all
low-mass white dwarfs possess hydrogen-rich atmospheres. This
conclusion is supported by the spectroscopic analyses of hotter ($\Te\
\gta12,000$~K) DA and DB stars \citep{bsl91, beauchamp96} which
revealed the existence of a low mass tail in the mass distribution of
DA stars which is not observed in the DB sample. These results are
illustrated in Figure \ref{fg:mass_distr} where we compare the mass
distributions of the hydrogen- and helium-rich atmosphere white dwarfs
in our sample (top panel) and in the hotter spectroscopic samples
(bottom panel); note that the photometric and spectroscopic samples do
not overlap in effective temperature. Since common envelope evolution
is required to produce degenerates with mass $\lta 0.5$~\msun --- the
Galaxy being too young to have produced them from single star
evolution --- we must conclude that this particular evolutionary
channel does not produce helium-rich atmosphere white dwarfs,
presumably because the objects which go through this close-binary
phase end up with hydrogen layers too massive to allow the DA to DB
conversion near $\Te\sim30,000$~K
\citep{beauchamp96}, or any possible convective mixing below
12,000~K. The fraction of low-mass, hydrogen-rich atmosphere white
dwarfs is larger in the parallax sample ($\sim 20$\%) than in the
spectroscopic sample ($\sim 10$\%). This is due to the fact that in
the parallax sample, unresolved double degenerate systems composed of
two normal 0.6 \msun\ white dwarfs, for example, will appear as a
single overluminous low-mass $\sim 0.3$ \msun\ star, while it will
still appear as a single 0.6 \msun\ object when analyzed
spectroscopically.

There are significant differences at high masses as well where the
trigonometric parallax sample contains more massive stars ($M\ \gta 1.0$
\msun) than the spectroscopic sample. Since the former sample is 
composed of older objects, it is possible that it contains several
cases of mergers which would result in a single white dwarf with
higher than average mass \citep{iben90}. Note that the most massive
star in our sample with $M\sim1.2$ \msun\ is ESO 439$-$26
(1136$-$286), discussed at length above.

The mean mass of the hydrogen-rich parallax subsample is $\langle
M\rangle=0.61$ \msun\ with a dispersion of $\sigma (M)=0.20$ \msun,
while we find for the helium-rich subsample a mean mass of $\langle
M\rangle=0.72$ \msun\ with a dispersion of $\sigma (M)=0.17$ \msun;
the combined sample yields $\langle M\rangle=0.65$
\msun, $\sigma (M)=0.20$ \msun. The lower mean value of the hydrogen-rich 
set is entirely due to the existence of the low-mass tail
component. These numbers can be compared with the values derived for
the spectroscopic analyses of hotter DA stars, $\langle M\rangle=0.59$
\msun\ and $\sigma (M)=0.13$ \msun\ \citep{bsl91, bergeronzz95}, and of
DB stars, $\langle M\rangle=0.59$ \msun\ and $\sigma (M)=0.06$ \msun\ 
\citep{beauchamp96}. Note that the dispersions of the spectroscopic 
mass distributions are considerably smaller due to the increased
sensitivity to $\logg$ of the spectroscopic technique over the
trigonometric parallax method.  While the mean masses obtained for the
cool hydrogen atmospheres and warmer DA stars agree well, those of the
helium atmospheres and DB stars differ by $\sim0.13$ \msun. As
discussed in BRL, these results can be somewhat reconciled if small
amounts of undetectable hydrogen is present in the cool helium-rich
atmospheres. Quantitative results not shown here indicate that the
masses derived from the pure helium models could be reduced by $\sim
0.07$ \msun\ were they analyzed with a $\nh=10^{-4}-10^{-3}$ model
grid.

It is worth mentioning that the spectroscopic analyses of DA and DB
stars discussed above are not sensitive to the presence of unknown
traces of hydrogen or helium. Indeed, DA stars in these studies are
restricted to a range of temperature above 15,000~K where the
assumption of pure hydrogen atmospheres is most certainly valid due to
the gravitational settling of helium \citep{bsl91}. Similarly, DB (and
DBA) stars have temperatures in excess of 12,000~K where the hydrogen
abundance is well determined, or at least constrained
\citep{beauchamp96}. Hence, the spectroscopic mass distributions
displayed in Figure
\ref{fg:mass_distr} are not sensitive to unknown abundances of trace
elements, unlike the results for cool ($\Te\ \lta 12,000$~K) DA stars
discussed in \S\ \ref{sbsc:compspec} and shown in Figure
\ref{fg:TspecvsTphot}, in which the presence of spectroscopically
invisible helium could overestimate the spectroscopic masses.

\subsection{Spectral Evolution}

Out of the 150 objects plotted in Figure \ref{fg:spec_evol}, 96 have a
hydrogen-rich atmosphere while 54 have a helium-rich atmosphere. Hence
36\% of all white dwarfs in our sample have helium-rich atmospheres, a
ratio which is significantly larger than the canonical value of 20\%
for hotter white dwarfs \citep{fw87}. The ratio determined here,
however, is not constant over the temperature range considered in this
analysis.  Figure \ref{fg:H_vs_He} shows the number of hydrogen-rich
and helium-rich stars as a function of effective temperature in 1000~K
bins using the results presented in Table 2; the corresponding
hydrogen- to helium-rich ratio is also shown.  The $\sim$5000-6000 K
bin shows a marked increase in the number of white dwarfs with
hydrogen-rich atmospheres, and a marked drop in the number with
helium-rich atmospheres. This is the non-DA gap first identified by
BRL. Our new results, with a much larger sample and with no overt
spectroscopic classification bias, confirms the reality of this
gap. Note also that there is no particular concentration of non-DA
stars below $\Te\sim 5000$~K, nor is there any particular
concentration of DA stars around 6000~K, contrary to the claim of
\citet{hansen99}. Quite the opposite, hydrogen- and helium-rich white
dwarfs in Figure
\ref{fg:spec_evol} are more or less homogeneously distributed in
effective temperature, with the exception of the non-DA gap itself.

We thus reaffirm our earlier conclusion that some as yet unidentified
physical mechanism is responsible for turning helium-rich atmosphere
white dwarfs into hydrogen-dominated atmospheres near $\Te\sim
6000$~K. The fact that the cooling timescales of these white dwarfs
may depend on their atmospheric composition, as observed by
\citet{hansen99}, has nothing to do with the conclusions reached
here. Indeed, if the existence of this gap is the result of the
spectral evolution of helium-rich into hydrogen-rich white dwarfs, the
only important factor is the mass-dependency of the mechanism through
which this transition occurs. If this mechanism --- still unknown
--- is only weakly mass-dependent, then helium-rich white dwarfs will
turn into DA stars near 6000~K, no matter what their cooling age might
be.  

An alternative explanation for the non-DA gap could be that the blue
end of the gap at 6000~K represents the end of the cooling sequence
for helium-rich white dwarfs, and that all non-DA stars observed below
the red end of the gap are the result of convectively mixed DA stars
at 5000~K. However, the cooling age of helium-rich white dwarfs near
6000~K is only around 4 Gyr, and since much older DA stars exist,
older and cooler helium-rich white dwarfs should exist as well.

The spectroscopic behavior seen in our sample around the gap is also
of interest.  To summarize, Figure \ref{fg:spec_evol} shows that:
hydrogen-rich white dwarfs are seen at all effective temperatures; DQ
stars are seen at $\Te\ \gta 6500$~K; DZ at $\Te\ \gta 6500$~K and
4500--5500~K; helium-rich DA's at 8000--9000~K; C$_2$H stars at
5500--6500~K; and the peculiar group whose featureless energy
distributions are better fit with pure hydrogen models at
6000--7500~K.

At temperatures just above and near the non-DA gap, non-DA white
dwarfs that are neither DQ nor DZ belong to the peculiar group of
white dwarfs whose featureless energy distributions are better fit
with pure hydrogen models. One exception directly above the gap is the
DC star G125$-$3 (1917$+$386) with $\Te = 6390$~K. According to
\citet{greenstein90}, this object could have barely visible, heavily
broadened C$_2$ bands. Their spectrum which is available to us reveals
three weak and broad absorption features which actually match those of
C$_2$H stars, and it is possible that this object belongs to this
spectral type, although better spectroscopic observations are required
before this result is confirmed. Near 7500~K, there are three
additional DC stars. The coolest object is Stein 2051B (0426$+$588)
with $\Te = 7120$~K, a true DC star according to \citet{wegner82} who
reported featureless optical as well as ultraviolet spectroscopic
observations. The second DC star is G195$-$19 (0912$+$536, $\Te =
7160$~K), a polarized $\sim100$ MG magnetic white dwarf
\citep{schmidtsmith95}, and the last one is LHS 2333 (1055$-$072, $\Te
= 7420$~K) whose fit has already been discussed in \S
\ref{sbsc:herich} and displayed in Figure \ref{fg:samplenonDA}. Another 
helium-rich object with only a slightly
higher effective temperature is GD 356 (1639$+$537, $\Te = 7510$~K), a
DAE white dwarf whose spectrum exhibits a hydrogen Zeeman triplet in
emission \citep{greenstein85}.

Most non-DA white dwarfs within the non-DA gap are C$_2$H
stars. G240$-$72 (1748$+$708) is considered in this spectral class as
well, following the interpretation of BRL that the observed 15\% deep
yellow sag near 5000 \AA\ is the result of C$_2$H molecular absorption
features in the $\sim 200$ MG magnetic field which characterizes this
star.  Near the red edge
of the gap are LHS 2710 (1313$-$198) and ESO 292$-$43
(2345$-$447) which have already been identified by BRL as being
peculiar in that the energy distributions cannot be matched by our models
(see their Figure 37 and the accompanying discussion). Most
importantly, {\it our new larger sample has not added any new non-DA
star within the gap}. During the course of our observational survey,
every time a previously classified non-DA star was found inside the
gap, subsequent high signal-to-noise spectroscopy would systematically
reveal the presence of \halpha. Note that the new magnetic candidate 
C$_2$H star LHS 2229 recently reported by \citet{schmidt99} has an
effective temperature of $\Te\sim 4400$~K when analyzed with our
fitting technique (assuming $\logg=8.0$). If this interpretation is
correct, C$_2$H stars may exist well below the red edge of the non-DA
gap.

The results shown in Figure \ref{fg:spec_evol} reveal that DQ and
C$_2$H stars do not overlap in effective temperature. Indeed, there
are no DQ stars observed below $\Te=6400$~K, while C$_2$H stars are
found only below this temperature. This
suggests that DQ stars probably turn into C$_2$H stars, and that the
existence of the non-DA gap and the nature of the C$_2$H white dwarfs
are somehow related. Since C$_2$H stars have been interpreted by
\citet{schmidt95} as the result of mixed H/He/C compositions, it is
the additional presence of hydrogen in a supposedly helium and carbon
mixture that would turn DQ stars into C$_2$H stars when they cool off
below 6400~K. The origin of the hydrogen source remains a mystery,
however, since the exotic model proposed by BRL has been discarded by
\citet{malo99} based on detailed model simulations. We just note that 
the same source of hydrogen, once identified, could be also
responsible for turning the remaining non-DA stars above $\sim 6000$~K
(DZ and peculiar) into DA stars below this temperature.

Peculiar white dwarfs whose energy distributions are better
reproduced with hydrogen models despite the absence of \halpha\ are
found over a range of effective temperature of 6200---7600~K, i.e.
all above the non-DA gap.  The hottest of these stars is LP 434$-$97 
(1154$+$186) at $\Te=7630$~K. Interestingly, immediately above this 
temperature are the two helium-rich DA stars in our sample,
L745$-$46A (0738$-$172) at $\Te=7710$~K, and Ross 640 (1626$+$368) at
$\Te=8640$~K. Given that there are no other such helium-rich DA stars
known, it is tempting to suggest that an evolutionary link exists
between these objects and the peculiar white dwarfs. Above 8000~K,
most non-DA stars are of the DQ type, although some are reported to be
featureless in the literature. The number of stars in this temperature
range is still too small, however, to draw meaningful conclusions
about any spectral evolution from one type to another.

The reappearance of non-DA stars below the gap has been interpreted by
BRL as the result of the convective mixing of the hydrogen atmosphere
with the deeper helium envelope, turning DA stars into non-DA stars.
The calculations reported by BRL showed indeed that envelope models
which include realistic model atmospheres as boundary conditions could
reduce the mixing temperature limit from the canonical value of
$\Te\sim 6000$~K down to 5000~K, which occurs for a hydrogen layer
mass of $M_{\rm H}/M_\star\sim10^{-6}$. Obviously, some DA stars have
thicker hydrogen layers since they did not mix, in particular the
(apparent or true) low-mass white dwarfs which are either unresolved
binaries, or the result of common envelope evolution. We note
that the coolest non-DA stars in our sample have hydrogen-rich
atmospheres despite the lack of any detectable hydrogen absorption
features. While \halpha\ can be detected almost to the end of the
disk cooling sequence, it will certainly not be detected in any
cooler white dwarfs discovered in the Galactic halo.

Our sample contains a higher ratio of non-DA to DA stars than is seen
in the temperature range 12,000-30,000~K, where the ratio is around
25\% \citep{fw87}. This increase can certainly be interpreted as the
result of convective mixing over a still unknown temperature range.
Since the 10,000--11,000~K bin in Figure \ref{fg:H_vs_He} already
shows a ratio near unity, and since the reappearance of non-DA stars
in large number below the non-DA gap has also been interpreted as the
result of convective mixing, we can conclude that the convective
mixing mechanism operate from $\Te\sim11,000$~K, all the way down to
the coolest possible mixing temperature near 5000~K. This implies that
DA stars must possess a large range of possible hydrogen layer
thicknesses, including very thick hydrogen layers above $q({\rm
H})=10^{-6}$.

Out of the 13 magnetic white dwarfs in our sample --- 8 with
hydrogen-rich and 5 with helium-rich atmospheres, 12 are located in a
narrow range of effective temperature between $\Te=5300$ and 8000~K;
the exception is GW$+$70 8247 (1900$+$705) at $\Te=12,070$~K. It is
possible that the lower boundary represents only a selection effect
since Zeeman splitting would make the detection of \halpha\ more
difficult in cool magnetic DA stars than in non-magnetic white
dwarfs. However, the spectrum of the coolest magnetic star in our
sample, LHS 1734 (0503$-$174), still exhibits strong
\halpha\ absorption features (see the spectrum in Fig.~5 of BRL), and
the detection of even weaker features is well within the detection
limit of our spectroscopic observations. Furthermore, while there are
no magnetic white dwarfs in the three dozen or so objects included in
our sample between $\Te\sim8000$ and 12,000 K, a few magnetic stars in
this particular range of effective temperature are known to exist
\citep{schmidtsmith95}.

All the features of the chemical distribution shown in Figure
\ref{fg:spec_evol} cannot be accounted for by convective mixing alone, 
however, or by any simple theoretical model. Hence we are left with
very few possible explanations, despite the abundance of observational
clues unveiled in BRL and in this analysis. Given these results, it is
worth repeating and developing an alternative explanation proposed by
BRL who questioned the stability of convectively mixed hydrogen and
helium convection zones. Consider a 0.6 \msun, pure helium atmosphere
white dwarf which accretes $\sim 10^{-8}$ \msun\ of hydrogen by the
time it reaches 6000~K. This value is totally plausible if one allows
a low accretion rate from the interstellar medium of $10^{-17}$
\msun\ yr$^{-1}$ over a period of $\sim1$ Gyr or so.
Since the mass of the helium convection zone is almost constant below
$\Te\sim 12,000~K$ ($M_{\rm He-conv}\sim 10^{-6}~M_{\ast}$), the
resulting abundance of $\nh\sim10^{-2}$ would make the star appear as
a non-DA white dwarf at 6000 K, \halpha\ being wiped out by van der
Waals broadening with neutral helium. It is reasonable to wonder
whether such a mixed hydrogen and helium convection zone would remain
convectively stable as the results shown in Figure 40 of BRL indicate
that a DA star with a hydrogen layer mass of $\sim 10^{-8}$ \msun\, or
slightly larger, would not have mixed in the first place, at least not
at $\Te\sim6000$~K. It is possible that convectively mixed hydrogen
and helium atmospheres return to a well separated hydrogen-dominated
atmosphere with a helium-rich envelope. Note that the chemical
separation needs to be complete only in the photospheric layers for
the atmosphere to appear hydrogen-rich.  Such ``born again'' DA stars
cannot retain a DA signature for long, however. As the star cools from
$\Te\sim6000$~K down to 5000~K, the bottom of the hydrogen atmosphere
further plunges from $q({\rm H})=10^{-8}$ to $10^{-6}$ (depending on
the mass of the star, see Fig.~4 of BRL), leading to another episode
of convective mixing, and supporting the trends seen in Figure
\ref{fg:spec_evol}.

If the separation of hydrogen and helium layers does occur, we
can speculate about what the energy distribution of these stars
might look like at the moment just before the separation
occurs. Obviously, the hydrogen and helium abundance profiles
throughout the atmosphere will be very inhomogeneous, even over the
disk surface. Perhaps this would explain why BRL \citep[see
also][]{malo99} were unable to fit the energy distributions of those
peculiar white dwarfs above 6000~K with pure hydrogen, pure helium, or
{\it homogeneously mixed} hydrogen and helium compositions. Models with
inhomogeneous abundance profiles could hold the key to these peculiar
stars.

We note finally that if the total mass of hydrogen is small, the
convectively mixed hydrogen and helium layer is stable. This can
certainly be the case for the C$_2$H stars with mixed atmospheres of
hydrogen, helium, and carbon, as well as Ross 640 and L745$-$46A, which
are simply too hot and thus too young to have accreted enough hydrogen
(these objects have $\nh\ \lta10^{-3}$).

\subsection{White Dwarf Cosmochronology}

The isochrones for our C/O core evolutionary models with thin and
thick hydrogen layers are shown in Figures \ref{fg:isochrone_CO_thin}
and \ref{fg:isochrone_CO_thick}, respectively. In these plots, we no
longer distinguish between objects with hydrogen- and helium-rich
atmospheres since the atmospheric composition of these stars may have
changed several times over their lifetime, and as such, it is not a
trivial task to associate a cooling age to a single object, as
discussed in the Introduction. Also shown are the corresponding
isochrones but with the main sequence lifetime added to the white
dwarf cooling age; here we simply assume
\citep{lrb98} $t_{\rm MS}=10(M_{\rm MS}/M_\odot)^{-2.5}$ Gyr and
$M_{\rm MS}/M_\odot=8\ln[(M_{\rm WD}/M_\odot)/0.4]$. As can be seen
from these results, white dwarfs with $M\ \lta0.48$ \msun\ cannot have
C/O cores, and yet have been formed from single star evolution within
the lifetime of this Galaxy. As discussed above, some of these
low-mass objects must be either unresolved double degenerates, or
single white dwarfs with helium cores. In the former case, the stellar
masses inferred from these figures are underestimated --- especially
if the unresolved components have comparable luminosities, and the
corresponding cooling ages derived here become meaningless. While some
of these low-mass white dwarfs may still be interpreted as having C/O
cores when the masses are corrected to allow for two objects in the
system, the lowest mass objects in our sample would still yield
``corrected'' masses too low to be C/O-core degenerates. The second
possibility corresponds to single (or binary) helium-core degenerates
whose core mass was truncated by Case B mass transfer before helium
ignition was reached. Such low-mass white dwarfs would be best
described with helium-core evolutionary models. However, since it is
impossible to distinguish between the two possibilities discussed
above for most of the low-mass stars in our sample, we refrain from
interpreting these objects any further.

In this temperature regime, low-mass (i.e.~large-radius) white dwarfs
evolve faster than their more massive counterparts, with the exception
of the most massive objects ($M\ \gta 1.0$ \msun) in which the onset
of crystallization reduces the cooling timescales considerably. These
begin to cool off much more rapidly than the lower mass stars to
produce the observed parabola-shaped isochrones.  With decreasing
effective temperature, crystallization gradually occurs in lower mass
models, and the turning point of these parabolas moves slowly towards
lower masses. Consequently, the inferred stellar ages are strongly
mass-dependent, a result which stresses the importance of determining
reliable masses through precise trigonometric parallax
measurements. The best example is provided by ESO 439$-$26
(1136$-$286), the most massive object in our sample, which could have
an inferred cooling age anywhere between 5 and 7.5 Gyr if its mass was
not known (assuming thin hydrogen layer models).

The effect of using different thicknesses for the outer hydrogen layer
varies as a function of the mass of the star as well. For instance,
the coolest white dwarfs near 0.6 \msun\ have cooling ages of roughly
8 Gyr with the thin hydrogen layer models, and 1 Gyr {\it older} with
the thick models. Out of the three oldest objects in our sample, two
have hydrogen-rich atmospheres, the coolest of which is ER 8
(1310$-$472) with $\Te = 4220$~K and mass near 0.6 \msun. If we make 
the assumption that this star
has a thick hydrogen envelope, the corresponding white dwarf cooling age
is $t_{\rm WD}=9.3$ Gyr ($t_{\rm WD+MS}\sim9.7$ Gyr).  The oldest
object in Figure \ref{fg:isochrone_CO_thick}, LP 701$-$29
(2251$-$070) with $\Te = 4580$~K, has a helium-rich atmosphere and it 
thus cannot have had a thick outer hydrogen layer, otherwise it would have
remained hydrogen-rich. Its more probable age derived from Figure
\ref{fg:isochrone_CO_thin} is actually much shorter, $t_{\rm WD}=7.8$ Gyr
($t_{\rm WD+MS}\sim7.9$ Gyr).  There are two objects in our sample cooler 
than ER 8 with hydrogen-rich atmospheres with apparently low masses of 
$\sim 0.4$ \msun, LHS 239 (0747$+$073B, $\Te=4210$~K) and LP 131$-$66
(1247$+$550, 4050~K).  If these objects are in fact an unresolved pair 
of normal mass white dwarfs then the total age of these systems is 
around 10~Gyr.

\section{Summary and Conclusions}

A detailed photometric and spectroscopic analysis of 152 cool white
dwarfs with trigonometric parallax measurements was presented. Optical
and infrared photometry were combined and compared with an improved
grid of model atmosphere calculations to determine the effective
temperatures and the solid angles for all objects, which combined with
the measured parallaxes yielded the stellar radii. The latter were
then converted into stellar masses through the mass-radius relations
obtained from new evolutionary models which best characterize cool
white dwarfs. These have carbon/oxygen cores, with both thin  
$M_{\rm H}/M_\star = q({\rm H})=10^{-10}$ and thick $q({\rm H})=10^{-4}$ 
hydrogen layers. This
parallax sample more than doubles the number of cool white dwarfs with
known masses compared to the sample presented in BRL.
Furthermore, this new sample is more representative of the population
of cool white dwarf stars as it contains, presumably, no bias in favor
of any particular spectral type, as opposed to the BRL sample which
favored non-DA stars.

One of the main results of our analysis is the confirmation of the
existence of a range in effective temperature between roughly 5000 and
6000~K in which almost all white dwarfs have hydrogen-rich
atmospheres. The only non-DA stars in this range of effective
temperature are the so-called C$_2$H stars which are believed to
represent the cooler stage of the DQ sequence, found only above the
gap. Clearly, the existence of this gap holds the key to our
understanding of the chemical evolution of cool white
dwarfs. Unfortunately, our understanding of the physical mechanisms
that could alter the atmospheric composition of cool white dwarfs as
they evolve are at best sketchy. After the ``phase transition'' model
proposed in BRL was convincingly discarded by \citet{malo99}, very few
alternatives exist.

It has become clear from our analysis, however, that convection plays
an important role, but that our understanding of what to expect from
the result of convective mixing needs to be revised. It is generally
believed that when the bottom of the thin hydrogen convection zone
deepens and reaches the underlying and more massive convective helium
envelope, convective mixing will occur, and that the resulting
atmospheric composition will simply be governed by the mass ratio of
the hydrogen and helium convection zones. Such arguments lead to
predicted abundances after convective mixing of $\nh=10^{-4}$,
$10^{-3}$, $10^{-2}$, and $10^{-1}$, for mixing temperatures of
$T_{\rm mix}\sim11,000$, 10,000, 8500, and 6500~K, respectively, the
mixing temperature depending on the exact thickness of the hydrogen
layer. Our analysis has demonstrated that such objects do not exist,
at least in our sample. Even the objects which possess mixed hydrogen
and helium atmospheres, and in particular Ross 640 and L745$-$46A,
have abundances that do not fit this simple convective mixing
picture. On the other hand, the definite increase of the ratio of
non-DA to DA stars in the temperature range considered in this
analysis with respect to that of hotter white dwarfs cannot easily be
explained in terms other than convective mixing. Also, convective
mixing has been called upon to explain the reappearance of non-DA
stars below the non-DA gap near 5000~K.

A different picture is now starting to emerge from our findings, and
those of other analyses, in which convective mixing {\it processes}
may lead to predicted chemical abundances that differ from the
standard model. First, our comparison of atmospheric parameters
determined from photometry and spectroscopy reveals that a mass
discrepancy exists in several DA stars, and that this discrepancy can
be resolved if we allow some amount of helium to be present in their
atmospheres. The required abundances close to $\nhe=1$ are clearly at
odds with the standard convective mixing model predictions, as noted
also by \citet{bergeron90}. Note that helium is spectroscopically
invisible in the temperature regime considered here and cannot be
detected directly. It is possible to conceive that helium is being
brought to the surface not by convective mixing, but perhaps by
convective overshooting of the helium convection zone into the
hydrogen layer. The bottom of the hydrogen convection zone would then
carry some of this helium to the surface, in a very similar fashion to
the way carbon is being brought to the photosphere of DQ stars, with the
exception that ordinary diffusion is responsible in this case for polluting
the bottom of the helium layer with carbon.

A variance of the convective model that could account for the
existence of the non-DA gap has also been proposed. The question
arises when one considers a helium-rich atmosphere on top of which
indefinite amounts of hydrogen are accumulated, presumably from
interstellar accretion. It is clear that the hydrogen abundance ratio
will gradually increase until some point where the homogeneously mixed
hydrogen and helium convection zone can no longer be sustained. A
stable chemical separation between the hydrogen and the helium layers
could then occur, leading to the predominance of DA stars below
$\Te\sim6000$~K. The implication of this proposed model is that cool
white dwarfs must accrete reasonable amounts of hydrogen from the
interstellar medium for this mechanism to work. Also, it was proposed
that the chemical separation could occur in a non-homogeneous fashion,
leading to chemically inhomogeneous atmospheres, similar to the hotter
DAB stars \citep[see][]{koester94, beauchamp93}. Such inhomogeneous
atmospheres could perhaps account for the peculiar objects 
identified in our sample whose energy distributions share both
characteristics of hydrogen and helium atmospheres.

Given the complexity of the chemical evolution of cool white dwarf
atmospheres revealed in our analysis, it is not clear what
evolutionary models should be used to derive cooling ages. Thin and
thick hydrogen layer models yield cooling ages which differ by 2
Gyr in some cases. Non-DA stars are probably better characterized by
thin hydrogen models; even if they were once DA stars, they probably
had thin hydrogen layers anyway in order to eventually mix. But cool
DA stars may have thin or thick hydrogen layers. Thick hydrogen layer
white dwarfs remain DA forever, but the fraction of such stars is
still unknown. The problem becomes even more complex when one attempts
to determine the age of the Galactic disk using the luminosity
function of cool white dwarfs. The theoretical luminosity functions
should in fact reflect the actual population of cool white dwarfs, and
should be a combination of thin and thick hydrogen layer
models. The problems are greatly simplified for hotter DA, DB,
or DO stars. From the results presented here, the
oldest white dwarfs in our sample have a minimum stellar age of 8.5
Gyr, but are more likely characterized by models with thick hydrogen
layers which yield an age around 9.5 Gyr.

We feel at this point that we have pushed both the observational and
theoretical aspects of this cool white dwarf study to their limits, and
that more work now needs to be done on the theoretical front to
explain the various features discovered in this
analysis. Several avenues of investigation have been proposed, most of
which are related to our poor understanding of convective mixing.

\acknowledgements We wish to acknowledge G.~Fontaine for calculating the
cooling models used in our analysis, P. Brassard for making his
evolutionary code available to us, and G.~D. Schmidt for his
spectropolarimetric observations of the magnetic white dwarf
candidates. We are deeply grateful to the various observatories for
awarding us so many nights for this project and to all the staff at
the observatories who were extremely helpful. We also wish to thank
the referee for pointing us out the results of Jorgensen et al.~of
which we were unaware. This work was supported in part by the NSERC
Canada and by the Fund FCAR (Qu\'ebec).  MTR received partial support
from Fondecyt grant 1980659 and Catedra Presidencial en Ciencias.

\clearpage
\hoffset -0.5truein
\begin{deluxetable}{p{2.3cm}lrrrllllclllcl}
\tabletypesize{\scriptsize}
\tablecolumns{15}
\tablewidth{0pt}
\rotate
\tablecaption{Observational Results}
\tablehead{
\colhead{} & 
\colhead{}& 
\colhead{$\pi$} & 
\colhead{$\sigma_\pi$} &
\colhead{$W$} & 
\colhead{} & 
\colhead{} & 
\colhead{} & 
\colhead{} & 
\colhead{N} &
\colhead{} & 
\colhead{} & 
\colhead{} & 
\colhead{N} \\
\colhead{WD} &
\colhead{Name} &
\colhead{(mas)} &
\colhead{(mas)} &
\colhead{(\AA)} &
\colhead{$V$} &
\colhead{$B-V$} &
\colhead{$V-R$} &
\colhead{$V-I$} &
\colhead{($BVRI$)} &
\colhead{$J$} &
\colhead{$J-H$} &
\colhead{$H-K$} &
\colhead{($JHK$)} &
\colhead{Notes}}
\startdata
0000$-$345\dotfill&LHS 1008      & 75.7& 9.0&0.0      &15.02   &$+$0.44  &$+$0.30  &$+$0.57  &1&14.17   &$+$0.15  &$+$0.15  &1&1, 2                \\
0009$+$501\dotfill&LHS 1038      & 90.6& 3.7&8.91     &14.36   &$+$0.45  &$+$0.28  &$+$0.59  &1&13.41   &$+$0.15  &$+$0.05  &1&1, 2, 3, 4          \\
0011$+$000\dotfill&G31$-$35      & 32.9& 4.8&34.8     &15.35   &$+$0.20  &$+$0.08  &$+$0.18  &1&15.21   &$+$0.08  &$+$0.01  &1&2                   \\
0011$-$134\dotfill&LHS 1044      & 51.3& 3.8&7.15     &15.89   &$+$0.63  &$+$0.33  &$+$0.67  &1&14.85   &$+$0.23  &$+$0.10  &1&1, 2, 4, 5          \\
0029$-$032\dotfill&LHS 1093      & 42.6& 1.0&0.0      &17.32   &$+$1.12  &$+$0.61  &$+$1.14  &1&15.56   &$+$0.19  &$+$0.02  &4&1, 2                \\
\\
0033$+$016\dotfill&G1$-$7        & 30.4& 4.0&44.2     &15.61   &$+$0.18  &$+$0.02  &$+$0.08  &1&15.63   &$+$0.03  &$-$0.03  &1&2                   \\
0038$-$226\dotfill&LHS 1126      &101.2&10.4&0.0      &14.50   &$+$0.70  &$+$0.42  &$+$0.79  &2&13.32   &$-$0.15  &$-$0.24  &2&1, 2                \\
0038$+$555\dotfill&G218$-$8      & 43.4& 2.0&0.0      &14.05   &$+$0.05  &$+$0.04  &$+$0.12  &1&13.97:  &$-$0.11: &$-$0.05: &1&2, 6                \\
0046$+$051\dotfill&vMa 2         &232.5& 1.9&0.0      &12.39   &$+$0.52  &$+$0.26  &$+$0.49  &2&11.69   &$+$0.08  &$+$0.09  &1&1, 2, 5             \\
0101$+$048\dotfill&G1$-$45       & 46.9& 3.8&18.0     &14.00   &$+$0.26  &$+$0.17  &$+$0.34  &1&13.51   &$+$0.12  &$+$0.01  &1&2                   \\
\\
0115$+$159\dotfill&LHS 1227      & 64.9& 3.0&0.0      &13.85   &$+$0.10  &$+$0.11  &$+$0.20  &1&13.72   &$+$0.00  &$-$0.02  &1&1, 2                \\
0117$-$145\dotfill&LHS 1233      & 34.8& 1.2&1.05     &16.96   &$+$0.89  &$+$0.51  &$+$0.98  &1&15.47   &$+$0.30  &$+$0.09  &1&2, 7                \\
0121$+$401\dotfill&G133$-$8      & 32.1& 5.5&4.17     &17.12   &$+$0.79  &$+$0.45  &$+$0.90  &1&15.64   &$+$0.21  &$+$0.02  &2&2, 7                \\
0126$+$101\dotfill&G2$-$40       & 28.4& 3.1&15.9     &14.40   &$+$0.24  &$+$0.12  &$+$0.28  &1&14.05   &$+$0.13  &$-$0.02  &1&2, 6                \\
0135$-$052\dotfill&L870$-$2      & 81.0& 2.8&11.5     &12.86   &$+$0.33  &$+$0.23  &$+$0.48  &1&12.12   &$+$0.18  &$+$0.02  &1&1, 2                \\
\\
0142$+$312\dotfill&G72$-$31      & 28.2& 4.3&28.0     &14.78   &$+$0.22  &$+$0.11  &$+$0.26  &1&14.38   &$+$0.05  &$-$0.05  &1&2                   \\
0208$+$396\dotfill&G74$-$7       & 59.8& 3.5&12.2     &14.51   &$+$0.33  &$+$0.25  &$+$0.47  &1&13.80   &$+$0.15  &$+$0.02  &1&1, 2, 5             \\
0213$+$427\dotfill&LHS 153       & 50.2& 4.1&3.64     &16.22   &$+$0.73  &$+$0.45  &$+$0.85  &1&14.98   &$+$0.25: &$+$0.21: &2&2, 5, 6             \\
0222$+$648\dotfill&LHS 1405      & 31.4& 0.8&0.0      &18.29   &$+$1.30  &$+$0.71  &$+$1.34  &1&16.35   &$+$0.29  &$+$0.08  &2&1, 2                \\
0230$-$144\dotfill&LHS 1415      & 64.0& 3.9&3.20     &15.77   &$+$0.69  &$+$0.43  &$+$0.84  &1&14.43   &$+$0.26  &$+$0.06  &1&1, 2                \\
\\
0243$-$026\dotfill&LHS 1442      & 47.1& 5.0&10.2     &15.54   &$+$0.40  &$+$0.26  &$+$0.53  &2&14.71   &$+$0.22  &$-$0.02  &2&1, 2                \\
0245$+$541\dotfill&LHS 1446      & 96.6& 3.1&0.73     &15.36   &$+$0.92  &$+$0.52  &$+$1.00  &1&13.89   &$+$0.23  &$+$0.06  &2&1, 2, 6             \\
0257$+$080\dotfill&LHS 5064      & 35.9& 3.5&7.09     &15.90   &$+$0.43  &$+$0.28  &$+$0.58  &1&15.01   &$+$0.17  &$+$0.04  &1&1, 2, 4             \\
0324$+$738\dotfill&LP 31$-$40    & 25.0& 3.8&0.0      &17.68   &$+$1.27  &$+$0.77  &$+$1.39  &1&\nodata &\nodata  &\nodata  &0&2                   \\
0326$-$273\dotfill&L587$-$77A    & 57.6&13.6&25.3     &14.00   &\nodata  &\nodata  &\nodata  &1&13.27   &$+$0.15  &$+$0.04  &1&2, 8                \\
\\
0341$+$182\dotfill&Wolf 219      & 52.6& 3.0&0.0      &15.19   &$+$0.33  &$+$0.28  &$+$0.54  &1&14.56   &$+$0.21  &$-$0.05  &1&1, 2, 5             \\
0357$+$081\dotfill&LHS 1617      & 56.1& 3.7&5.97     &15.92   &$+$0.70  &$+$0.41  &$+$0.83  &1&14.59   &$+$0.26  &$+$0.07  &2&1, 2                \\
0426$+$588\dotfill&Stein 2051B   &180.6& 0.8&0.0      &12.43   &$+$0.30  &$+$0.26  &$+$0.57  &1&11.84   &$+$0.11  &$+$0.05  &1&2, 5, 6             \\
0433$+$270\dotfill&G39$-$27      & 60.2& 2.9&4.02     &15.81   &$+$0.67  &$+$0.41  &$+$0.80  &1&14.61   &$+$0.29  &$+$0.10  &1&1, 2                \\
0435$-$088\dotfill&L879$-$14     &105.2& 2.6&0.0      &13.75   &$+$0.33  &$+$0.32  &$+$0.57  &2&13.00   &$+$0.15  &$+$0.06  &1&1, 2, 5             \\
\\
0440$+$510\dotfill&G175$-$46     & 21.6& 5.2&19.0     &15.95   &$+$0.24  &$+$0.13  &$+$0.26  &1&15.60   &$+$0.10  &$-$0.03  &1&2                   \\
0503$-$174\dotfill&LHS 1734      & 45.6& 4.0&3.82     &16.01   &$+$0.72  &$+$0.44  &$+$0.90  &1&14.55   &$+$0.22  &$+$0.10  &2&1, 2, 4             \\
0518$+$333\dotfill&G86$-$B1B     & 15.3& 2.2&20.1     &16.06   &$+$0.22  &$+$0.19  &$+$0.41  &1&15.52   &$+$0.19  &$-$0.01  &1&2                   \\
0548$-$001\dotfill&G99$-$37      & 90.3& 2.8&0.0      &14.56   &$+$0.52  &$+$0.34  &$+$0.61  &1&13.73   &$+$0.10  &$+$0.00  &2&1, 2, 4             \\
0551$+$468\dotfill&LHS 1801      & 32.3& 1.5&2.18     &17.23   &$+$0.84  &$+$0.47  &$+$0.94  &1&15.84   &$+$0.29  &$+$0.02  &2&2, 7                \\
\\
0552$-$041\dotfill&LP 658$-$2    &155.0& 2.1&0.0      &14.47   &$+$1.01  &$+$0.50  &$+$0.98  &2&13.02   &$+$0.12  &$+$0.08  &2&1, 2, 5             \\
0553$+$053\dotfill&G99$-$47      &125.0& 3.6&8.23     &14.16   &$+$0.62  &$+$0.38  &$+$0.75  &2&12.96   &$+$0.19  &$+$0.11  &2&1, 2, 4, 5          \\
0618$+$067\dotfill&LHS 1838      & 44.2& 4.2&5.52     &16.41   &$+$0.56  &$+$0.36  &$+$0.74  &2&15.29   &$+$0.24  &$+$0.05  &2&1, 2                \\
0644$+$025\dotfill&G108$-$26     & 54.2& 5.5&15.3     &15.71   &$+$0.35  &$+$0.23  &$+$0.46  &2&15.00   &$+$0.15  &$-$0.08  &2&1, 2                \\
0648$+$641\dotfill&LP 58$-$53    & 30.0& 5.2&6.24     &16.65   &$+$0.56  &$+$0.36  &$+$0.74  &1&15.46   &$+$0.27  &$+$0.07  &1&2                   \\
\\
0651$-$479\dotfill&ESO 207$-$21  & 16.1& 1.6&0.0      &19.54   &$+$1.21  &$+$0.65  &$+$1.24  &3&17.70   &$+$0.23  &$+$0.02  &2&1, 9                \\
0654$+$027\dotfill&G108$-$42     & 26.0& 3.7&0.0      &16.17   &$+$0.12  &$+$0.15  &$+$0.20  &1&15.98   &$+$0.00  &$+$0.00  &1&2                   \\
0657$+$320\dotfill&LHS 1889      & 53.5& 0.9&1.60     &16.62   &$+$1.02  &$+$0.53  &$+$1.01  &1&14.99   &$+$0.22  &$+$0.08  &5&1, 2                \\
0659$-$064\dotfill&LHS 1892      & 81.0&24.2&8.31     &15.43   &$+$0.43  &$+$0.30  &$+$0.60  &1&14.58   &$+$0.29  &$+$0.05  &1&2, 5                \\
0706$+$377\dotfill&G87$-$29      & 41.2& 2.4&0.0      &15.64   &$+$0.30  &$+$0.26  &$+$0.48  &1&15.00   &$+$0.12  &$+$0.06  &1&1, 2                \\
\\
0727$+$482A\dotfill&G107$-$70A    & 90.0& 1.0&1.41     &15.26   &$+$0.98  &$+$0.53  &$+$1.02  &1&13.66   &$+$0.24  &$+$0.09  &3&1, 2, 5, 7          \\
0727$+$482B\dotfill&G107$-$70B    & 90.0& 1.0&1.41     &15.56   &$+$0.98  &$+$0.53  &$+$1.02  &1&13.96   &$+$0.24  &$+$0.09  &3&1, 2, 5, 7          \\
0738$-$172\dotfill&L745$-$46A    &112.4& 2.7&4.88     &13.06   &$+$0.24  &$+$0.18  &$+$0.34  &1&12.65   &$+$0.04  &$+$0.09  &3&2, 5                \\
0743$-$340\dotfill&vB 3          & 57.5& 3.1&0.0      &16.60   &$+$1.21  &$+$0.64  &$+$1.21  &2&14.85   &$+$0.14  &$+$0.16  &2&1, 10               \\
0747$+$073A\dotfill&LHS 240       & 54.7& 0.7&0.41     &16.63   &$+$1.06  &$+$0.58  &$+$1.08  &2&14.96   &$+$0.23  &$+$0.01  &2&1, 2, 5             \\
\\
0747$+$073B\dotfill&LHS 239       & 54.7& 0.7&0.0      &16.96   &$+$1.21  &$+$0.65  &$+$1.26  &2&15.05   &$+$0.15  &$+$0.04  &2&1, 2, 5             \\
0751$+$578\dotfill&G193$-$78     & 28.4& 3.8&0.0      &15.08   &$+$0.09  &$+$0.08  &$+$0.17  &1&14.94   &$+$0.00  &$-$0.02  &1&2                   \\
0752$-$676\dotfill&BPM 4729      &141.2& 8.4&4.70     &13.95   &$+$0.65  &$+$0.37  &$+$0.75  &1&12.79   &$+$0.27  &$+$0.09  &2&1, 2                \\
0752$+$365\dotfill&G90$-$28      & 29.5& 4.3&13.2     &16.09   &$+$0.29  &$+$0.18  &$+$0.40  &1&15.50   &$+$0.15  &$+$0.00  &1&2                   \\
0802$+$386\dotfill&LP 257$-$28   & 24.0& 3.0&0.0      &15.56   &$+$0.04  &$+$0.05  &$+$0.08  &1&15.60   &$+$0.02  &$-$0.06  &1&2                   \\
\\
0816$+$387\dotfill&G111$-$71     & 25.2& 2.5&13.1     &16.51   &$+$0.30  &$+$0.22  &$+$0.44  &1&15.87   &$+$0.15  &$-$0.01  &1&2                   \\
0827$+$328\dotfill&LHS 2022      & 44.9& 3.8&13.4     &15.73   &$+$0.32  &$+$0.22  &$+$0.47  &1&15.01   &$+$0.16  &$+$0.01  &1&2                   \\
0839$-$327\dotfill&L532$-$81     &112.7& 9.7&26.7     &11.90   &$+$0.25  &\nodata  &\nodata  &0&11.59   &$+$0.04  &$+$0.00  &1&2, 8                \\
0856$+$331\dotfill&G47$-$18      & 48.8& 3.4&0.0      &15.16   &$+$0.03  &$+$0.13  &$+$0.19  &1&15.12   &$+$0.03  &$-$0.02  &1&1, 2                \\
0912$+$536\dotfill&G195$-$19     & 97.3& 1.9&0.0      &13.84   &$+$0.35  &$+$0.20  &$+$0.33  &2&13.22   &$+$0.07  &$+$0.06  &1&1, 2, 4, 5          \\
\\
0913$+$442\dotfill&G116$-$16     & 34.6& 4.0&22.0     &15.37   &$+$0.22  &$+$0.15  &$+$0.30  &1&14.96   &$+$0.12  &$-$0.03  &1&2                   \\
0930$+$294\dotfill&G117$-$25     & 31.2& 4.6&21.2     &15.98   &$+$0.24  &$+$0.15  &$+$0.32  &1&15.51   &$+$0.11  &$-$0.05  &1&2                   \\
0946$+$534\dotfill&G195$-$42     & 43.5& 3.5&0.0      &15.18   &$+$0.17  &$+$0.13  &$+$0.28  &1&14.90   &$+$0.03  &$-$0.01  &2&1, 2                \\
0955$+$247\dotfill&G49$-$33      & 40.9& 4.5&20.1     &15.06   &$+$0.24  &$+$0.12  &$+$0.32  &1&14.66   &$+$0.07  &$-$0.06  &1&2                   \\
1012$+$083\dotfill&G43$-$38      & 34.5& 3.9&8.68     &16.03   &$+$0.41  &$+$0.27  &$+$0.57  &1&15.20   &$+$0.21  &$+$0.05  &1&2                   \\
\\
1019$+$637\dotfill&LP 62$-$147   & 61.2& 3.6&11.5     &14.70   &$+$0.38  &$+$0.25  &$+$0.53  &1&13.83   &$+$0.20  &$-$0.02  &1&2                   \\
1039$+$145\dotfill&G44$-$32      & 22.2& 3.5&0.0      &16.51   &$+$0.28  &$+$0.22  &$+$0.41  &1&15.93   &$+$0.07  &$+$0.08  &2&1, 2                \\
1043$-$188\dotfill&LHS 290       & 56.9& 6.5&0.0      &15.52   &$+$0.57  &$+$0.49  &\nodata  &1&14.62   &$+$0.21  &$+$0.05  &1&2, 5                \\
1055$-$072\dotfill&LHS 2333      & 82.3& 3.5&0.0      &14.33   &$+$0.30  &$+$0.20  &$+$0.42  &1&13.81   &$+$0.10  &$+$0.02  &1&2, 5                \\
1108$+$207\dotfill&LHS 2364      & 38.3& 2.7&0.0      &17.70   &$+$0.97  &$+$0.53  &$+$1.07  &2&15.91   &$+$0.22  &$+$0.07  &2&1, 2, 5             \\
\\
1115$-$029\dotfill&LHS 2392      & 26.3& 5.1&0.0      &15.34   &$+$0.08  &$+$0.09  &$+$0.17  &1&15.23   &$-$0.04: &$-$0.02: &1&2                   \\
1121$+$216\dotfill&Ross 627      & 74.4& 2.8&14.0     &14.21   &$+$0.31  &$+$0.20  &$+$0.45  &1&13.58   &$+$0.18  &$+$0.00  &1&2, 5                \\
1124$-$296\dotfill&ESO 439$-$80  & 16.4& 1.7&29.3     &15.02   &\nodata  &$+$0.07  &$+$0.26  &1&14.90   &$+$0.02  &$+$0.08  &1&1, 9                \\
1136$-$286\dotfill&ESO 439$-$26  & 24.5& 1.0&0.0      &20.52   &$+$1.03  &$+$0.64  &$+$1.14  &3&18.64   &\nodata  &\nodata  &1&1, 11               \\
1142$-$645\dotfill&LHS 43        &218.3& 6.7&0.0      &11.49   &$+$0.17  &$+$0.16  &$+$0.29  &1&11.19   &$+$0.07  &$+$0.03  &1&2                   \\
\\
1147$+$255\dotfill&LP 375$-$51   & 19.5& 2.9&43.9     &15.66   &$+$0.15  &$+$0.06  &$+$0.13  &1&15.53   &$+$0.04  &$-$0.02  &1&2                   \\
1154$+$186\dotfill&LP 434$-$97   & 33.5& 3.5&0.0      &15.61   &$+$0.26  &$+$0.18  &$+$0.37  &1&15.15   &$+$0.12  &$+$0.01  &1&2                   \\
1208$+$576\dotfill&LHS 2522      & 48.9& 4.6&4.87     &15.78   &$+$0.57  &$+$0.37  &$+$0.74  &1&14.64   &$+$0.25  &$+$0.07  &1&2                   \\
1215$+$323\dotfill&G148$-$B4B    & 32.2& 3.6&0.0      &17.00   &$+$0.33  &$+$0.23  &$+$0.60  &2&16.39   &$+$0.15  &$-$0.13: &1&2                   \\
1236$-$495\dotfill&LTT 4816      & 61.0& 9.4&45.5     &13.80   &$+$0.18  &$-$0.02  &$-$0.03  &1&13.92   &$+$0.02  &$-$0.08  &1&2                   \\
\\
1244$+$149\dotfill&G61$-$17      & 15.7& 4.2&39.6     &15.89   &$+$0.17  &$+$0.03  &$+$0.06  &1&15.84   &$-$0.02  &$+$0.02  &1&2                   \\
1247$+$550\dotfill&LP 131$-$66   & 39.6& 0.7&0.0      &17.79   &$+$1.44  &$+$0.76  &$+$1.45  &1&15.72   &$+$0.05  &$+$0.04  &2&1, 2, 5             \\
1257$+$037\dotfill&LHS 2661      & 60.3& 3.8&5.07     &15.84   &$+$0.66  &$+$0.38  &$+$0.76  &2&14.56   &$+$0.23  &$+$0.08  &2&1, 2, 5             \\
1257$+$278\dotfill&G149$-$28     & 28.9& 4.1&20.4     &15.41   &$+$0.23  &$+$0.15  &$+$0.28  &1&14.99   &$+$0.08  &$+$0.00: &1&2                   \\
1300$+$263\dotfill&LHS 2673      & 28.4& 3.3&0.0      &18.77   &$+$1.24  &$+$0.68  &$+$1.28  &2&16.89   &$+$0.18  &$+$0.01  &2&5, 12               \\
\\
1310$-$472\dotfill&ER 8          & 66.5& 2.4&0.0      &17.13   &$+$1.39  &$+$0.72  &$+$1.41  &2&15.21   &$+$0.10  &$+$0.08  &1&1, 2                \\
1313$-$198\dotfill&LHS 2710      & 40.6& 1.8&0.0      &17.14   &$+$0.95  &$+$0.43  &$+$0.82  &2&15.87   &$+$0.17  &$+$0.14  &4&1, 9                \\
1325$+$581\dotfill&G199$-$71     & 27.0& 5.5&9.74     &16.70   &$+$0.40  &$+$0.28  &$+$0.54  &1&15.82   &$+$0.14  &$+$0.03  &1&2                   \\
1328$+$307\dotfill&G165$-$7      & 33.4& 5.3&0.0      &16.03   &$+$0.70  &$+$0.29  &$+$0.43  &1&15.50   &$+$0.14  &$+$0.02  &1&2                   \\
1334$+$039\dotfill&Wolf 489      &121.4& 3.4&1.19     &14.63   &$+$0.95  &$+$0.51  &$+$1.01  &1&13.06   &$+$0.26  &$+$0.10  &2&1, 2, 5             \\
\\
1344$+$106\dotfill&LHS 2800      & 49.9& 3.6&11.6     &15.12   &$+$0.38  &$+$0.22  &$+$0.48  &1&14.38   &$+$0.18  &$+$0.01  &1&2, 5                \\
1345$+$238\dotfill&LP 380$-$5    & 82.9& 2.2&0.46     &15.71   &$+$1.15  &$+$0.59  &$+$1.13  &2&13.92   &$+$0.25  &$+$0.08  &2&1, 2, 5             \\
1418$-$088\dotfill&G124$-$26     & 24.6& 3.8&17.9     &15.38   &$+$0.28  &$+$0.17  &$+$0.38  &1&14.81   &$+$0.12  &$+$0.00  &1&2                   \\
1444$-$174\dotfill&LHS 378       & 69.0& 4.0&0.0      &16.44   &$+$1.03  &$+$0.49  &$+$1.01  &1&14.94   &$+$0.15  &$+$0.11  &2&1, 2, 5             \\
1455$+$298\dotfill&LHS 3007      & 28.9& 4.1&13.0     &15.60   &$+$0.31  &$+$0.22  &$+$0.42  &1&14.86   &$+$0.13  &$+$0.01  &1&2                   \\
\\
1503$-$070\dotfill&GD 175        & 38.5& 5.5&9.08     &15.88   &$+$0.40  &$+$0.28  &$+$0.55  &1&15.07   &$+$0.14  &$+$0.02  &1&2, 4                \\
1544$-$377\dotfill&L481$-$60     & 74.6&10.1&33.8     &12.97:  &$+$0.18: &$+$0.22: &$+$0.60: &1&\nodata &\nodata  &\nodata  &0&2, 13               \\
1606$+$422\dotfill&Case 2        & 22.2& 3.5&48.8     &13.82   &$+$0.11  &$-$0.03  &$-$0.06  &1&13.92   &$+$0.00  &$-$0.09  &1&2                   \\
1609$+$135\dotfill&LHS 3163      & 54.5& 4.7&25.9     &15.11   &$+$0.20  &$+$0.10  &$+$0.24  &1&14.77   &$+$0.01  &$+$0.01  &1&2                   \\
1625$+$093\dotfill&G138$-$31     & 42.8& 3.7&10.8     &16.14   &$+$0.34  &$+$0.27  &$+$0.52  &3&15.34   &$+$0.22  &$+$0.06  &3&1, 2                \\
\\
1626$+$368\dotfill&Ross 640      & 62.7& 2.0&8.99     &13.83   &$+$0.19  &$+$0.08  &$+$0.17  &2&13.58   &$+$0.01  &$-$0.01  &1&2                   \\
1633$+$433\dotfill&G180$-$63     & 66.2& 3.0&9.11     &14.84   &$+$0.43  &$+$0.27  &$+$0.56  &1&13.95   &$+$0.19  &$+$0.03  &1&2                   \\
1633$+$572\dotfill&G225$-$68     & 69.2& 2.5&0.0      &14.99   &$+$0.50  &$+$0.31  &$+$0.61  &1&14.03   &$+$0.06  &$-$0.02  &2&1, 2, 5             \\
1635$+$137\dotfill&G138$-$47     & 25.5& 4.6&11.1     &16.94   &$+$0.38  &$+$0.27  &$+$0.47  &2&16.11   &$+$0.15  &$-$0.02: &1&2, 7                \\
1637$+$335\dotfill&G180$-$65     & 35.0& 3.2&34.3     &14.65   &$+$0.18  &$+$0.06  &$+$0.14  &1&14.56   &$+$0.06  &$-$0.04  &1&2                   \\
\\
1639$+$537\dotfill&GD 356        & 47.4& 3.5&-4.85    &15.05   &$+$0.29  &$+$0.21  &$+$0.39  &2&14.54   &$+$0.08  &$+$0.04  &1&2, 4                \\
1655$+$215\dotfill&LHS 3254      & 43.0& 3.1&26.4     &14.13   &$+$0.21  &$+$0.10  &$+$0.21  &1&13.89   &$+$0.09  &$-$0.05  &1&2                   \\
1656$-$062\dotfill&LP 686$-$32   & 30.7& 0.9&4.21     &17.14   &$+$0.79  &$+$0.34  &$+$0.72  &2&15.89   &$+$0.25  &$+$0.08  &1&1, 9                \\
1705$+$030\dotfill&G139$-$13     & 57.0& 5.4&0.0      &15.20   &$+$0.43  &$+$0.24  &$+$0.46  &1&14.62   &$+$0.12  &$+$0.02  &1&2                   \\
1716$+$020\dotfill&G19$-$20      & 28.1& 2.6&60.2     &14.36   &$+$0.12  &$-$0.04  &$-$0.04  &1&14.68   &$+$0.03  &$-$0.06: &1&2                   \\
\\
1733$-$544\dotfill&L270$-$137    & 45.6& 0.8&7.35     &15.70   &$+$0.46  &$+$0.29  &$+$0.53  &2&14.89   &$+$0.34  &$+$0.09  &1&1, 9                \\
1736$+$052\dotfill&G140$-$2      & 23.4& 3.9&38.8     &15.92   &$+$0.23  &$+$0.12  &$+$0.26  &1&15.62   &$+$0.06  &$+$0.07  &1&2                   \\
1748$+$708\dotfill&G240$-$72     &164.7& 2.4&0.0      &14.13   &$+$0.48  &$+$0.53  &$+$1.05  &1&12.77   &$+$0.07  &$+$0.20  &1&1, 2, 4, 5          \\
1756$+$827\dotfill&LHS 56        & 63.9& 2.9&12.1     &14.34   &$+$0.35  &$+$0.22  &$+$0.47  &2&\nodata &\nodata  &\nodata  &0&2, 5                \\
1811$+$327A\dotfill&G206$-$17     & 19.5& 4.3&10.7     &16.40   &$+$0.30  &$+$0.20  &$+$0.41  &1&15.71   &$+$0.15  &$+$0.02  &1&2                   \\
\\
1811$+$327B\dotfill&G206$-$18     & 19.5& 4.3&6.82     &17.07   &$+$0.44  &$+$0.28  &$+$0.59  &1&16.08   &$+$0.14  &$+$0.12: &1&2, 6                \\
1818$+$126\dotfill&G141$-$2      & 24.5& 5.5&5.08     &16.09   &$+$0.51  &$+$0.33  &$+$0.67  &1&15.07   &$+$0.17  &$+$0.03  &1&1, 2                \\
1820$+$609\dotfill&G227$-$28     & 78.2& 4.1&0.82     &15.69   &$+$0.98  &$+$0.54  &$+$1.05  &1&13.96   &$+$0.23  &$+$0.08  &2&2, 7                \\
1824$+$040\dotfill&G21$-$15      & 18.2& 2.3&33.8     &13.89   &$+$0.09  &$-$0.05  &$-$0.07  &1&14.07   &$-$0.07  &$+$0.00  &1&2, 6                \\
1826$-$045\dotfill&G21$-$16      & 34.9& 3.8&23.8     &14.58   &$+$0.21  &$+$0.11  &$+$0.26  &1&14.44:  &$+$0.00: &$+$0.05: &1&2, 6                \\
\\
1829$+$547\dotfill&G227$-$35     & 66.8& 5.6&0.0      &15.57   &$+$0.49  &$+$0.29  &$+$0.60  &1&14.76   &$+$0.15  &$+$0.11  &1&2, 4                \\
1831$+$197\dotfill&G184$-$12     & 17.9& 4.8&0.0      &16.41   &$+$0.29  &$+$0.23  &$+$0.37  &3&15.93   &$+$0.11  &$+$0.01  &1&2                   \\
1840$+$042\dotfill&GD 215        & 40.2& 3.4&24.3     &14.79   &$+$0.22  &$+$0.13  &$+$0.27  &1&14.53:  &$+$0.07: &$-$0.04: &1&2                   \\
1855$+$338\dotfill&G207$-$9      & 30.5& 4.4&54.8     &14.64   &$+$0.17  &$+$0.01  &$-$0.02  &1&14.74   &$+$0.02  &$-$0.05  &1&2                   \\
1900$+$705\dotfill&GW+70 8247    & 77.0& 2.3&0.0      &13.25   &$+$0.06  &$+$0.01  &$+$0.02  &1&\nodata &\nodata  &\nodata  &0&2, 4                \\
\\
1917$+$386\dotfill&G125$-$3      & 85.5& 3.4&0.0      &14.61   &$+$0.45  &$+$0.30  &$+$0.57  &1&13.77   &$+$0.08  &$+$0.10  &1&1, 2                \\
1953$-$011\dotfill&LHS 3501      & 87.8& 2.9&13.8     &13.69   &$+$0.28  &$+$0.19  &$+$0.38  &2&13.12   &$+$0.10  &$+$0.00  &1&2, 4                \\
2002$-$110\dotfill&LHS 483       & 57.7& 0.8&0.0      &16.95   &$+$1.16  &$+$0.59  &$+$1.09  &2&15.32   &$+$0.21  &$+$0.02  &2&1, 2, 5, 6          \\
2011$+$065\dotfill&G24$-$9       & 44.7& 1.9&0.0      &15.78   &$+$0.38  &$+$0.27  &$+$0.53  &2&14.94   &$+$0.15  &$+$0.04  &2&1, 2                \\
2048$+$263\dotfill&G187$-$8      & 49.8& 3.4&1.80     &15.63   &$+$0.92  &$+$0.52  &$+$1.00  &1&14.12   &$+$0.29  &$+$0.04  &2&1, 2                \\
\\
2054$-$050\dotfill&vB 11         & 64.6& 5.1&0.0      &16.69   &$+$1.20  &$+$0.66  &$+$1.32  &1&14.82   &$+$0.21  &$+$0.07  &1&1, 2, 5             \\
2059$+$190\dotfill&G144$-$51     & 26.1& 4.4&7.91     &16.36   &$+$0.43  &$+$0.26  &$+$0.54  &1&15.52   &$+$0.16  &$+$0.02  &2&1, 2                \\
2059$+$247\dotfill&G187$-$16     & 36.1& 4.2&7.75     &16.57   &$+$0.59  &$+$0.31  &$+$0.64  &1&15.45   &$+$0.16  &$+$0.03  &1&2, 7                \\
2059$+$316\dotfill&G187$-$15     & 29.0& 3.5&0.0      &15.04   &$+$0.11  &$+$0.07  &$+$0.14  &1&14.94   &$-$0.03  &$-$0.01  &1&2                   \\
2105$-$820\dotfill&L24$-$52      & 58.6& 8.8&35.5     &13.61   &$+$0.21  &$+$0.05  &$+$0.14  &1&13.52   &$-$0.01  &$-$0.05  &1&2                   \\
\\
2107$-$216\dotfill&LHS 3636      & 42.2& 1.5&6.45     &16.80   &$+$0.59  &$+$0.35  &$+$0.77  &1&15.63   &$+$0.18: &$+$0.05: &2&1, 9                \\
2111$+$261\dotfill&G187$-$32     & 31.4& 3.7&18.2     &14.68   &$+$0.24  &$+$0.16  &$+$0.32  &1&14.15   &$+$0.07  &$-$0.01  &1&2, 6                \\
2136$+$229\dotfill&G126$-$18     & 23.8& 3.1&31.9     &15.25   &$+$0.17  &$+$0.06  &$+$0.14  &1&15.04   &$+$0.08  &$-$0.13  &1&2, 6                \\
2140$+$207\dotfill&LHS 3703      & 79.9& 3.2&0.0      &13.24   &$+$0.13  &$+$0.14  &$+$0.26  &1&12.95   &$+$0.02  &$-$0.02  &1&2, 6                \\
2154$-$512\dotfill&BPM 27606     & 68.5&10.6&0.0      &14.74   &$+$0.19  &$+$0.44  &$+$0.61  &1&13.47   &$+$0.00  &$+$0.18  &1&1, 2, 13            \\
\\
2207$+$142\dotfill&G18$-$34      & 39.2& 4.4&12.8     &15.60   &$+$0.29  &$+$0.20  &$+$0.42  &1&14.99   &$+$0.18  &$-$0.03  &1&2, 6                \\
2246$+$223\dotfill&G67$-$23      & 52.5& 4.1&49.2     &14.39   &$+$0.17  &$+$0.05  &$+$0.14  &1&14.28   &$-$0.03  &$-$0.06  &1&2                   \\
2248$+$293\dotfill&G128$-$7      & 47.8& 4.2&4.70     &15.54   &$+$0.66  &$+$0.40  &$+$0.79  &1&14.24   &$+$0.23  &$+$0.07  &2&1, 2, 5             \\
2251$-$070\dotfill&LP 701$-$29   &123.7& 4.3&0.0      &15.71   &$+$1.84  &$+$0.61  &$+$1.15  &2&13.86   &$+$0.23  &$+$0.16  &2&1, 2, 5             \\
2253$-$081\dotfill&G156$-$64     & 36.7& 5.3&8.45     &16.48   &$+$0.40  &$+$0.27  &$+$0.56  &1&15.59   &$+$0.12  &$+$0.11: &2&2                   \\
\\
2311$-$068\dotfill&G157$-$34     & 39.8& 4.7&0.0      &15.40   &$+$0.23  &$+$0.18  &$+$0.35  &1&14.98   &$+$0.05  &$+$0.03  &2&2, 6                \\
2312$-$024\dotfill&LHS 3917      & 37.5& 5.9&0.0      &16.31   &$+$0.51  &$+$0.26  &$+$0.49  &1&15.70   &$+$0.17  &$-$0.05  &1&1, 2                \\
2316$-$064\dotfill&LHS 542       & 32.2& 3.7&0.0      &18.15   &$+$1.08  &$+$0.62  &$+$1.16  &1&16.38   &$+$0.24  &$+$0.04  &2&5, 12               \\
2329$+$267\dotfill&G128$-$72     & 25.9& 4.7&32.4     &15.33   &$+$0.27  &$+$0.06  &$+$0.17  &2&15.13   &$+$0.10  &$-$0.15  &1&2                   \\
2345$-$447\dotfill&ESO 292$-$43  & 38.1& 2.2&0.0      &17.87   &$+$0.79  &$+$0.42  &$+$0.84  &1&16.66   &$+$0.07  &$+$0.26  &2&1, 9                \\
\\
2347$+$292\dotfill&LHS 4019      & 46.5& 4.1&4.12     &15.76   &$+$0.59  &$+$0.35  &$+$0.72  &1&14.59   &$+$0.24  &$+$0.11  &1&1, 2, 6             \\
2352$+$401\dotfill&G171$-$27     & 38.7& 5.6&0.0      &14.94   &$+$0.19  &$+$0.19  &$+$0.32  &1&14.57   &$+$0.05  &$+$0.02  &1&1, 2                \\
\enddata
\tablecomments{
(1) In common with \citet{brl}; 
(2) $\pi$ from Yale Parallax Catalog; 
(3) Spectrum from \citet{schmidt94}; 
(4) Magnetic; 
(5) In common with \citet{lrb98}; 
(6) Spectrum from \citet{greenstein86}; 
(7) Newly identified DA white dwarf; 
(8) $V$ and $B$--$V$ from \citet{mccook99}; 
(9) $\pi$ from \citet{ruiz96}; 
(10) $\pi$ from \citet{ruiz89}; 
(11) $\pi$ from \citet{ruizetal95}; 
(12) $\pi$ from \citet{ldm88}; 
(13) Not analyzed (see text). }
\end{deluxetable}

\clearpage
\hoffset 0.0truein
\begin{deluxetable}{p{2.3cm}rlcllllc}
\tabletypesize{\footnotesize}
\tablecolumns{9}
\tablewidth{0pt}
\tablecaption{Atmospheric Parameters of Cool White Dwarfs}
\tablehead{
\colhead{} &
\colhead{$\Te$} &
\colhead{} &
\colhead{} &
\colhead{} &
\colhead{} &
\colhead{} &
\colhead{Age$^a$} &
\colhead{}\\
\colhead{WD} &
\colhead{(K)} &
\colhead{log $g$} &
\colhead{Comp} &
\colhead{$M/$\msun} &
\colhead{$M_V$} &
\colhead{log $L/$\lsun} &
\colhead{(Gyr)} &
\colhead{Notes}}
\startdata
0000$-$345\dotfill& 6240 (140)                        &8.31 (0.16)    &He&0.77 (0.11)    &14.41 (0.26)   &$-$3.85 (0.11) &4.18 (0.89)    &1                   \\
0009$+$501\dotfill& 6540 (150)                        &8.23 (0.06)    &H &0.74 (0.04)    &14.15 (0.09)   &$-$3.71 (0.03) &2.97 (0.38)    &2                   \\
0011$+$000\dotfill& 9610 (260)                        &8.40 (0.19)    &H &0.85 (0.12)    &12.94 (0.32)   &$-$3.15 (0.12) &1.29 (0.52)    &                    \\
0011$-$134\dotfill& 6010 (120)                        &8.20 (0.11)    &H &0.71 (0.07)    &14.44 (0.16)   &$-$3.84 (0.07) &3.65 (0.79)    &2                   \\
0029$-$032\dotfill& 4770 ( 50)                        &8.07 (0.04)    &He&0.61 (0.02)    &15.47 (0.05)   &$-$4.17 (0.02) &7.07 (0.18)    &                    \\
\\
0033$+$016\dotfill&10700 (350)                        &8.66 (0.16)    &H &1.02 (0.09)    &13.02 (0.29)   &$-$3.14 (0.12) &1.67 (0.45)    &                    \\
0038$-$226\dotfill& 5400 (170)                        &7.91 (0.17)    &He&0.52 (0.10)    &14.53 (0.22)   &$-$3.87 (0.09) &3.51 (1.21)    &3                   \\
0038$+$555\dotfill&10900 (630)                        &8.14 (0.07)    &He&0.67 (0.04)    &12.24 (0.10)   &$-$2.77 (0.04) &0.62 (0.05)    &4                   \\
0046$+$051\dotfill& 6770 (200)                        &8.40 (0.01)    &He&0.83 (0.01)    &14.22 (0.02)   &$-$3.77 (0.01) &3.67 (0.04)    &5                   \\
0101$+$048\dotfill& 8080 (200)                        &7.55 (0.14)    &H &0.37 (0.05)    &12.36 (0.18)   &$-$2.96 (0.07) &0.63 (0.07)    &6                   \\
\\
0115$+$159\dotfill& 9800 (360)                        &8.38 (0.06)    &He&0.82 (0.04)    &12.91 (0.10)   &$-$3.10 (0.04) &1.17 (0.16)    &4                   \\
0117$-$145\dotfill& 5150 (110)                        &7.75 (0.06)    &H &0.44 (0.03)    &14.67 (0.07)   &$-$3.86 (0.03) &2.63 (0.27)    &                    \\
0121$+$401\dotfill& 5340 (120)                        &7.90 (0.29)    &H &0.52 (0.16)    &14.65 (0.38)   &$-$3.88 (0.15) &2.93 (1.87)    &                    \\
0126$+$101\dotfill& 8500 (200)                        &7.20 (0.17)    &H &0.25 (0.04)    &11.67 (0.24)   &$-$2.68 (0.10) &0.43 (0.06)    &6                   \\
0135$-$052\dotfill& 7140 (160)                        &7.19 (0.06)    &H &0.25 (0.01)    &12.40 (0.07)   &$-$3.00 (0.03) &0.66 (0.03)    &6                   \\
\\
0142$+$312\dotfill& 8660 (210)                        &7.51 (0.25)    &H &0.35 (0.10)    &12.03 (0.33)   &$-$2.82 (0.13) &0.52 (0.11)    &6                   \\
0208$+$396\dotfill& 7310 (180)                        &8.01 (0.09)    &H &0.60 (0.05)    &13.39 (0.13)   &$-$3.38 (0.05) &1.38 (0.17)    &                    \\
0213$+$427\dotfill& 5600 (160)                        &8.12 (0.12)    &H &0.66 (0.08)    &14.72 (0.18)   &$-$3.91 (0.07) &3.89 (1.09)    &                    \\
0222$+$648\dotfill& 4520 ( 40)                        &8.00 (0.05)    &He&0.57 (0.03)    &15.77 (0.06)   &$-$4.23 (0.02) &7.08 (0.46)    &                    \\
0230$-$144\dotfill& 5480 (120)                        &8.11 (0.09)    &H &0.65 (0.06)    &14.80 (0.13)   &$-$3.95 (0.05) &4.28 (0.94)    &                    \\
\\
0243$-$026\dotfill& 6820 (160)                        &8.18 (0.15)    &H &0.70 (0.10)    &13.90 (0.23)   &$-$3.61 (0.09) &2.27 (0.78)    &                    \\
0245$+$541\dotfill& 5280 (120)                        &8.28 (0.05)    &H &0.76 (0.03)    &15.28 (0.07)   &$-$4.11 (0.03) &6.93 (0.30)    &                    \\
0257$+$080\dotfill& 6680 (150)                        &7.96 (0.16)    &H &0.57 (0.09)    &13.68 (0.21)   &$-$3.51 (0.08) &1.60 (0.38)    &2                   \\
0324$+$738\dotfill& 4650 ( 50)                        &7.10 (0.42)    &He&0.19 (0.10)    &14.67 (0.33)   &$-$3.77 (0.14) &1.97 (0.74)    &6                   \\
0326$-$273\dotfill& 7200 (200)                        &7.53 (0.40)    &H &0.35 (0.15)    &12.80 (0.52)   &$-$3.15 (0.21) &0.82 (0.30)    &6                   \\
\\
0341$+$182\dotfill& 6900 (170)                        &8.16 (0.09)    &He&0.67 (0.05)    &13.80 (0.12)   &$-$3.58 (0.05) &2.24 (0.35)    &4                   \\
0357$+$081\dotfill& 5490 (130)                        &8.02 (0.11)    &H &0.60 (0.06)    &14.66 (0.14)   &$-$3.90 (0.06) &3.40 (0.87)    &                    \\
0426$+$588\dotfill& 7120 (180)                        &8.17 (0.01)    &He&0.68 (0.00)    &13.71 (0.01)   &$-$3.54 (0.01) &2.08 (0.02)    &                    \\
0433$+$270\dotfill& 5620 (110)                        &8.14 (0.07)    &H &0.67 (0.05)    &14.71 (0.10)   &$-$3.92 (0.04) &4.07 (0.69)    &                    \\
0435$-$088\dotfill& 6620 (160)                        &8.08 (0.04)    &He&0.62 (0.02)    &13.86 (0.05)   &$-$3.61 (0.02) &2.22 (0.17)    &4                   \\
\\
0440$+$510\dotfill& 8710 (220)                        &7.96 (0.38)    &H &0.57 (0.21)    &12.62 (0.53)   &$-$3.05 (0.22) &0.82 (0.39)    &                    \\
0503$-$174\dotfill& 5300 (120)                        &7.60 (0.16)    &H &0.37 (0.07)    &14.31 (0.19)   &$-$3.74 (0.07) &1.86 (0.33)    &2                   \\
0518$+$333\dotfill& 7780 (180)                        &7.14 (0.24)    &H &0.24 (0.06)    &11.98 (0.31)   &$-$2.81 (0.12) &0.52 (0.09)    &6                   \\
0548$-$001\dotfill& 6400 (140)                        &8.32 (0.04)    &He&0.78 (0.03)    &14.34 (0.07)   &$-$3.81 (0.02) &3.90 (0.22)    &2, 4                \\
0551$+$468\dotfill& 5380 (120)                        &8.01 (0.08)    &H &0.59 (0.04)    &14.78 (0.10)   &$-$3.92 (0.04) &3.75 (0.73)    &                    \\
0552$-$041\dotfill& 5060 ( 60)                        &8.31 (0.02)    &He&0.77 (0.01)    &15.42 (0.03)   &$-$4.21 (0.02) &6.92 (0.05)    &5                   \\
0553$+$053\dotfill& 5790 (110)                        &8.20 (0.05)    &H &0.71 (0.03)    &14.65 (0.06)   &$-$3.90 (0.02) &4.11 (0.35)    &2                   \\
0618$+$067\dotfill& 5940 (120)                        &8.27 (0.14)    &H &0.76 (0.09)    &14.64 (0.21)   &$-$3.90 (0.09) &4.27 (0.90)    &                    \\
0644$+$025\dotfill& 7410 (180)                        &8.66 (0.12)    &H &1.01 (0.07)    &14.38 (0.22)   &$-$3.78 (0.08) &3.79 (0.23)    &                    \\
0648$+$641\dotfill& 5780 (120)                        &7.75 (0.30)    &H &0.45 (0.14)    &14.04 (0.38)   &$-$3.66 (0.15) &1.74 (0.65)    &                    \\
\\
0651$-$479\dotfill& 4660 ( 40)                        &8.02 (0.16)    &He&0.58 (0.10)    &15.57 (0.22)   &$-$4.19 (0.09) &7.00 (1.24)    &                    \\
0654$+$027\dotfill& 9450 (340)                        &8.51 (0.18)    &He&0.91 (0.12)    &13.24 (0.31)   &$-$3.26 (0.12) &1.79 (0.56)    &                    \\
0657$+$320\dotfill& 4990 (130)                        &8.07 (0.03)    &H &0.62 (0.02)    &15.26 (0.04)   &$-$4.09 (0.02) &6.55 (0.27)    &                    \\
0659$-$064\dotfill& 6520 (150)                        &8.71 (0.36)    &H &1.04 (0.22)    &14.97 (0.67)   &$-$4.04 (0.27) &4.81 (0.60)    &                    \\
0706$+$377\dotfill& 6990 (180)                        &8.14 (0.09)    &He&0.66 (0.05)    &13.71 (0.13)   &$-$3.55 (0.05) &2.09 (0.32)    &4                   \\
\\
0727$+$482A\dotfill& 5020 (120)                        &7.92 (0.02)    &H &0.53 (0.01)    &15.03 (0.02)   &$-$4.00 (0.01) &4.60 (0.22)    &6                   \\
0727$+$482B\dotfill& 5000 (130)                        &8.12 (0.02)    &H &0.66 (0.01)    &15.33 (0.02)   &$-$4.12 (0.01) &7.02 (0.14)    &6                   \\
0738$-$172\dotfill& 7710 (220)                        &8.09 (0.03)    &He&0.63 (0.02)    &13.31 (0.05)   &$-$3.35 (0.02) &1.45 (0.05)    &                    \\
0743$-$340\dotfill& 4740 ( 50)                        &7.97 (0.09)    &He&0.55 (0.05)    &15.40 (0.12)   &$-$4.14 (0.04) &6.36 (0.96)    &                    \\
0747$+$073A\dotfill& 4850 ( 50)                        &8.04 (0.02)    &He&0.59 (0.01)    &15.32 (0.03)   &$-$4.13 (0.01) &6.71 (0.15)    &                    \\
\\
0747$+$073B\dotfill& 4210 ( 90)                        &7.69 (0.03)    &H &0.40 (0.01)    &15.65 (0.03)   &$-$4.19 (0.01) &5.27 (0.23)    &                    \\
0751$+$578\dotfill& 9770 (360)                        &8.01 (0.21)    &He&0.58 (0.12)    &12.35 (0.29)   &$-$2.89 (0.12) &0.71 (0.20)    &                    \\
0752$-$676\dotfill& 5730 (110)                        &8.21 (0.09)    &H &0.72 (0.06)    &14.70 (0.13)   &$-$3.93 (0.06) &4.36 (0.70)    &                    \\
0752$+$365\dotfill& 7700 (190)                        &8.19 (0.21)    &H &0.71 (0.13)    &13.44 (0.32)   &$-$3.40 (0.13) &1.60 (0.69)    &                    \\
0802$+$386\dotfill&11070 (480)                        &8.31 (0.18)    &He&0.78 (0.11)    &12.46 (0.27)   &$-$2.85 (0.11) &0.76 (0.18)    &5                   \\
\\
0816$+$387\dotfill& 7570 (190)                        &8.19 (0.14)    &H &0.71 (0.09)    &13.52 (0.22)   &$-$3.43 (0.09) &1.67 (0.51)    &                    \\
0827$+$328\dotfill& 7270 (180)                        &8.39 (0.11)    &H &0.85 (0.07)    &13.99 (0.18)   &$-$3.63 (0.08) &3.06 (0.58)    &                    \\
0839$-$327\dotfill& 8930 (230)                        &7.70 (0.14)    &H &0.44 (0.07)    &12.16 (0.19)   &$-$2.86 (0.07) &0.57 (0.08)    &6                   \\
0856$+$331\dotfill&10390 (410)                        &8.84 (0.07)    &He&1.11 (0.04)    &13.60 (0.15)   &$-$3.33 (0.06) &2.07 (0.08)    &4                   \\
0912$+$536\dotfill& 7160 (190)                        &8.28 (0.03)    &He&0.75 (0.02)    &13.78 (0.04)   &$-$3.59 (0.02) &2.54 (0.16)    &2                   \\
\\
0913$+$442\dotfill& 8490 (220)                        &8.19 (0.17)    &H &0.71 (0.11)    &13.06 (0.25)   &$-$3.23 (0.10) &1.22 (0.34)    &                    \\
0930$+$294\dotfill& 8330 (220)                        &8.38 (0.20)    &H &0.84 (0.13)    &13.45 (0.32)   &$-$3.38 (0.12) &1.94 (0.74)    &                    \\
0946$+$534\dotfill& 8760 (290)                        &8.45 (0.11)    &He&0.87 (0.07)    &13.37 (0.18)   &$-$3.35 (0.07) &1.97 (0.41)    &4                   \\
0955$+$247\dotfill& 8670 (220)                        &8.27 (0.15)    &H &0.77 (0.10)    &13.12 (0.24)   &$-$3.24 (0.09) &1.31 (0.40)    &                    \\
1012$+$083\dotfill& 6750 (150)                        &8.02 (0.18)    &H &0.60 (0.10)    &13.72 (0.25)   &$-$3.53 (0.09) &1.72 (0.52)    &                    \\
\\
1019$+$637\dotfill& 6780 (160)                        &7.98 (0.09)    &H &0.58 (0.05)    &13.63 (0.13)   &$-$3.50 (0.05) &1.59 (0.22)    &                    \\
1039$+$145\dotfill& 7280 (200)                        &7.93 (0.26)    &He&0.53 (0.14)    &13.24 (0.35)   &$-$3.36 (0.14) &1.32 (0.47)    &1                   \\
1043$-$188\dotfill& 6190 (200)                        &8.09 (0.17)    &He&0.63 (0.11)    &14.30 (0.25)   &$-$3.74 (0.10) &2.85 (1.03)    &3                   \\
1055$-$072\dotfill& 7420 (200)                        &8.42 (0.06)    &He&0.85 (0.04)    &13.91 (0.09)   &$-$3.62 (0.04) &3.01 (0.22)    &                    \\
1108$+$207\dotfill& 4650 (160)                        &8.07 (0.11)    &H &0.62 (0.07)    &15.62 (0.15)   &$-$4.21 (0.06) &7.71 (0.98)    &                    \\
1115$-$029\dotfill&10120 (490)                        &8.13 (0.30)    &He&0.66 (0.18)    &12.44 (0.43)   &$-$2.90 (0.17) &0.76 (0.31)    &4                   \\
1121$+$216\dotfill& 7490 (180)                        &8.20 (0.05)    &H &0.72 (0.03)    &13.57 (0.08)   &$-$3.45 (0.03) &1.76 (0.22)    &                    \\
1124$-$296\dotfill& 9440 (250)                        &7.10 (0.15)    &H &0.23 (0.03)    &11.09 (0.23)   &$-$2.43 (0.10) &0.30 (0.04)    &6                   \\
1136$-$286\dotfill& 4490 ( 80)                        &9.02 (0.05)    &He&1.19 (0.02)    &17.47 (0.09)   &$-$4.94 (0.03) &4.96 (0.27)    &                    \\
1142$-$645\dotfill& 8490 (270)                        &8.27 (0.05)    &He&0.75 (0.03)    &13.19 (0.07)   &$-$3.28 (0.02) &1.44 (0.13)    &4                   \\
\\
1147$+$255\dotfill& 9790 (270)                        &7.92 (0.23)    &H &0.55 (0.13)    &12.11 (0.33)   &$-$2.82 (0.13) &0.57 (0.17)    &                    \\
1154$+$186\dotfill& 7630 (220)                        &8.06 (0.16)    &He&0.61 (0.10)    &13.23 (0.23)   &$-$3.35 (0.09) &1.44 (0.32)    &1                   \\
1208$+$576\dotfill& 5880 (120)                        &7.95 (0.15)    &H &0.56 (0.09)    &14.23 (0.20)   &$-$3.73 (0.08) &2.18 (0.60)    &                    \\
1215$+$323\dotfill& 7100 (210)                        &8.68 (0.13)    &He&1.02 (0.08)    &14.54 (0.24)   &$-$3.88 (0.10) &3.91 (0.10)    &1                   \\
1236$-$495\dotfill&11550 (470)                        &8.63 (0.18)    &H &1.00 (0.11)    &12.73 (0.34)   &$-$2.98 (0.13) &1.24 (0.44)    &7                   \\
\\
1244$+$149\dotfill&10280 (310)                        &7.88 (0.42)    &H &0.53 (0.21)    &11.87 (0.60)   &$-$2.71 (0.24) &0.48 (0.24)    &                    \\
1247$+$550\dotfill& 4050 ( 70)                        &7.57 (0.03)    &H &0.35 (0.01)    &15.78 (0.04)   &$-$4.20 (0.01) &4.63 (0.31)    &                    \\
1257$+$037\dotfill& 5590 (110)                        &8.16 (0.09)    &H &0.68 (0.06)    &14.74 (0.14)   &$-$3.94 (0.05) &4.32 (0.88)    &                    \\
1257$+$278\dotfill& 8540 (240)                        &7.97 (0.22)    &H &0.58 (0.13)    &12.71 (0.31)   &$-$3.09 (0.12) &0.87 (0.26)    &                    \\
1300$+$263\dotfill& 4320 (100)                        &8.07 (0.18)    &H &0.62 (0.12)    &16.04 (0.25)   &$-$4.34 (0.11) &8.73 (1.47)    &                    \\
\\
1310$-$472\dotfill& 4220 ( 80)                        &8.12 (0.05)    &H &0.65 (0.04)    &16.24 (0.08)   &$-$4.41 (0.03) &9.34 (0.34)    &                    \\
1313$-$198\dotfill& 5330 (100)                        &8.33 (0.06)    &He&0.79 (0.04)    &15.18 (0.10)   &$-$4.14 (0.04) &6.47 (0.19)    &                    \\
1325$+$581\dotfill& 6810 (160)                        &8.14 (0.31)    &H &0.68 (0.19)    &13.86 (0.45)   &$-$3.58 (0.17) &2.08 (1.21)    &                    \\
1328$+$307\dotfill& 7320 (230)                        &8.21 (0.24)    &He&0.71 (0.15)    &13.65 (0.35)   &$-$3.51 (0.14) &2.02 (0.87)    &5                   \\
1334$+$039\dotfill& 5030 (120)                        &7.95 (0.05)    &H &0.55 (0.03)    &15.05 (0.06)   &$-$4.01 (0.02) &4.90 (0.57)    &                    \\
\\
1344$+$106\dotfill& 7110 (170)                        &8.10 (0.11)    &H &0.65 (0.07)    &13.61 (0.16)   &$-$3.48 (0.06) &1.67 (0.31)    &                    \\
1345$+$238\dotfill& 4590 (150)                        &7.76 (0.05)    &H &0.44 (0.02)    &15.30 (0.06)   &$-$4.07 (0.02) &4.57 (0.51)    &                    \\
1418$-$088\dotfill& 7810 (190)                        &7.43 (0.26)    &H &0.32 (0.09)    &12.34 (0.34)   &$-$2.96 (0.13) &0.63 (0.12)    &6                   \\
1444$-$174\dotfill& 4960 ( 60)                        &8.37 (0.08)    &He&0.81 (0.05)    &15.63 (0.13)   &$-$4.29 (0.05) &7.19 (0.07)    &                    \\
1455$+$298\dotfill& 7310 (170)                        &7.66 (0.23)    &H &0.41 (0.11)    &12.90 (0.31)   &$-$3.19 (0.12) &0.89 (0.21)    &6                   \\
\\
1503$-$070\dotfill& 6990 (160)                        &8.17 (0.21)    &H &0.70 (0.13)    &13.81 (0.31)   &$-$3.56 (0.12) &2.02 (0.91)    &2                   \\
1606$+$422\dotfill&11320 (570)                        &7.12 (0.23)    &H &0.25 (0.05)    &10.55 (0.35)   &$-$2.11 (0.13) &0.18 (0.04)    &6                   \\
1609$+$135\dotfill& 9080 (250)                        &8.75 (0.10)    &H &1.07 (0.06)    &13.79 (0.19)   &$-$3.50 (0.07) &2.71 (0.17)    &                    \\
1625$+$093\dotfill& 6870 (170)                        &8.44 (0.12)    &H &0.88 (0.07)    &14.30 (0.19)   &$-$3.76 (0.07) &3.74 (0.48)    &                    \\
1626$+$368\dotfill& 8640 (280)                        &8.03 (0.05)    &He&0.60 (0.03)    &12.82 (0.07)   &$-$3.12 (0.03) &1.02 (0.07)    &                    \\
\\
1633$+$433\dotfill& 6650 (150)                        &8.14 (0.07)    &H &0.68 (0.04)    &13.94 (0.10)   &$-$3.63 (0.04) &2.28 (0.34)    &                    \\
1633$+$572\dotfill& 6180 (240)                        &8.09 (0.06)    &He&0.63 (0.03)    &14.19 (0.08)   &$-$3.74 (0.03) &2.84 (0.37)    &3                   \\
1635$+$137\dotfill& 6990 (190)                        &8.29 (0.26)    &H &0.78 (0.16)    &13.97 (0.40)   &$-$3.63 (0.15) &2.77 (1.15)    &                    \\
1637$+$335\dotfill& 9940 (280)                        &8.13 (0.13)    &H &0.68 (0.08)    &12.37 (0.20)   &$-$2.91 (0.08) &0.74 (0.14)    &                    \\
1639$+$537\dotfill& 7510 (210)                        &8.14 (0.11)    &He&0.67 (0.07)    &13.43 (0.16)   &$-$3.42 (0.07) &1.64 (0.27)    &2                   \\
1655$+$215\dotfill& 9180 (230)                        &7.87 (0.12)    &H &0.53 (0.06)    &12.30 (0.16)   &$-$2.91 (0.07) &0.64 (0.09)    &                    \\
1656$-$062\dotfill& 5520 (100)                        &8.03 (0.04)    &H &0.60 (0.03)    &14.58 (0.06)   &$-$3.89 (0.03) &3.34 (0.38)    &                    \\
1705$+$030\dotfill& 7050 (180)                        &8.35 (0.13)    &He&0.80 (0.09)    &13.98 (0.21)   &$-$3.66 (0.08) &3.11 (0.62)    &5                   \\
1716$+$020\dotfill&13470 (900)                        &8.11 (0.13)    &H &0.67 (0.08)    &11.60 (0.20)   &$-$2.37 (0.09) &0.32 (0.06)    &                    \\
1733$-$544\dotfill& 6520 (150)                        &8.13 (0.03)    &H &0.67 (0.02)    &13.99 (0.04)   &$-$3.65 (0.01) &2.37 (0.14)    &                    \\
\\
1736$+$052\dotfill& 8890 (230)                        &8.11 (0.25)    &H &0.67 (0.15)    &12.77 (0.37)   &$-$3.10 (0.14) &0.97 (0.35)    &                    \\
1748$+$708\dotfill& 5590 ( 90)                        &8.36 (0.02)    &He&0.81 (0.01)    &15.21 (0.03)   &$-$4.07 (0.02) &5.86 (0.06)    &2, 3?               \\
1756$+$827\dotfill& 7270 (330)                        &7.98 (0.07)    &H &0.58 (0.04)    &13.37 (0.10)   &$-$3.38 (0.04) &1.35 (0.13)    &                    \\
1811$+$327A\dotfill& 7380 (180)                        &7.65 (0.37)    &H &0.41 (0.16)    &12.85 (0.49)   &$-$3.17 (0.19) &0.86 (0.33)    &                    \\
1811$+$327B\dotfill& 6480 (160)                        &7.75 (0.37)    &H &0.45 (0.18)    &13.52 (0.49)   &$-$3.45 (0.20) &1.32 (0.60)    &                    \\
\\
1818$+$126\dotfill& 6340 (130)                        &7.28 (0.41)    &H &0.26 (0.12)    &13.04 (0.50)   &$-$3.26 (0.20) &0.93 (0.26)    &6                   \\
1820$+$609\dotfill& 4780 (140)                        &7.83 (0.09)    &H &0.48 (0.05)    &15.16 (0.11)   &$-$4.03 (0.04) &4.68 (0.95)    &                    \\
1824$+$040\dotfill&12240 (680)                        &7.00 (0.16)    &H &0.23 (0.03)    &10.19 (0.28)   &$-$1.89 (0.11) &0.12 (0.03)    &6                   \\
1826$-$045\dotfill& 9480 (350)                        &7.94 (0.17)    &H &0.57 (0.09)    &12.29 (0.24)   &$-$2.89 (0.10) &0.64 (0.14)    &                    \\
1829$+$547\dotfill& 6280 (140)                        &8.50 (0.11)    &H &0.90 (0.07)    &14.69 (0.18)   &$-$3.96 (0.08) &4.76 (0.24)    &2, 8                \\
\\
1831$+$197\dotfill& 7590 (210)                        &7.61 (0.48)    &He&0.37 (0.19)    &12.67 (0.60)   &$-$3.12 (0.24) &0.80 (0.40)    &4                   \\
1840$+$042\dotfill& 9090 (340)                        &8.19 (0.12)    &H &0.71 (0.08)    &12.81 (0.18)   &$-$3.11 (0.08) &1.02 (0.18)    &                    \\
1855$+$338\dotfill&11240 (430)                        &8.19 (0.20)    &H &0.72 (0.13)    &12.06 (0.32)   &$-$2.73 (0.12) &0.58 (0.18)    &7                   \\
1900$+$705\dotfill&12070 (990)                        &8.58 (0.03)    &He&0.95 (0.02)    &12.68 (0.06)   &$-$2.88 (0.03) &0.94 (0.09)    &2                   \\
1917$+$386\dotfill& 6390 (140)                        &8.28 (0.05)    &He&0.75 (0.04)    &14.27 (0.09)   &$-$3.79 (0.03) &3.68 (0.32)    &3?                  \\
\\
1953$-$011\dotfill& 7920 (200)                        &8.23 (0.05)    &H &0.74 (0.03)    &13.41 (0.07)   &$-$3.38 (0.03) &1.63 (0.16)    &2                   \\
2002$-$110\dotfill& 4800 ( 50)                        &8.31 (0.02)    &He&0.77 (0.01)    &15.76 (0.03)   &$-$4.31 (0.01) &7.35 (0.04)    &                    \\
2011$+$065\dotfill& 6400 (150)                        &8.13 (0.06)    &He&0.65 (0.04)    &14.03 (0.09)   &$-$3.69 (0.03) &2.73 (0.38)    &1                   \\
2048$+$263\dotfill& 5200 (110)                        &7.31 (0.12)    &H &0.26 (0.04)    &14.12 (0.15)   &$-$3.64 (0.06) &1.53 (0.15)    &6                   \\
2054$-$050\dotfill& 4620 ( 40)                        &8.09 (0.12)    &He&0.62 (0.08)    &15.74 (0.17)   &$-$4.24 (0.07) &7.49 (0.56)    &                    \\
\\
2059$+$190\dotfill& 6840 (160)                        &7.86 (0.28)    &H &0.51 (0.15)    &13.44 (0.37)   &$-$3.42 (0.15) &1.33 (0.48)    &                    \\
2059$+$247\dotfill& 6070 (130)                        &8.17 (0.17)    &H &0.69 (0.11)    &14.36 (0.25)   &$-$3.80 (0.10) &3.32 (1.12)    &                    \\
2059$+$316\dotfill&10080 (390)                        &8.08 (0.18)    &He&0.63 (0.11)    &12.35 (0.26)   &$-$2.87 (0.11) &0.72 (0.17)    &4                   \\
2105$-$820\dotfill&10200 (290)                        &8.23 (0.21)    &H &0.75 (0.13)    &12.45 (0.33)   &$-$2.93 (0.13) &0.81 (0.27)    &                    \\
2107$-$216\dotfill& 5830 (140)                        &8.40 (0.05)    &H &0.85 (0.03)    &14.93 (0.08)   &$-$4.02 (0.03) &5.24 (0.23)    &                    \\
\\
2111$+$261\dotfill& 8120 (190)                        &7.42 (0.19)    &H &0.32 (0.07)    &12.16 (0.26)   &$-$2.88 (0.11) &0.57 (0.08)    &6                   \\
2136$+$229\dotfill& 9480 (250)                        &7.85 (0.20)    &H &0.52 (0.11)    &12.13 (0.28)   &$-$2.84 (0.12) &0.58 (0.14)    &                    \\
2140$+$207\dotfill& 8860 (300)                        &8.07 (0.06)    &He&0.62 (0.04)    &12.75 (0.09)   &$-$3.09 (0.04) &1.00 (0.07)    &4                   \\
2207$+$142\dotfill& 7620 (190)                        &8.24 (0.16)    &H &0.75 (0.10)    &13.57 (0.24)   &$-$3.45 (0.10) &1.86 (0.64)    &                    \\
2246$+$223\dotfill&10330 (300)                        &8.57 (0.09)    &H &0.97 (0.06)    &12.99 (0.17)   &$-$3.14 (0.07) &1.56 (0.33)    &                    \\
2248$+$293\dotfill& 5580 (110)                        &7.53 (0.15)    &H &0.35 (0.06)    &13.94 (0.19)   &$-$3.61 (0.08) &1.50 (0.21)    &6                   \\
2251$-$070\dotfill& 4580 ( 60)                        &8.38 (0.05)    &He&0.82 (0.03)    &16.17 (0.08)   &$-$4.43 (0.03) &7.75 (0.01)    &5                   \\
2253$-$081\dotfill& 6770 (180)                        &8.41 (0.19)    &H &0.86 (0.13)    &14.30 (0.32)   &$-$3.76 (0.12) &3.75 (0.94)    &                    \\
2311$-$068\dotfill& 7980 (240)                        &8.27 (0.17)    &He&0.75 (0.11)    &13.40 (0.26)   &$-$3.39 (0.11) &1.74 (0.58)    &4                   \\
2312$-$024\dotfill& 6840 (160)                        &8.41 (0.22)    &He&0.84 (0.14)    &14.18 (0.34)   &$-$3.75 (0.14) &3.62 (0.85)    &5                   \\
\\
2316$-$064\dotfill& 4720 ( 50)                        &8.17 (0.17)    &He&0.68 (0.11)    &15.69 (0.25)   &$-$4.25 (0.10) &7.30 (0.52)    &                    \\
2329$+$267\dotfill& 9400 (240)                        &8.02 (0.28)    &H &0.61 (0.16)    &12.40 (0.40)   &$-$2.95 (0.15) &0.73 (0.27)    &                    \\
2345$-$447\dotfill& 5400 ( 80)                        &8.72 (0.07)    &He&1.04 (0.04)    &15.77 (0.13)   &$-$4.38 (0.05) &6.07 (0.24)    &                    \\
2347$+$292\dotfill& 5810 (120)                        &7.82 (0.15)    &H &0.48 (0.08)    &14.10 (0.19)   &$-$3.69 (0.08) &1.89 (0.37)    &                    \\
2352$+$401\dotfill& 8260 (260)                        &7.99 (0.23)    &He&0.57 (0.13)    &12.88 (0.32)   &$-$3.17 (0.12) &1.07 (0.32)    &4                   \\
\\
\enddata
\tablenotetext{a}{White dwarf cooling age only, not including the main sequence lifetime}
\tablecomments{
(1) Energy distribution better reproduced with pure hydrogen models (yet no \halpha); 
(2) Magnetic; 
(3) C$_2$H star; 
(4) DQ star; 
(5) DZ star; 
(6) Known or suspected double degenerate binary; 
(7) ZZ Ceti star; 
(8) Helium solution adopted despite hydrogen-rich atmosphere (see text). }
\end{deluxetable}

\clearpage
\figcaption[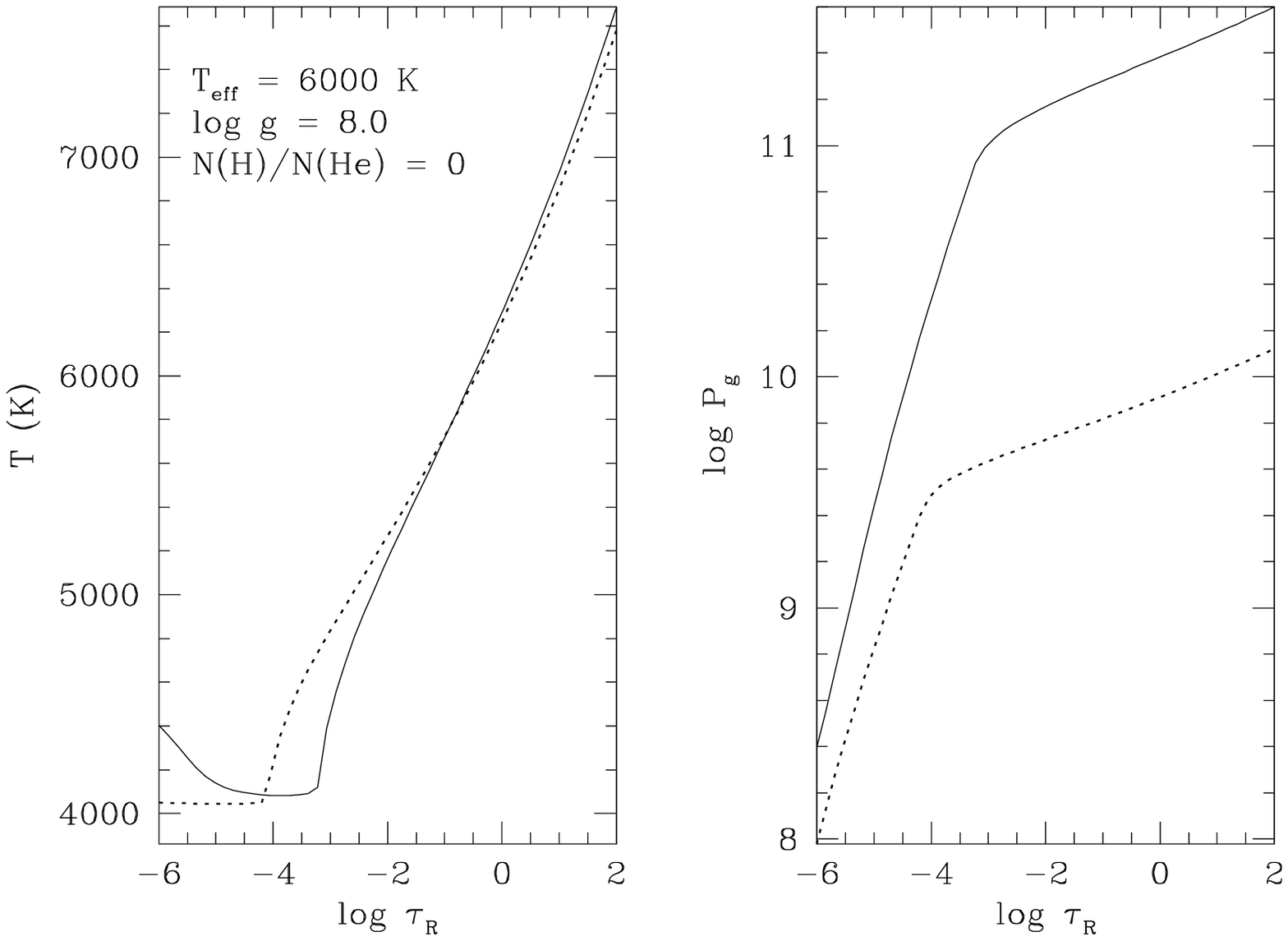] {Comparison of 
temperature (in K) and gas pressure (in dyn cm$^{-2}$) structures as a
function of Rosseland mean optical depth for two models with
$\Te=6000$~K, $\logg=8.0$, and $\nh=0$. One of these models has its
electron density increased artificially by a factor of 1000 ({\it
dotted lines}) with respect to the reference model ({\it solid
lines}).\label{fg:comp_tp}}

\figcaption[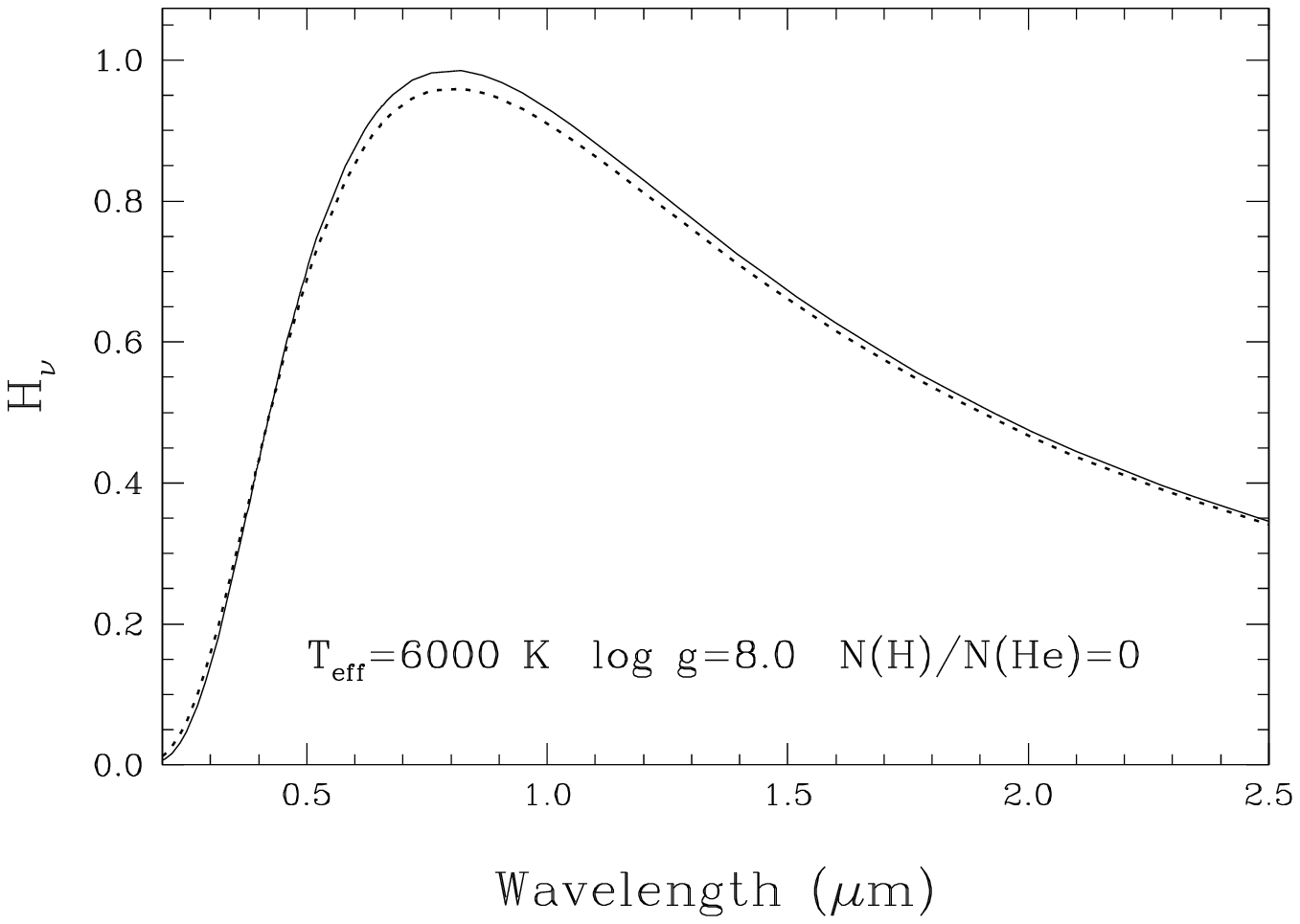] {Comparison of 
the Eddington fluxes (in units of 10$^{-5}$ ergs cm$^{-2}$ s$^{-1}$
Hz$^{-1}$ ster$^{-1}$) between the two models described in Figure
\ref{fg:comp_tp}.\label{fg:compflux}}

\figcaption[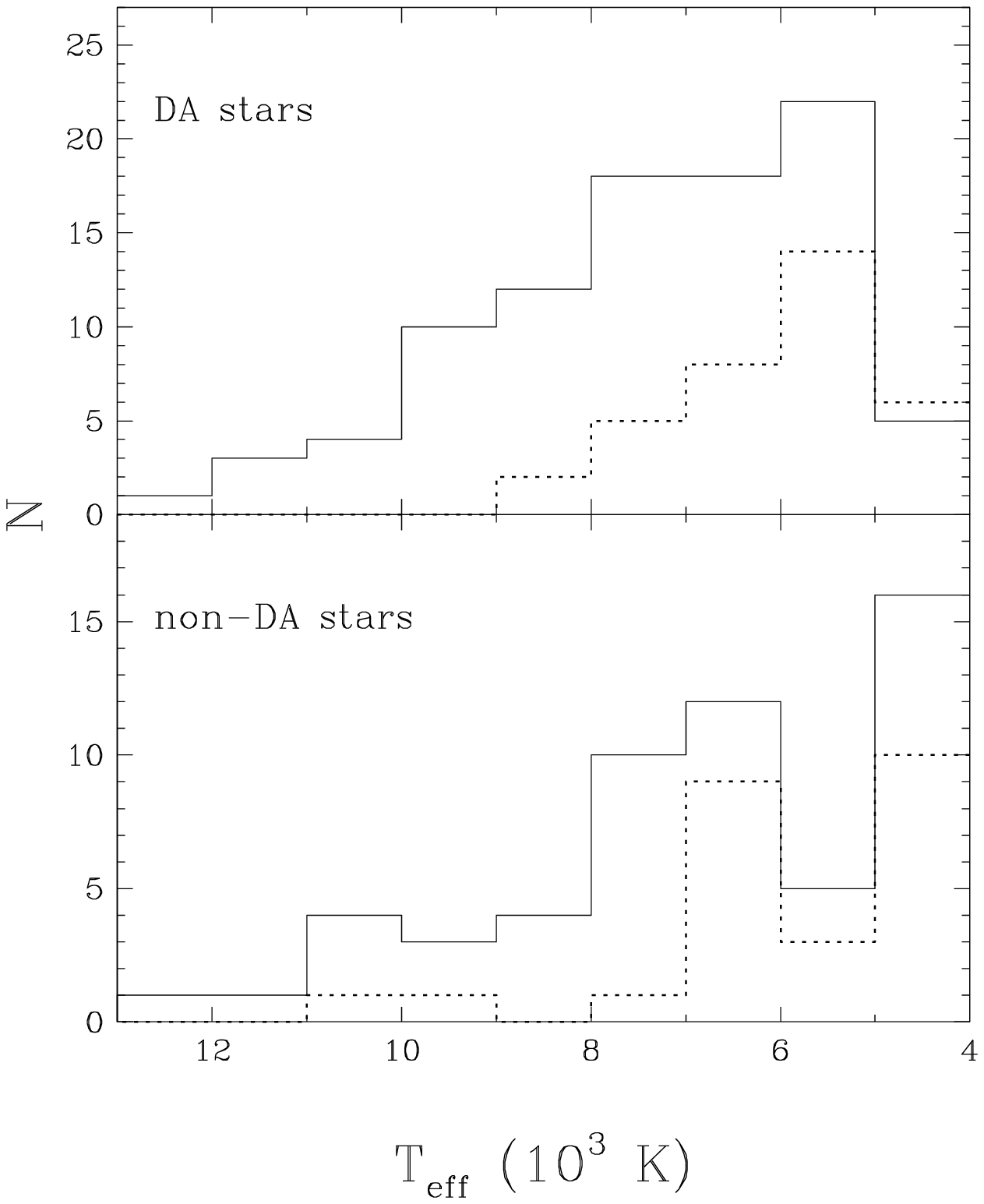] {Number of DA and non-DA stars 
as a function of effective temperature in 1000~K bins for the sample
presented in this paper ({\it solid line}) compared to the parallax
subsample of BRL ({\it dotted line}). The DA/non-DA classification is
based on the presence or absence of \halpha\ in the spectrum, and not
on the dominant constituent of the atmosphere.  These spectral types
are based on our own spectroscopic observations.\label{fg:DAvsnonDA}}

\figcaption[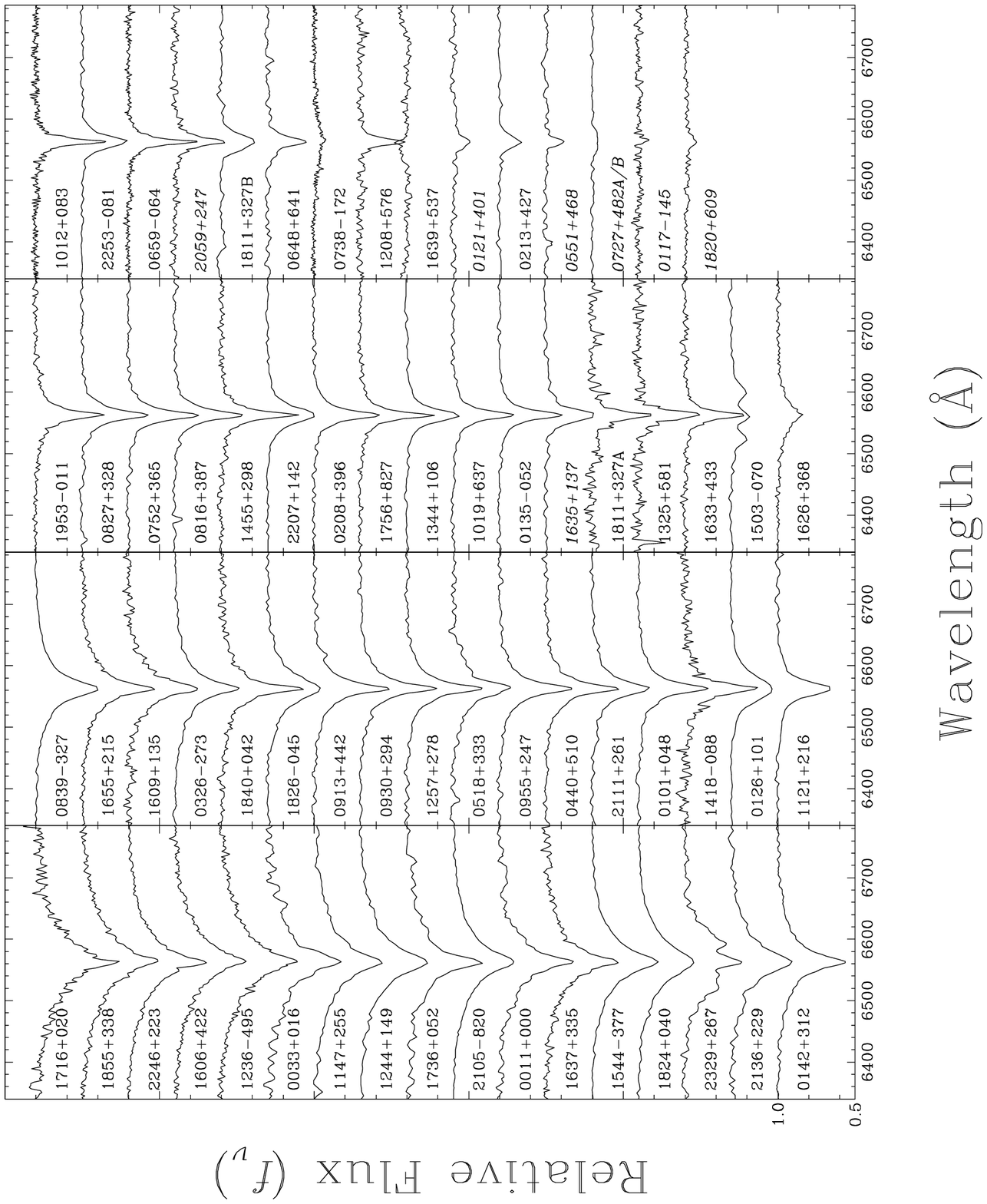] {White dwarfs in our sample whose spectra
show \halpha; the spectra already displayed in BRL are not reproduced
here. All spectra are normalized to a continuum set to unity, and
offset vertically from each other by a factor of 0.3. Equivalent
widths decrease from upper left to bottom right. The names of newly
identified DA stars are shown in italics.\label{fg:plotspecDA}}

\figcaption[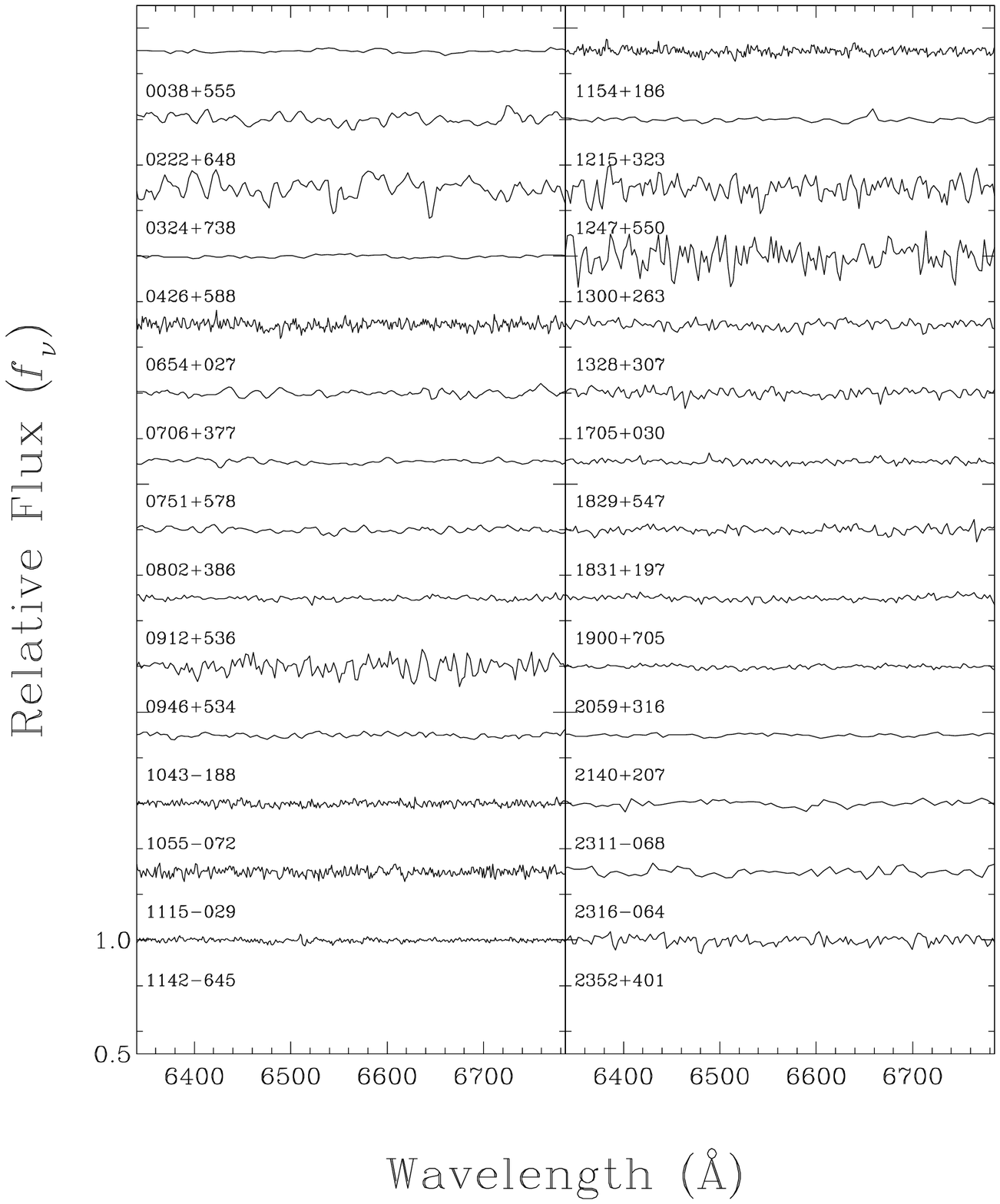] {Same as Figure \ref{fg:plotspecDA} but for 
our featureless spectra near the \halpha\ region. Spectra are shown in order
of right ascension.\label{fg:plotspecnonDA}}

\figcaption[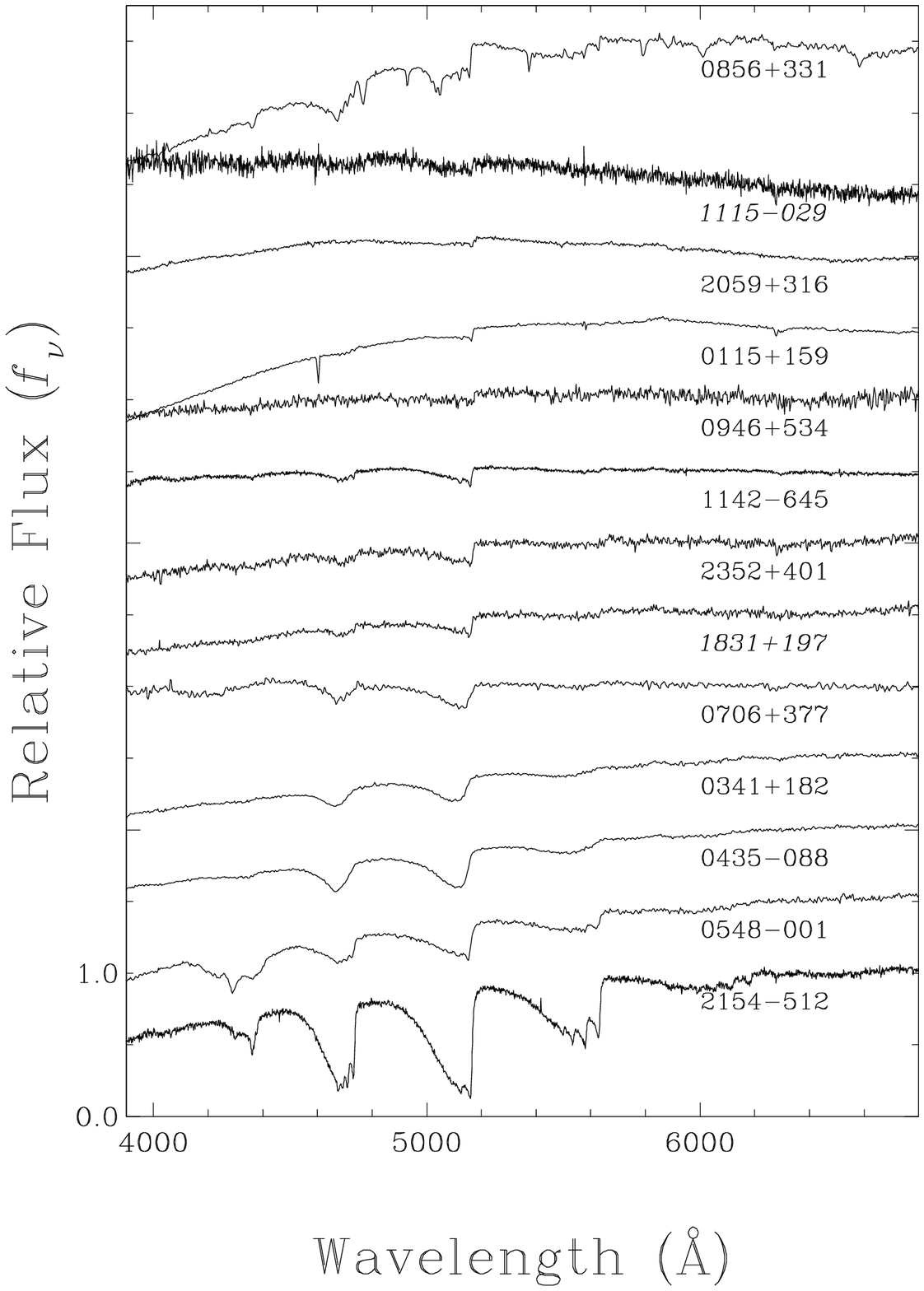] {Our blue spectroscopic observations of DQ stars.
Spectra are shown in order of decreasing $\Te$.  All spectra are
normalized at 6250 \AA\ and are offset from each other by a factor of
0.5; the spectrum of 1115$-$029 is blown up by a factor of 2 to make
the weak C$_2$ feature at 5150 \AA\ more obvious. The names of newly
identified DQ stars are shown in italics.\label{fg:plotspecDQ}}

\figcaption[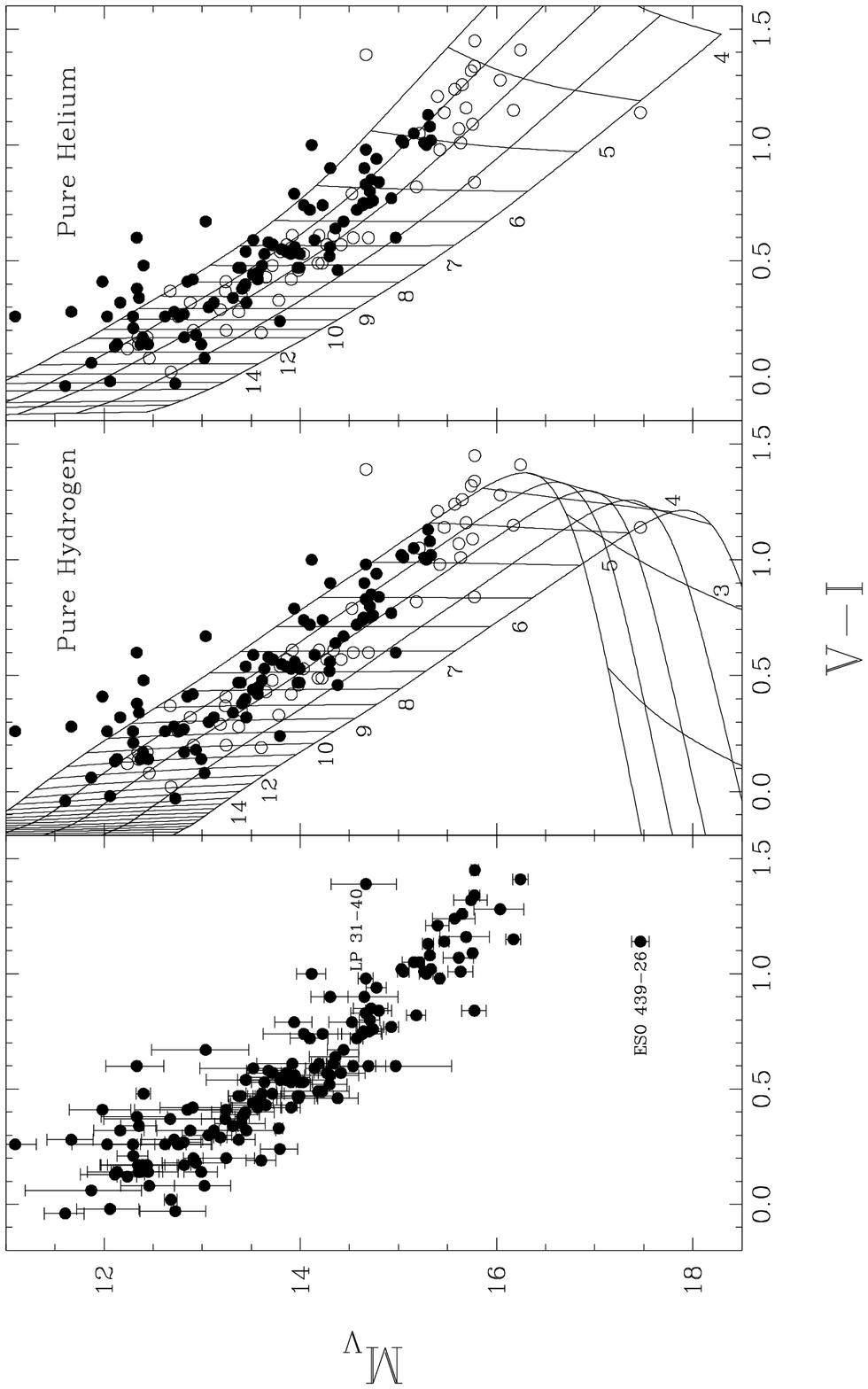] {$M_V$ vs ($V$--$I$) color-magnitude diagram
for the data set from Table 1. In the two rightmost panels, the 
data set is split into DA stars ({\it filled circles}) and non-DA
stars ({\it open circles}), based on the presence or absence of \halpha. 
The pure hydrogen and pure helium model
sequences are superimposed on the observed data; temperatures are
indicated in units of $10^3$~K, and $M=0.4$, 0.6, 0.8, 1.0, and 1.2
\msun\ from top to bottom. The stars marked in the left panel are 
discussed in the text.\label{fg:Mv:vmi}}

\figcaption[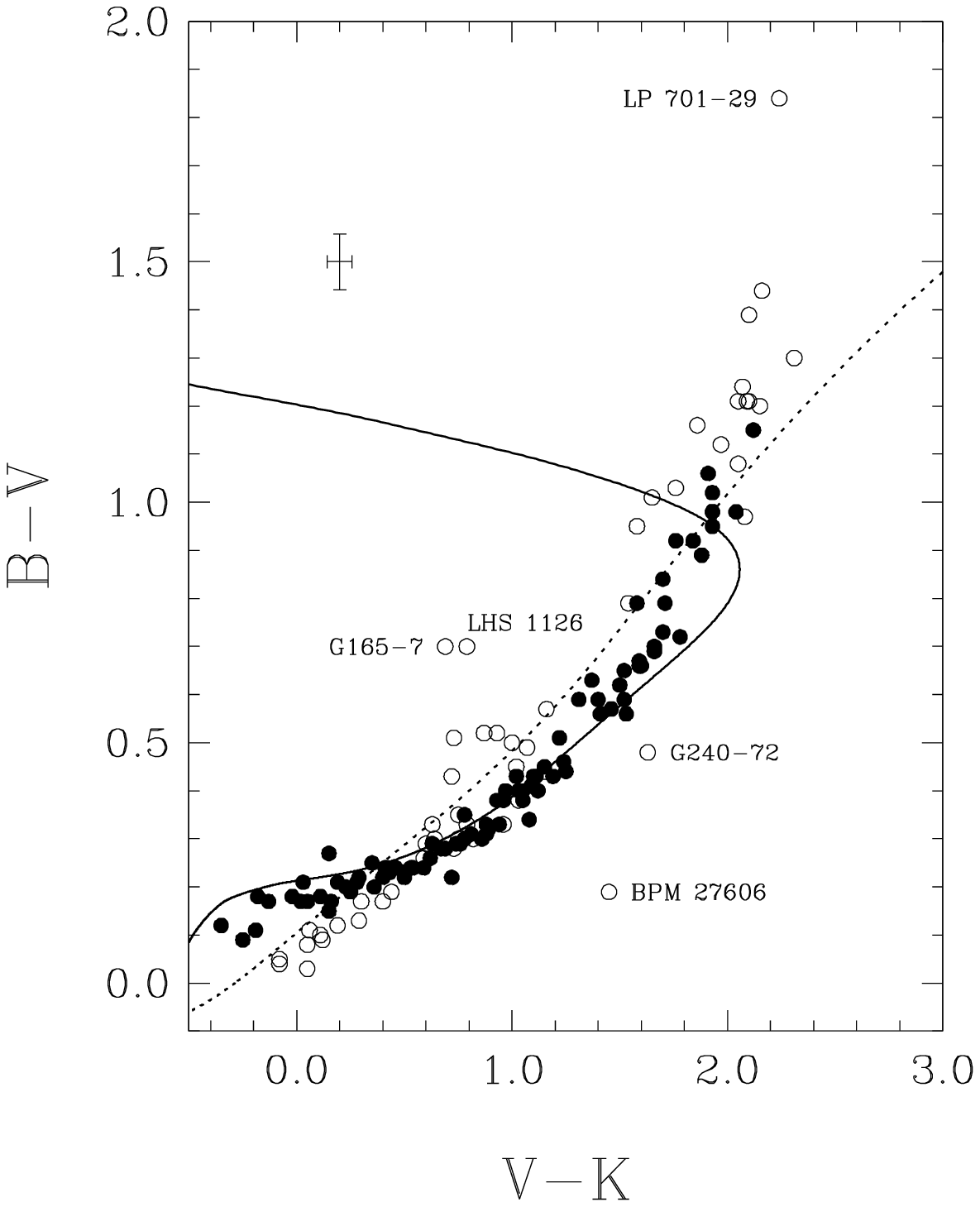] {($B$--$V$, $V$--$K$) two-color diagram for the
data set from Table 1; DA and non-DA stars are represented by filled
and open circles, respectively, and the cross indicates the size of
the error bars. The objects marked are discussed in the text. The 0.6
\msun\ pure hydrogen ({\it solid line}) and pure helium ({\it dashed
line}) model sequences are superimposed on the observed
sequences.\label{fg:bmv:vmk}}

\figcaption[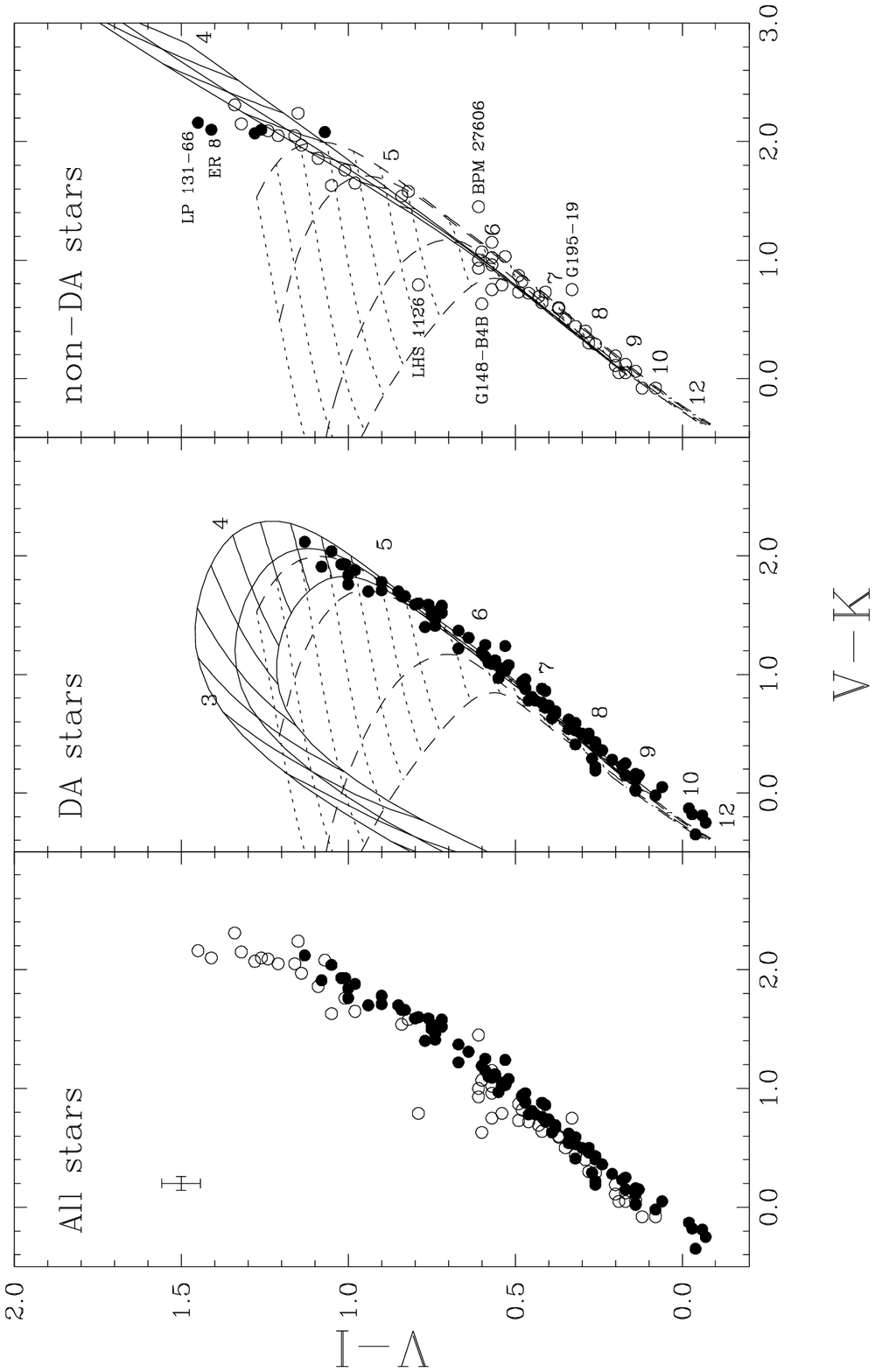] {($V$--$I$, $V$--$K$) two-color diagram for the
data set from Table 1 (left panel); DA and non-DA stars are
represented by filled and open circles, respectively, and the cross
indicates the size of the error bars. In the middle and right panels,
the data set is divided into DA and non-DA stars, respectively.  The
objects labeled in the right panel are discussed in the text. The pure
hydrogen (DA panel) and pure helium (non-DA panel) model sequences are
superimposed on the observed data ({\it solid lines}); temperatures
are indicated in units of $10^3$~K, and log $g=7.0$, 8.0, and 9.0 from
right to left (DA panel) and from top to bottom (non-DA panel). Also
shown in both panels are the mixed hydrogen/helium models with
$\nhe=0.1$, 1, 10, and 100, from right to left ({\it dotted} and {\it
dashed lines}), for log $g=8.0$. 
In the right panel, filled and open circles indicate
non-DA stars with hydrogen- and helium-rich compositions,
respectively. Note the presence of a non-DA gap in the 5000-6000~K
region.\label{fg:vmi:vmk}}

\figcaption[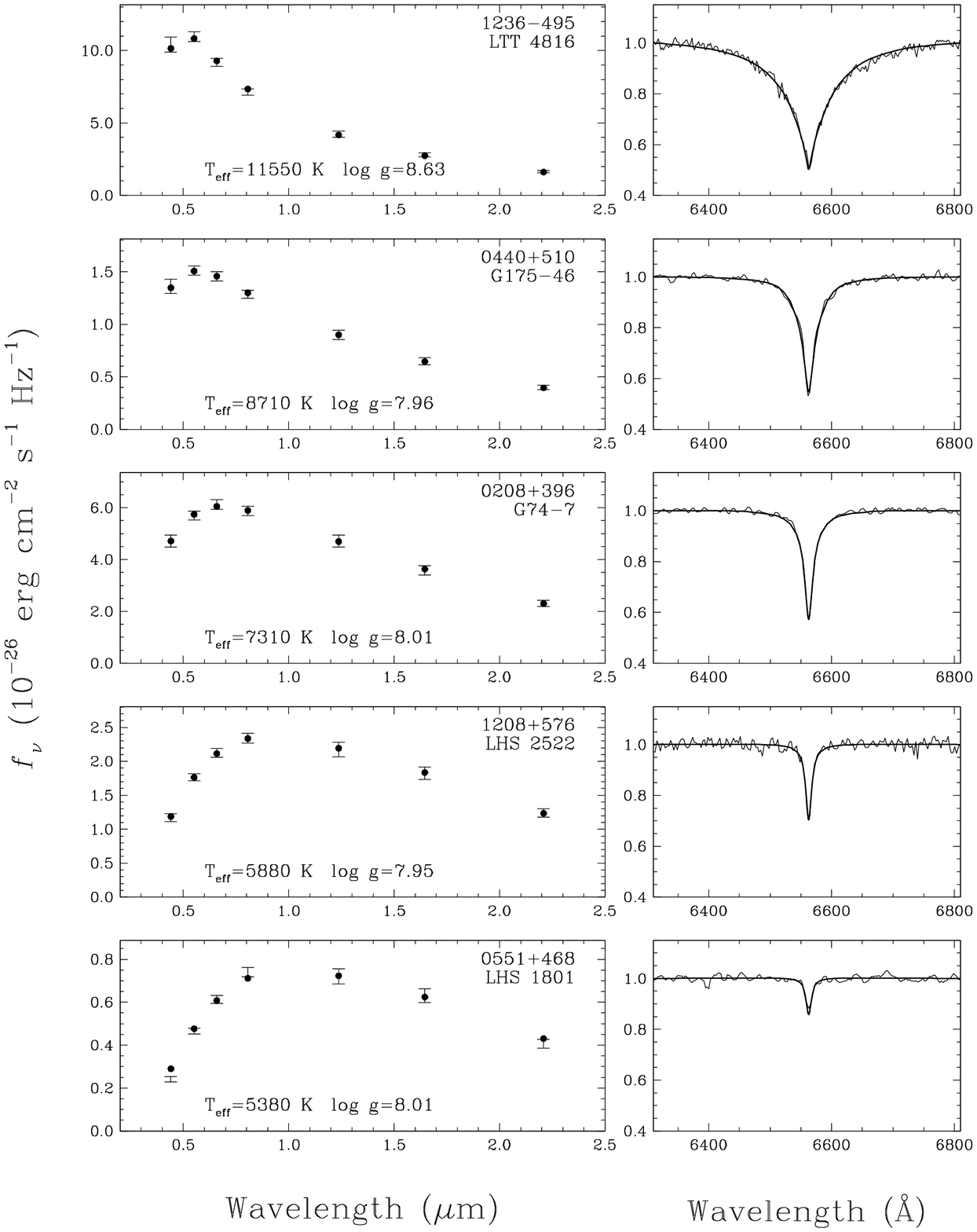] {Fits to the energy distributions of DA stars
with pure hydrogen models. Here and in the following Figures, the
$BVRI$ and $JHK$ photometric observations are represented by error
bars while the model fluxes are shown as filled circles. In the right
panels are shown the observed normalized spectra together with the
synthetic line profiles interpolated at the parameters obtained from
the energy distribution fits (left panels).\label{fg:sampleDA}}

\figcaption[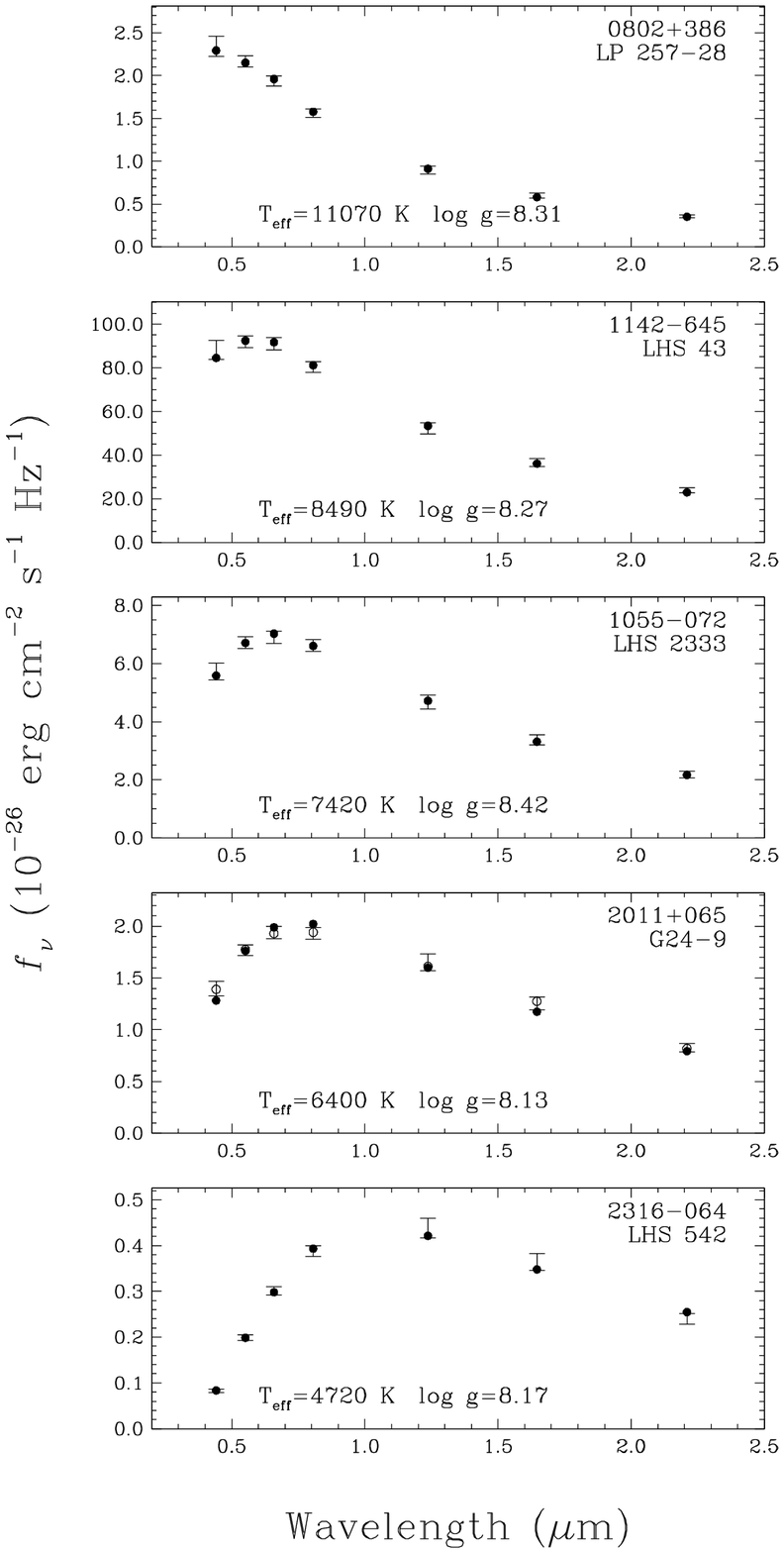] {Fits to the energy distributions of non-DA 
stars with pure helium models ({\it filled circles}). All objects have
featureless spectra near the \halpha\ region. G24$-$9 belongs to this
group of peculiar objects whose energy distributions are better
reproduced by hydrogen-rich models (shown here as {\it open
circles}).\label{fg:samplenonDA}}

\figcaption[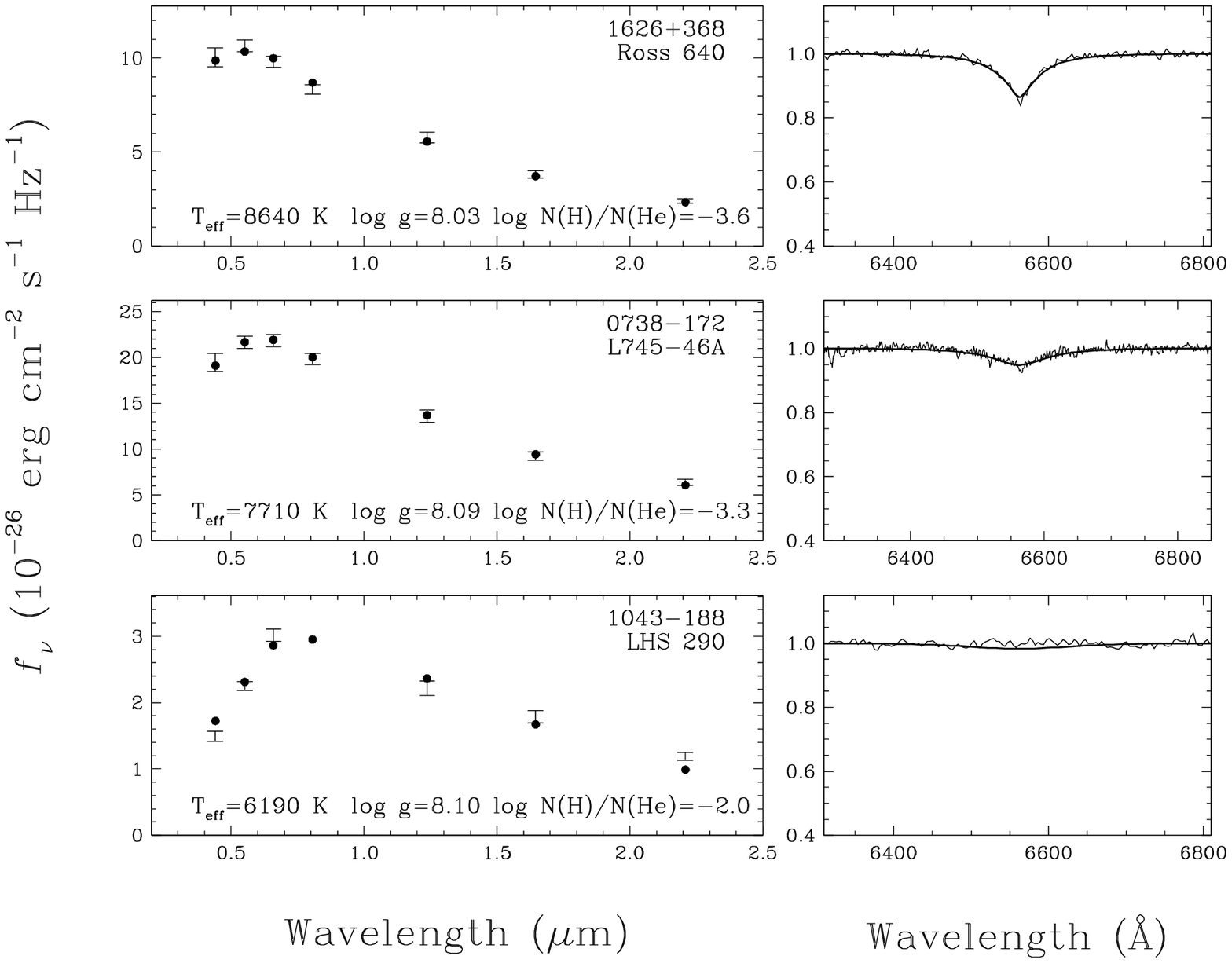] {Fits to the energy distributions of white dwarfs 
with mixed hydrogen and helium atmospheric compositions. The
hydrogen-to-helium abundance ratio is determined (or constrained in
the case of LHS 290) by fitting the \halpha\ line profile.\label{fg:sampleHe}}

\figcaption[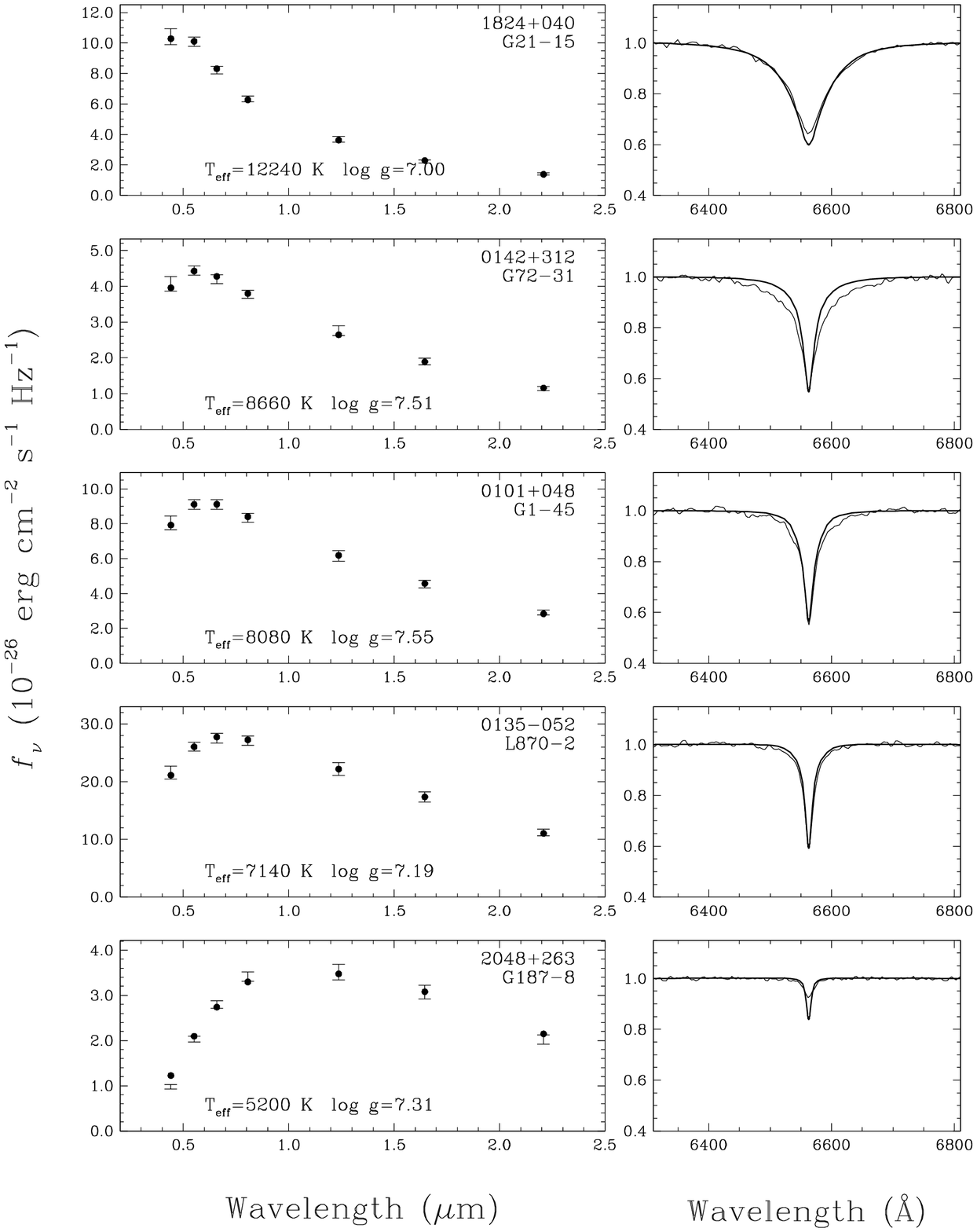] {Fits to the energy distributions of confirmed,
or suspected, unresolved double degenerates. In the right panels are
shown the observed normalized spectra together with the synthetic line
profiles interpolated at the parameters obtained from the energy
distribution fits (left panels).\label{fg:sampledoubleDA}}

\figcaption[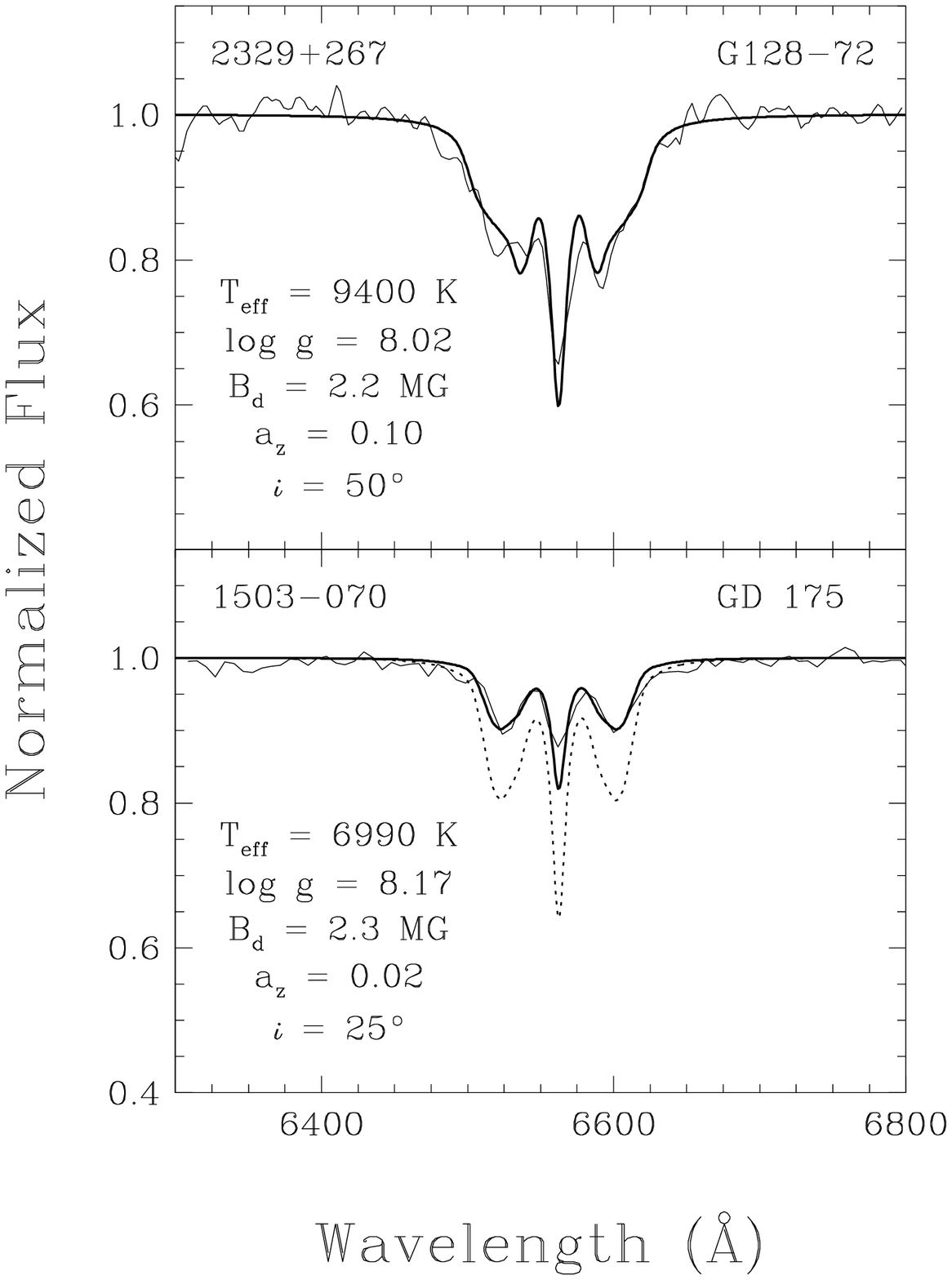] {Our best fits to two magnetic white dwarfs whose
spectrum exhibits Zeeman splitting at \halpha; all spectra are
normalized to a continuum set to unity.  The effective temperatures
and surface gravities are taken from our fits to the energy
distributions. The magnetic parameters are given in each
panel. G128$-$72 has been identified recently by \citet{moran98},
while GD 175 has been discovered in our survey. For GD 175, the model
spectrum shown by the dotted line corresponds to our best fit with a
single magnetic DA star, while the thick solid line represents our
best fit under the assumption that an additional featureless DC white
dwarf contributes equally to the total continuum flux near the
\halpha\ region.\label{fg:figMAG}}

\figcaption[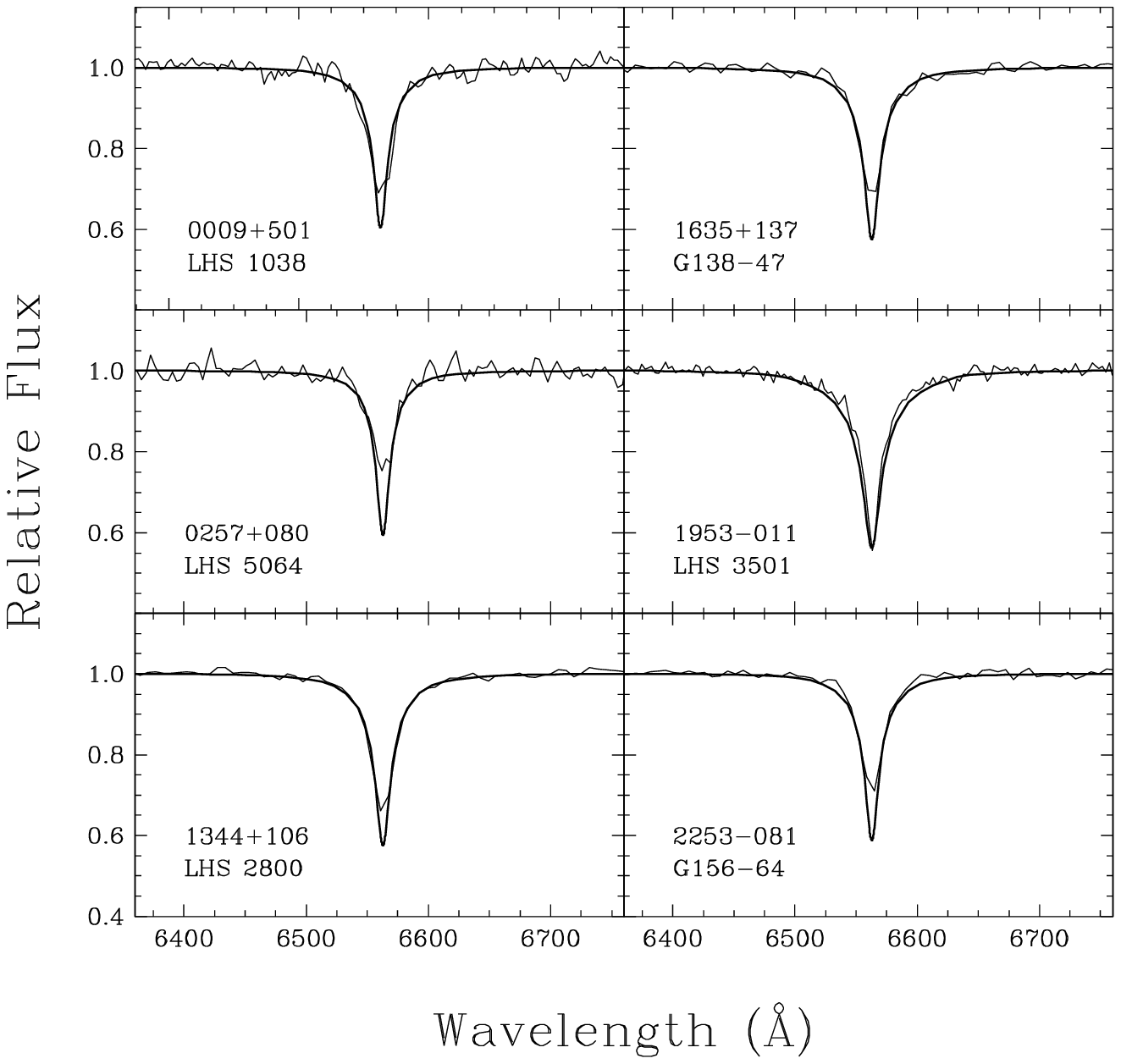] {Our fits at \halpha\ to weakly magnetic,
or suspected to be magnetic, DA white dwarfs using non-magnetic pure
hydrogen model spectra. All spectra are normalized to a continuum set
to unity, and the atmospheric parameters are those obtained from the
energy distributions. All objects except LHS 3501 have been observed
in spectropolarimetry, but only LHS 1038 and LHS 5064 have shown a
Zeeman polarization feature. LHS 3501 is a known magnetic white dwarf
with a $\sim 500$ kG magnetic field \citet{maxted00}.\label{fg:sampleMAG}}

\figcaption[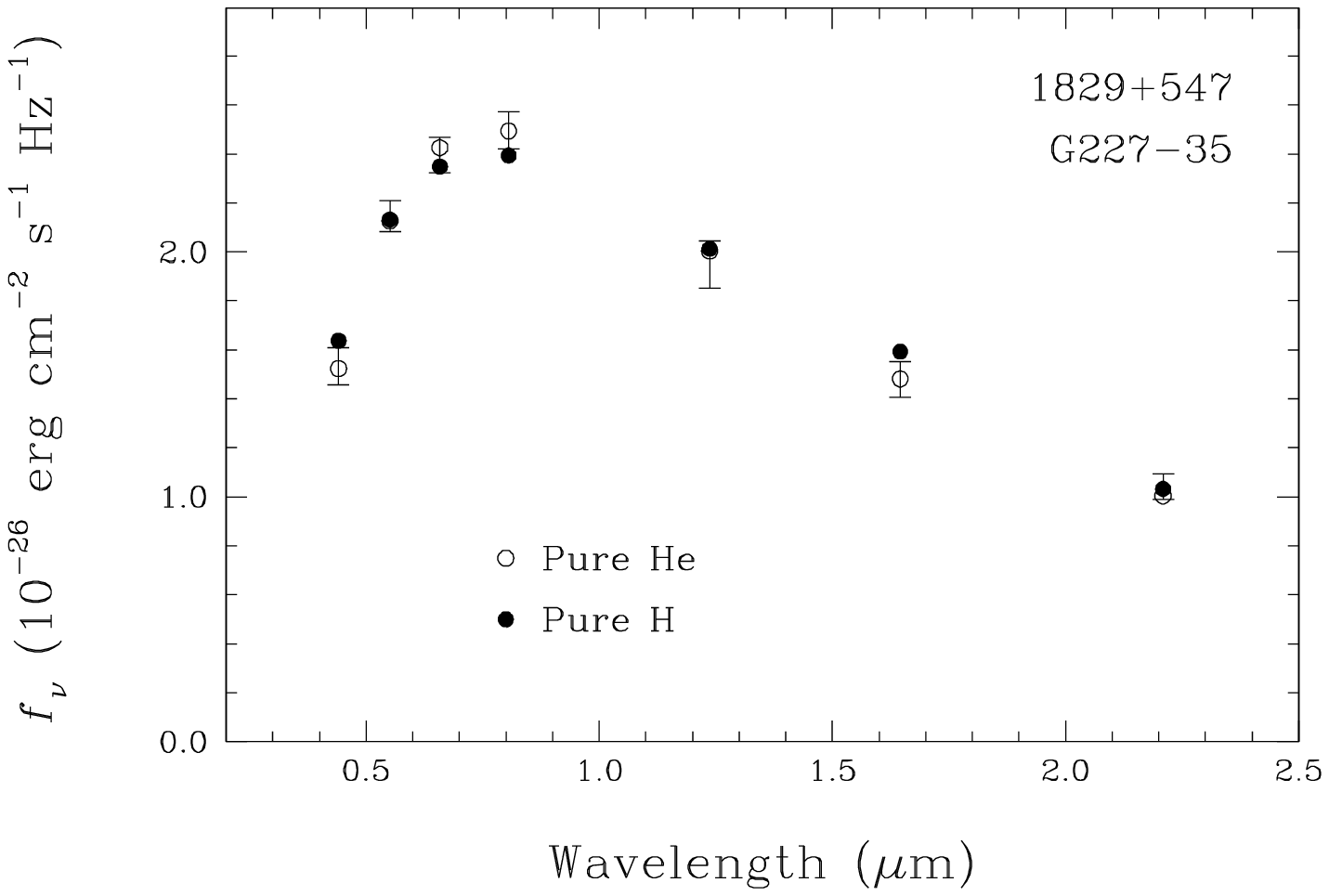] {A comparison of our best fit to the energy 
distribution of G227$-$35 with a pure helium model ({\it open
circles}) at $\Te=6280$~K, $\logg=8.50$, and with a pure hydrogen
model ({\it filled circles}) at $\Te=6750$~K, $\logg=8.64$. Even
though the quality of the fit is superior with the pure helium model,
G227$-$35 has most likely a hydrogen-rich atmospheric composition in a
strong magnetic field according to \citet{putney95}.\label{fg:figG227-35}}

\figcaption[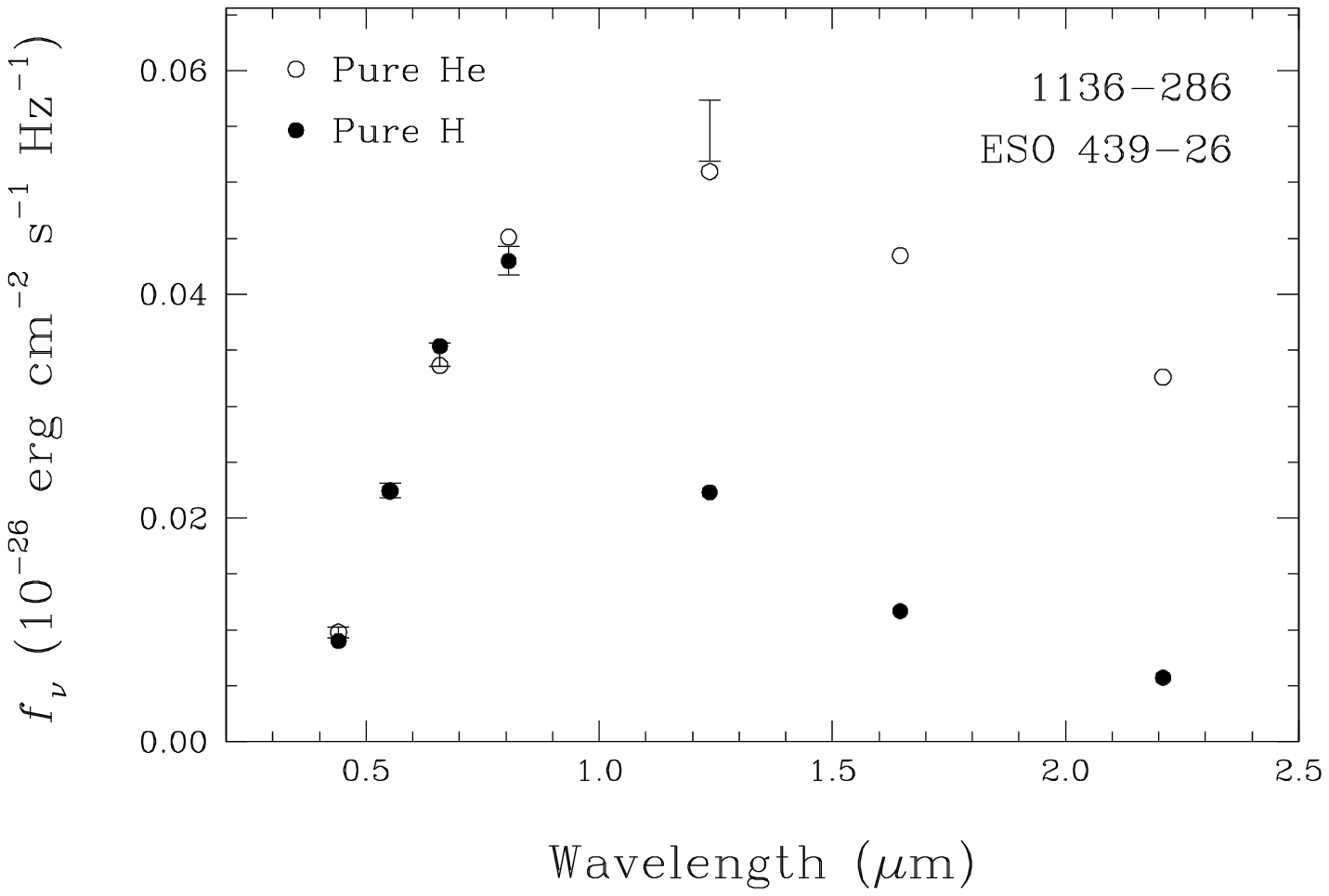] {A comparison of our best fit to the energy 
distribution of ESO 439$-$26 with a pure helium model ({\it open
circles}) at $\Te=4490$~K, $\logg=9.02$, and our best fit using only
the $V$ and $I$ bandpasses with a pure hydrogen model ({\it filled
circles}) at $\Te=3150$~K, $\logg=8.29$. Because of the infrared $J$
magnitude, the pure hydrogen solution can be easily ruled
out.\label{fg:figESO439-26}}

\figcaption[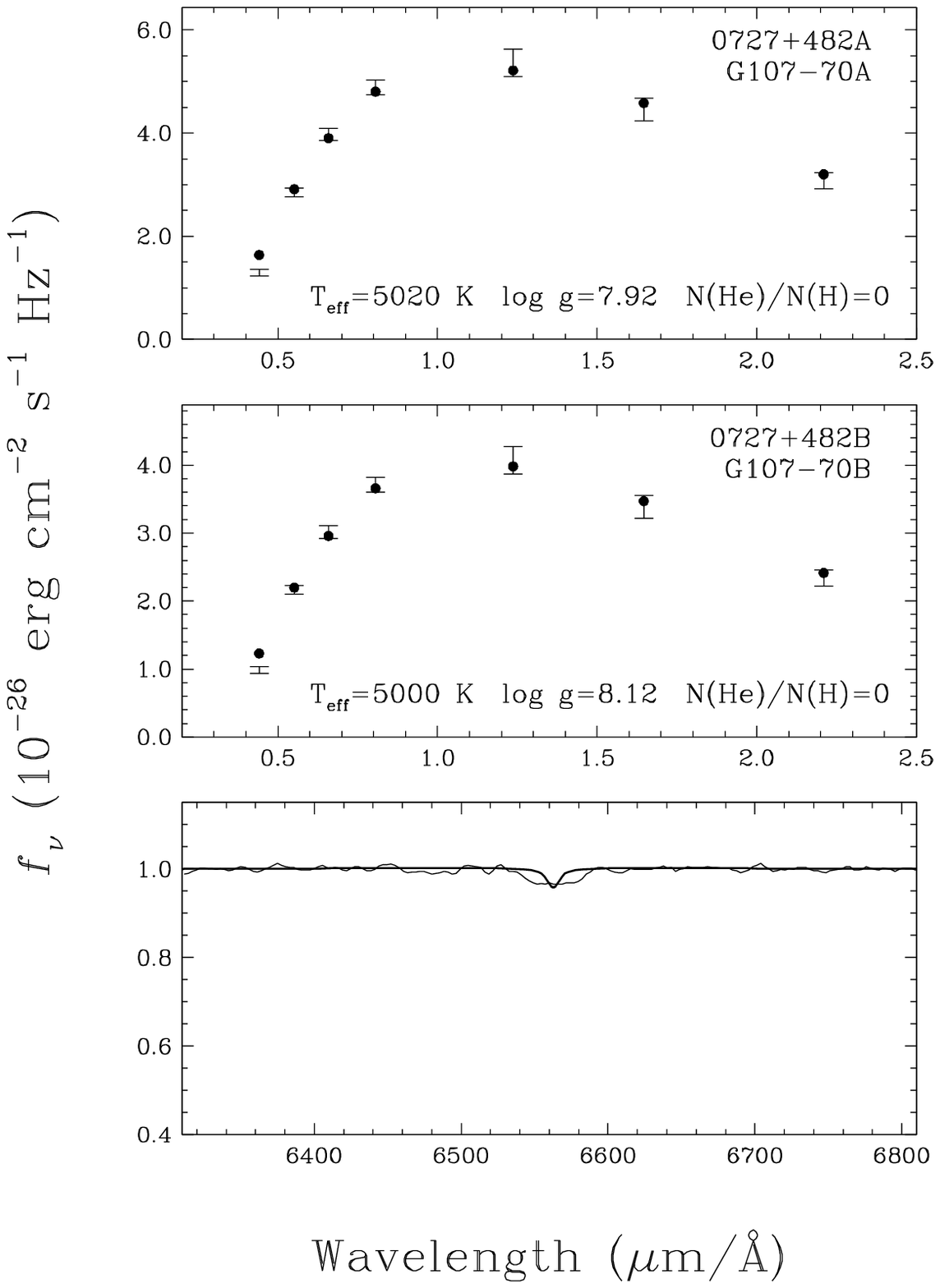] {Our best fits to the energy distributions of 
G107$-$70A and B. The combined \halpha\ line profile of both components
is compared with the predicted line profile obtained as the sum of the
individual model spectra interpolated at the parameters obtained from
the fits to the energy distributions, and properly weighted by the
respective luminosity.\label{fg:figG107-70}}

\figcaption[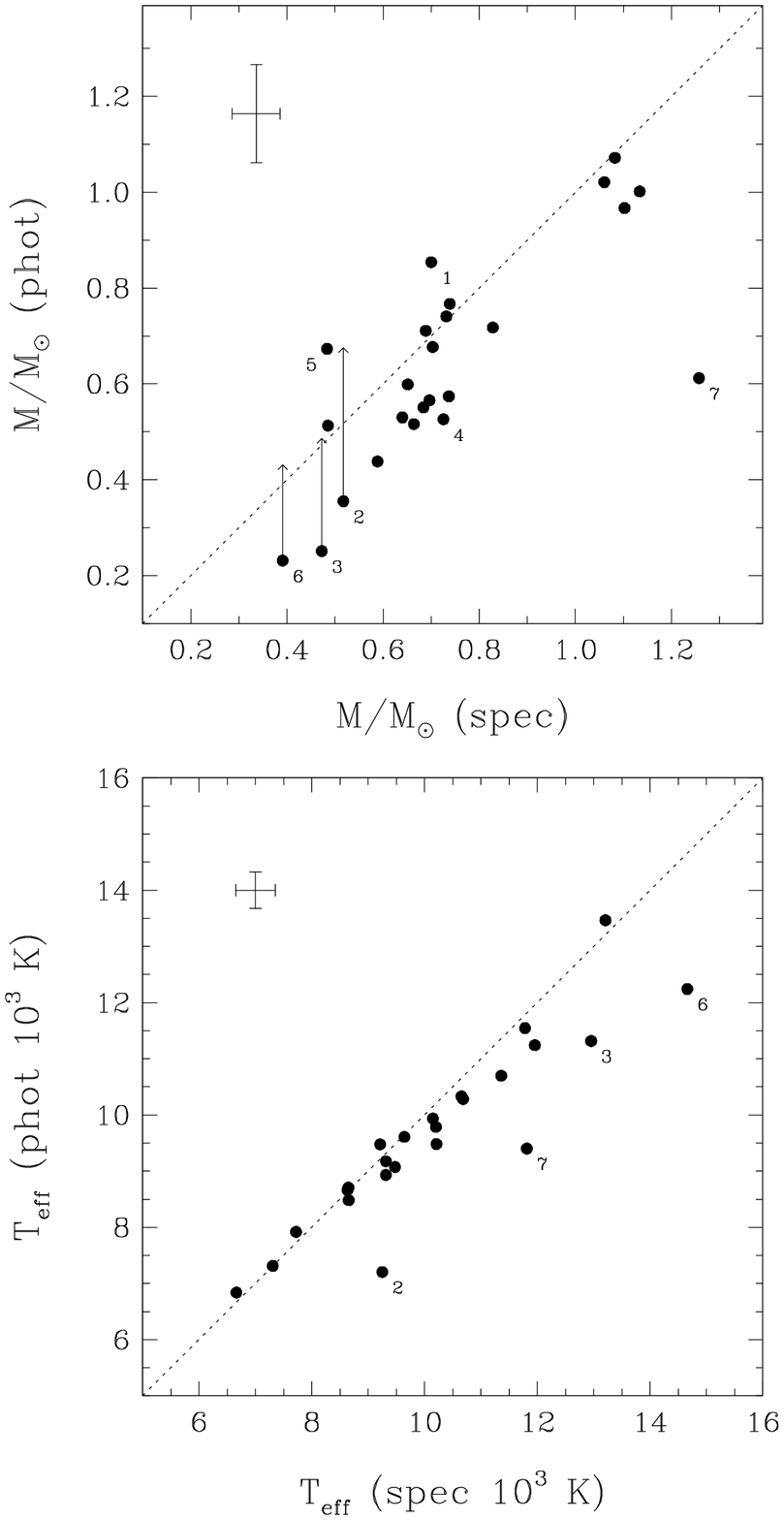] {Comparison of effective temperatures and
stellar masses determined from the photometric technique used in this
analysis ({\it phot}), and from the spectroscopic technique of fitting
simultaneously the Balmer line profiles ({\it spec}), for 24 DA stars.  
The dashed line
in each panel indicates the 1:1 correspondence. The average
uncertainties are shown in each panel as well. The labels correspond
to (1) 0011$+$000, (2) 0326$-$273, (3) 1606$+$422, (4) 1655$+$215, (5)
1716$+$020, (6) 1824$+$040, (7) 2329$+$267, and these objects are
discussed in the text. The three vectors in the upper panel indicate
the photometric masses obtained under the assumption that two
identical white dwarfs contribute to the total flux received at
Earth.\label{fg:TspecvsTphot}}

\figcaption[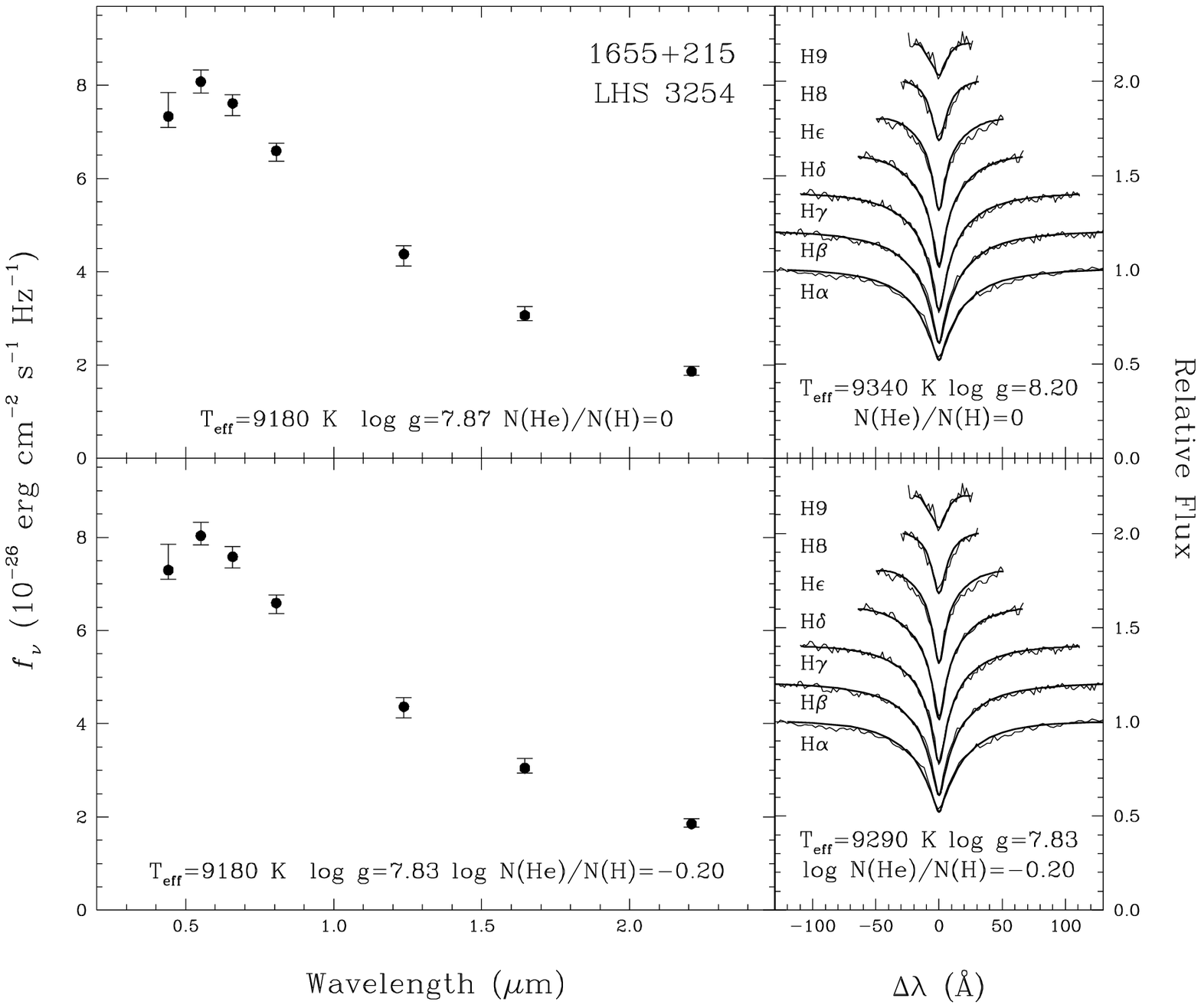] {Our best photometric and spectroscopic fits 
to LHS 3254 (1655$+$215), object 4 in Fig.~17, 
using pure hydrogen models (top panels). The
line profiles are normalized to a continuum set to unity and are
offset vertically from each other by a factor of 0.2. The discrepancy
in log $g$ (and hence mass) between both techniques can be resolved if
some amount of helium is allowed in the atmosphere (bottom
panels). From a qualitative point of view, both the pure hydrogen and
mixed H/He solutions are indistinguishable.\label{fg:exampleHe}}

\figcaption[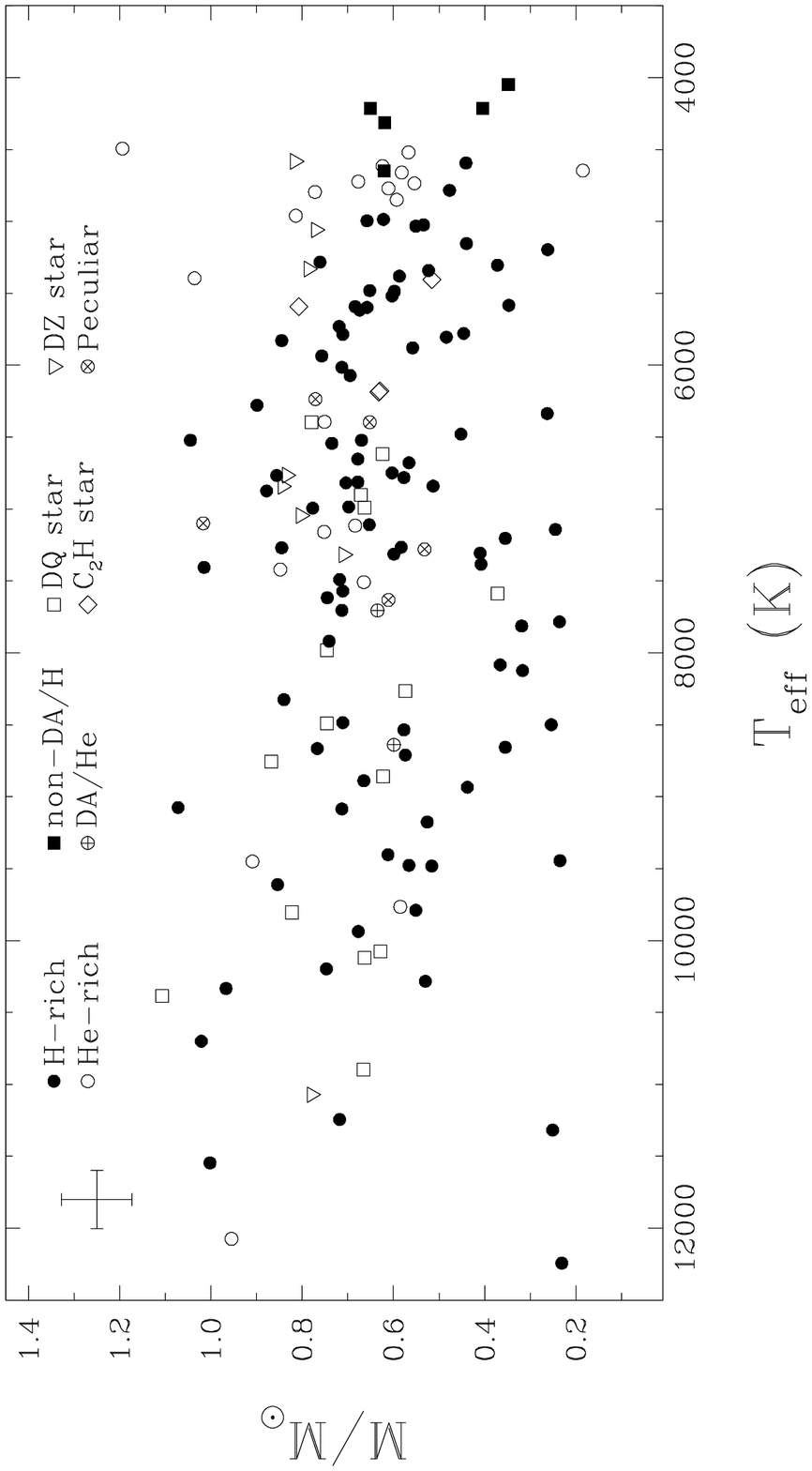] {Masses of all white dwarfs in our current
trigonometric 
parallax sample as a function of effective temperature. Not seen here
is the hottest DA star in our sample, G19$-$20 (1716$+$020), with
$\Te=13,470$~K and $M=0.67$ \msun. Filled and open symbols represent
hydrogen-rich and helium-rich atmospheric compositions, respectively,
while additional information about the composition of or spectroscopic
features seen in some of these objects is provided by the different
symbols explained in the legend. Note that there are two C$_2$H stars
which overlap near $\Te\sim6200$~K and $M\sim0.6$ \msun. The cross in
the upper left corner represents the average error in $\Te$ of 200~K
and in mass of 0.077
\msun.\label{fg:spec_evol}}

\figcaption[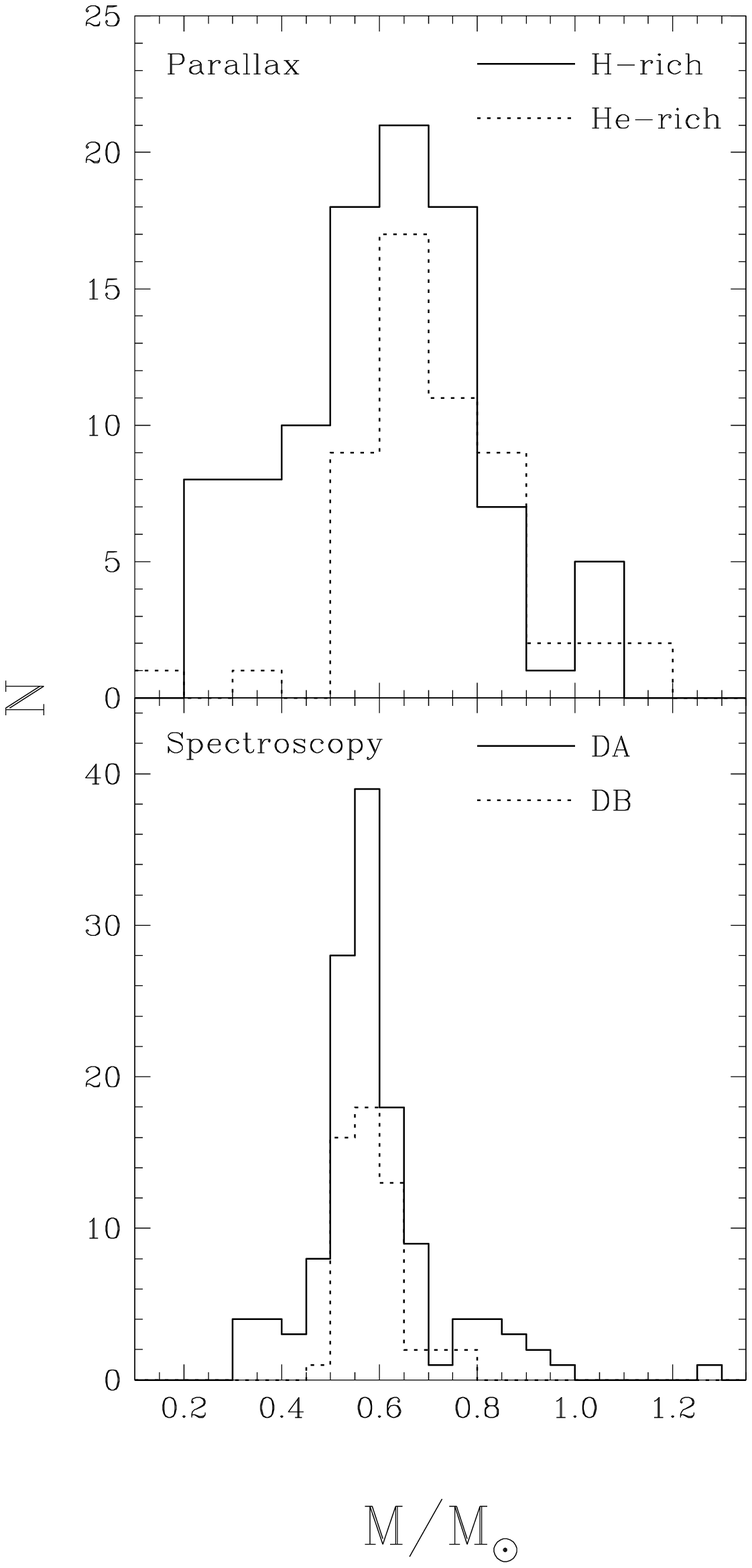] {Top panel: Mass distributions for the
hydrogen- and helium-rich atmosphere white dwarfs in our parallax
sample. The mean mass of the hydrogen-rich subsample is $\langle
M\rangle=0.61$ \msun\ with a dispersion of $\sigma (M)=0.20$ \msun,
and the corresponding values for the helium-rich subsample are $\langle
M\rangle=0.72$ \msun\ and $\sigma (M)=0.17$ \msun. Bottom panel: Mass
distributions for hotter DA and DB stars determined from spectroscopic
analyses. The mean mass and dispersion for the DA stars are $\langle
M\rangle=0.59$ \msun, $\sigma=0.13$ \msun, and for the DB stars
$\langle M\rangle=0.59$ \msun, $\sigma=0.06$ \msun.\label{fg:mass_distr}}

\figcaption[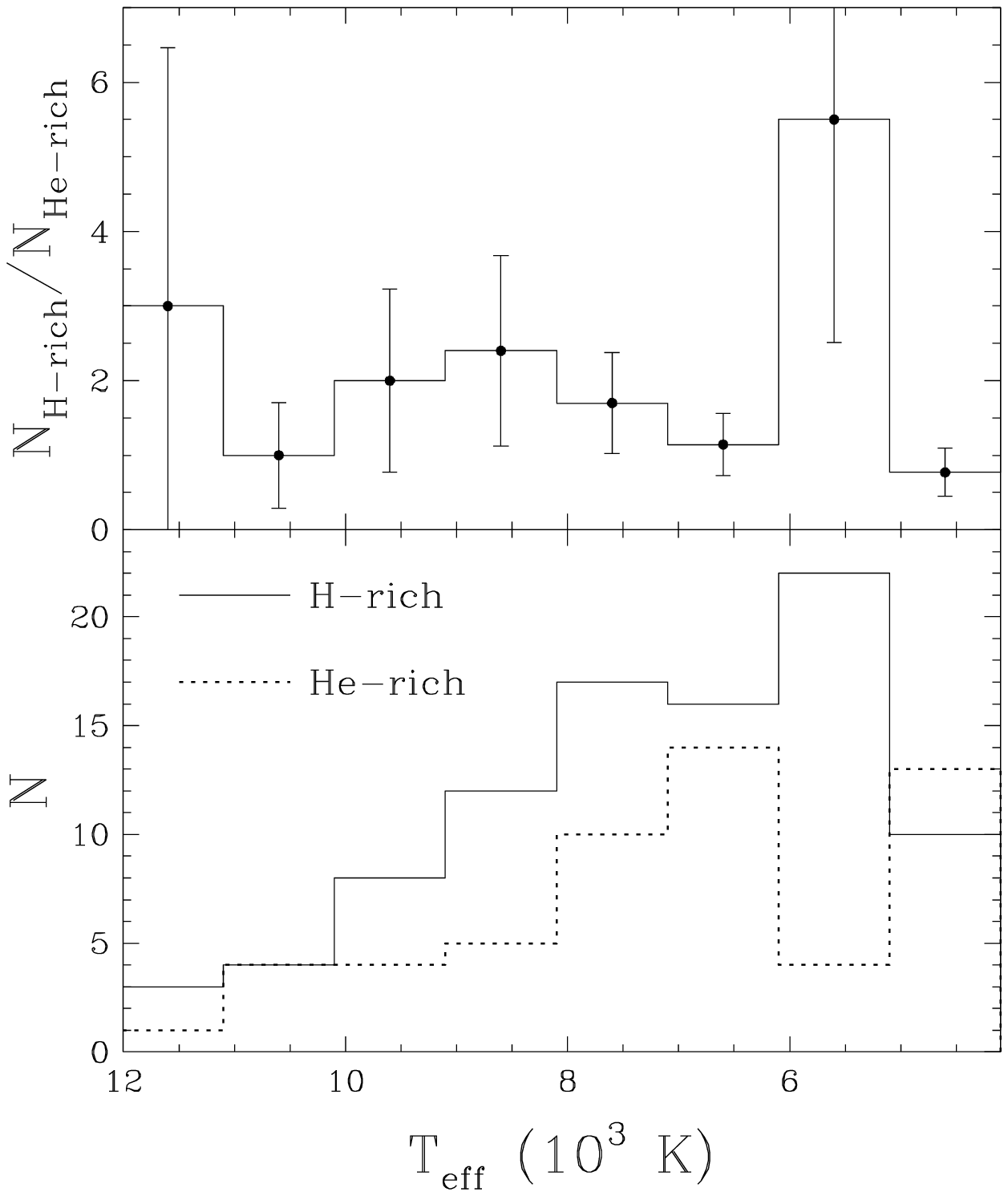] {Bottom panel: Number of hydrogen-rich 
({\it solid line}) and helium-rich ({\it dashed line}) stars as a
function of effective temperature in 1000~K bins for the results
presented in Table 2. Top panel: The corresponding hydrogen- to
helium-rich ratio; the error bars represent the Poisson statistics
of each bin.\label{fg:H_vs_He}}

\figcaption[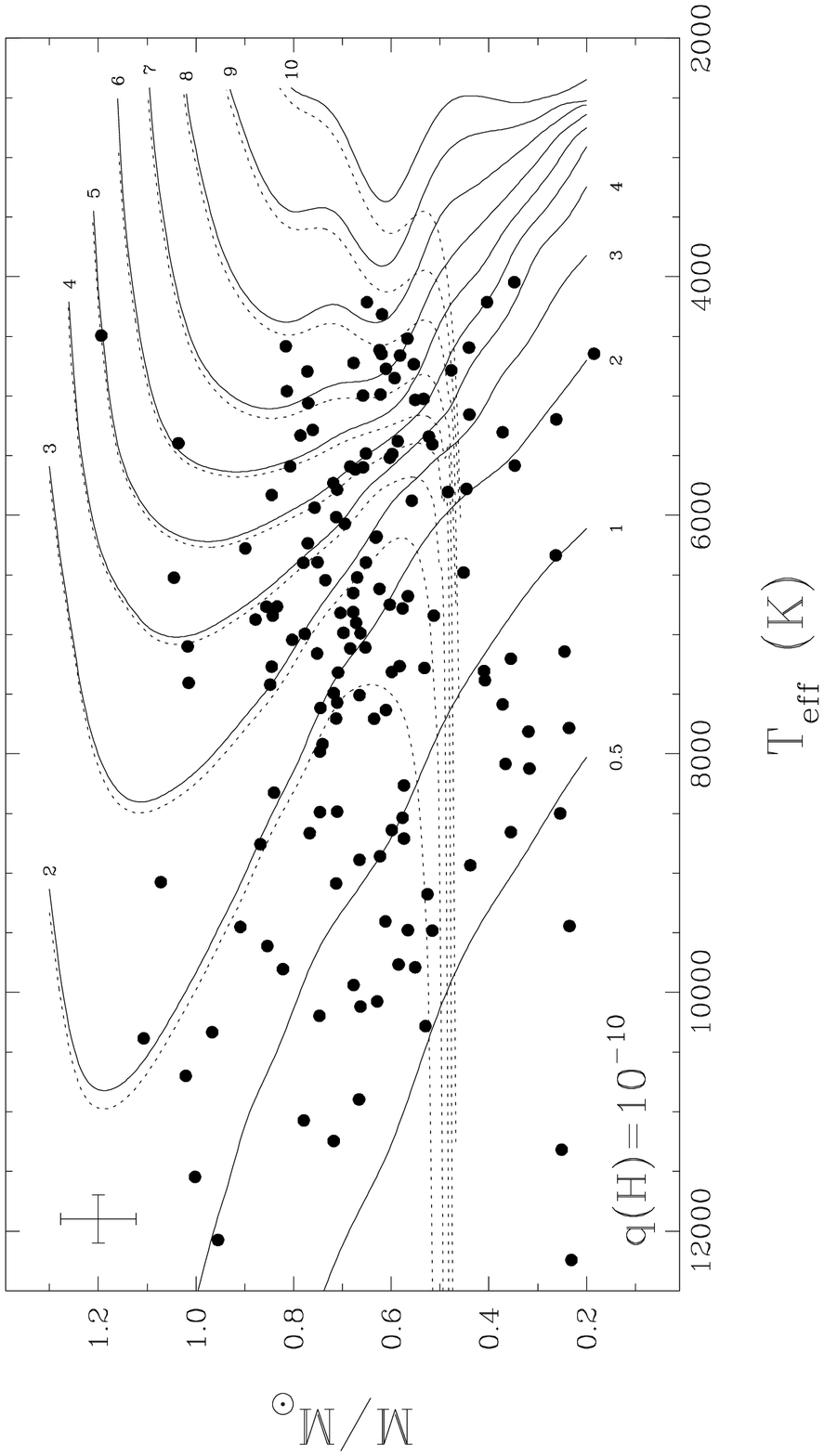] {Masses of white dwarfs in our 
trigonometric parallax sample as a function of effective temperature
together with the isochrones from the new cooling sequences with C/O
core compositions, $q({\rm He})=10^{-2}$, and $q({\rm H})=10^{-10}$
({\it solid lines}). The isochrones are labeled in units of $10^9$
years. Also shown are the corresponding isochrones with the main
sequence lifetime taken into account ({\it dotted lines}). The cross
in the upper left corner represents the average error in $\Te$ of
200~K and in mass of 0.077 \msun.\label{fg:isochrone_CO_thin}}

\figcaption[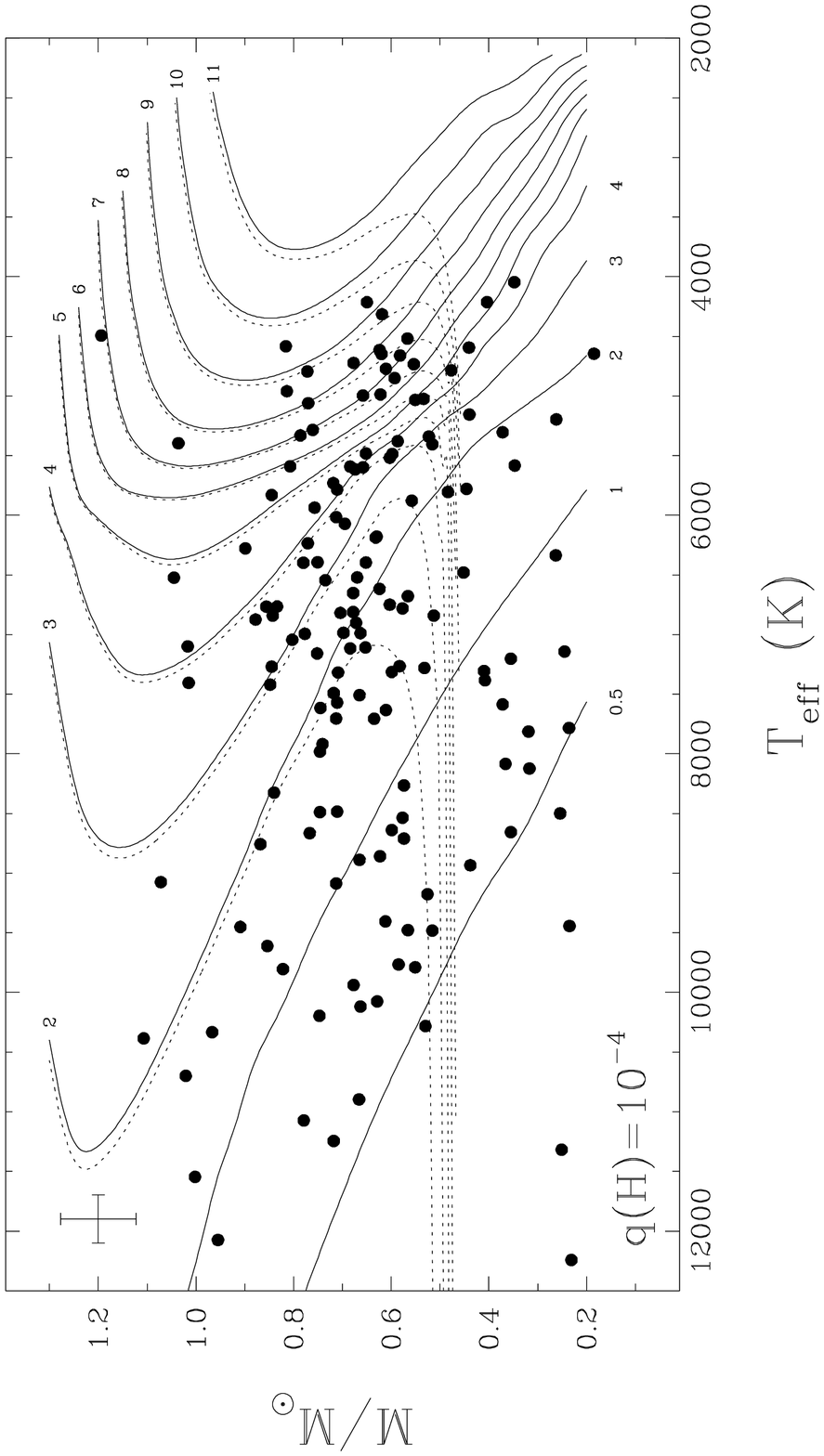] {Same as Figure 
\ref{fg:isochrone_CO_thin} but for cooling sequences with C/O core 
compositions, $q({\rm He})=10^{-2}$, and $q({\rm
H})=10^{-4}$.\label{fg:isochrone_CO_thick}}

\clearpage
\begin{figure}[p]
\plotone{f1.eps}
\begin{flushright}
Figure \ref{fg:comp_tp}
\end{flushright}
\end{figure}

\clearpage
\begin{figure}[p]
\plotone{f2.eps}
\begin{flushright}
Figure \ref{fg:compflux}
\end{flushright}
\end{figure}

\clearpage
\begin{figure}[p]
\plotone{f3.eps}
\begin{flushright}
Figure \ref{fg:DAvsnonDA}
\end{flushright}
\end{figure}

\clearpage
\begin{figure}[p]
\plotone{f4.eps}
\begin{flushright}
Figure \ref{fg:plotspecDA}
\end{flushright}
\end{figure}

\clearpage
\begin{figure}[p]
\plotone{f5.eps}
\begin{flushright}
Figure \ref{fg:plotspecnonDA}
\end{flushright}
\end{figure}

\clearpage
\begin{figure}[p]
\plotone{f6.eps}
\begin{flushright}
Figure \ref{fg:plotspecDQ}
\end{flushright}
\end{figure}

\clearpage
\begin{figure}[p]
\plotone{f7.eps}
\begin{flushright}
Figure \ref{fg:Mv:vmi}
\end{flushright}
\end{figure}

\clearpage
\begin{figure}[p]
\plotone{f8.eps}
\begin{flushright}
Figure \ref{fg:bmv:vmk}
\end{flushright}
\end{figure}

\clearpage
\begin{figure}[p]
\plotone{f9.eps}
\begin{flushright}
Figure \ref{fg:vmi:vmk}
\end{flushright}
\end{figure}

\clearpage
\begin{figure}[p]
\plotone{f10.eps}
\begin{flushright}
Figure \ref{fg:sampleDA}
\end{flushright}
\end{figure}

\clearpage
\begin{figure}[p]
\plotone{f11.eps}
\begin{flushright}
Figure \ref{fg:samplenonDA}
\end{flushright}
\end{figure}

\clearpage
\begin{figure}[p]
\plotone{f12.eps}
\begin{flushright}
Figure \ref{fg:sampleHe}
\end{flushright}
\end{figure}

\clearpage
\begin{figure}[p]
\plotone{f13.eps}
\begin{flushright}
Figure \ref{fg:sampledoubleDA}
\end{flushright}
\end{figure}

\clearpage
\begin{figure}[p]
\plotone{f14.eps}
\begin{flushright}
Figure \ref{fg:figMAG}
\end{flushright}
\end{figure}

\clearpage
\begin{figure}[p]
\plotone{f15.eps}
\begin{flushright}
Figure \ref{fg:sampleMAG}
\end{flushright}
\end{figure}

\clearpage
\begin{figure}[p]
\plotone{f16.eps}
\begin{flushright}
Figure \ref{fg:figG227-35}
\end{flushright}
\end{figure}

\clearpage
\begin{figure}[p]
\plotone{f17.eps}
\begin{flushright}
Figure \ref{fg:figESO439-26}
\end{flushright}
\end{figure}

\clearpage
\begin{figure}[p]
\plotone{f18.eps}
\begin{flushright}
Figure \ref{fg:figG107-70}
\end{flushright}
\end{figure}

\clearpage
\begin{figure}[p]
\plotone{f19.eps}
\begin{flushright}
Figure \ref{fg:TspecvsTphot}
\end{flushright}
\end{figure}

\clearpage
\begin{figure}[p]
\plotone{f20.eps}
\begin{flushright}
Figure \ref{fg:exampleHe}
\end{flushright}
\end{figure}

\clearpage
\begin{figure}[p]
\plotone{f21.eps}
\begin{flushright}
Figure \ref{fg:spec_evol}
\end{flushright}
\end{figure}

\clearpage
\begin{figure}[p]
\plotone{f22.eps}
\begin{flushright}
Figure \ref{fg:mass_distr}
\end{flushright}
\end{figure}

\clearpage
\begin{figure}[p]
\plotone{f23.eps}
\begin{flushright}
Figure \ref{fg:H_vs_He}
\end{flushright}
\end{figure}

\clearpage
\begin{figure}[p]
\plotone{f24.eps}
\begin{flushright}
Figure \ref{fg:isochrone_CO_thin}
\end{flushright}
\end{figure}

\clearpage
\begin{figure}[p]
\plotone{f25.eps}
\begin{flushright}
Figure \ref{fg:isochrone_CO_thick}
\end{flushright}
\end{figure}

\end{document}